\documentclass[prd,twocolumn,nopacs,nofootinbib,floatfix,superscriptaddress]{revtex4-1}

\usepackage[utf8]{inputenc}
\usepackage[T1]{fontenc}
\usepackage{graphicx}
\usepackage{amsmath}
\usepackage{amsfonts}
\usepackage{amssymb}
\usepackage[version=4]{mhchem}
\usepackage{stmaryrd}
\usepackage{graphicx}
\usepackage{amsbsy}
\usepackage{mathrsfs}
\usepackage{latexsym}
\usepackage{comment}
\usepackage{natbib}
\usepackage{bm}
\usepackage{subfigure} 
\usepackage{color}
\usepackage{wasysym}
\usepackage{mathbbol}
\usepackage{bigints}
\allowdisplaybreaks
\usepackage[normalem]{ulem}
\usepackage[dvipsnames]{xcolor}
\usepackage{multirow}
\usepackage{physics}
\usepackage[export]{adjustbox}
\usepackage[justification=raggedright]{caption}
\usepackage{subfigure}

\usepackage{hyperref,url}
\hypersetup{
	colorlinks=true,       % false: boxed links; true: colored links
	linkcolor=blue,          % color of internal links
	citecolor=blue,        % color of links to bibliography
	urlcolor=blue           % color of external links
}
\def\nn{\nonumber} 

\def\f{\frac}
\def\l{\left}
\def\r{\right}
\def\d{{\rm d}}
\def\sinc{{\rm sinc}\,}
\begin{document}

%%%%%%%%%%%%%%%%%%%%%%%%%%%%%%%%%%%%%%%%%%%%%%%%%%%%%%%%%%%%%%%%%%%%%%%%%%%%%%
\title{Gravitational wave detection via photon-graviton scattering and quantum interference}
%Quantum Gravitational-Wave Detection from Two-Photon Interference}

%
\author{K. Hari }
\email{hari.k@iitb.ac.in}
\author{S. Shankaranarayanan}
\email{shanki@iitb.ac.in}

\affiliation{Department of Physics,  Indian Institute of Technology Bombay, Mumbai 400076, India}
\begin{abstract}
\noindent 
We present a fully quantum field-theoretic framework for gravitational wave (GW) detection, describing the interaction as inelastic photon-graviton scattering. We demonstrate that for \emph{a coherent graviton state}, this microscopic model explicitly reproduces the standard classical geometric-optics time delay, providing a rigorous consistency test of the correspondence limit. We propose a detection scheme based on Hong-Ou-Mandel interference, where the GW-induced spectral shifts render entangled photon pairs distinguishable. This encodes the gravitational signal into the modulation of photon coincidence rates --- a quantum observable robust against laser intensity noise. Furthermore, we show that non-classical graviton states, such as squeezed states, induce unique decoherence signatures, establishing theoretical bounds on gravitational decoherence and defining the operational limits for probing the quantum nature of gravity.
\end{abstract}
\maketitle

%\noindent \underline{\emph{Introduction:}} 
\section{Introduction}
Almost all precision measurements involving electromagnetic (EM) radiation rely on its interaction with matter --- scattering processes that lead to absorption, emission, or phase shifts~\cite{Milonni:1994xx,Cohen-Tannoudji:1987oxb}. Canonical examples, such as Compton or Thomson scattering, are rigorously understood within quantum theory~\cite{Peskin:1995ev,Mandel:1995seg}, where both the radiation and the target are treated as quantum dynamical degrees of freedom.

In contrast, the detection of gravitational waves (GWs) is almost exclusively analyzed within a semi-classical framework. The EM field is treated as a classical probe of a Riemannian geometry, operationally described by variations in proper distance or optical path length between test masses~\cite{Saulson:2017jlf}. While this geometric optics approximation has been validated by the LIGO-Virgo-KAGRA collaboration~\cite{LIGOScientific:2016ao}, it suffers from fundamental limitations~\cite{Gefen:2022ztc,Yu:2022upc}. It is susceptible to gauge-dependent interpretations if applied naively~\cite{Finn:2008np,Rakhmanov:2004eh}, becomes inadequate in high-frequency regimes where the phonon/optical wavelength is comparable to the GW wavelength~\cite{Aggarwal:2020olq}, and, crucially, lacks a microscopic account of how linearized gravity couples to photons.

In this work, we bridge this gap by establishing a fully field-theoretic framework for GW detection. We treat the interaction \emph{not} as a classical metric perturbation stretching a baseline, but as a scattering process between quantized EM fields and gravitons. This approach addresses foundational questions regarding the coupling of gravity to quantum systems and the detectability of gravitons, a subject of longstanding theoretical interest~\cite{Rothman:2006fp,PhysRevD.10.461,Boughn:2006st,Hogan:2007pk, Dyson:2013hbl,Bringmann:2023gba,Grafe:2023ngy,Tobar:2023ksi,Carney:2023nzz,Arani:2026jyz}. Within our framework, the GW-induced signal emerges from an energy exchange --- analogous to the Stokes and anti-Stokes lines in Raman scattering --- where the absorption or emission of gravitons induces minute frequency shifts in the photon field~\cite{Schutzhold:2025vti,Pang:2018eec,Arani:2021oim,Arani:2023ljw,Arani:2025nfk}. To validate this microscopic description, we model the GW as a coherent state of gravitons and explicitly demonstrate that the cumulative phase shift acquired by the photon field strictly reproduces the standard geometric-optics time delay. To our knowledge, this provides the first explicit field-theoretic verification that a coherent graviton state correctly recovers the phenomenological General Relativistic response, serving as a consistency test for the quantized interaction model.

{Crucially, this microscopic perspective opens new avenues for detection beyond classical intensity interferometry. We propose a detection scheme based on Hong-Ou-Mandel (HOM) interference~\cite{Hong:1987}, a quintessential quantum phenomenon with no classical analogue. By exploiting the extreme sensitivity of two-photon quantum interference to distinguishability, we show that the \emph{gravitational phase} can be read out via changes in photon coincidence rates. This architecture shifts the detection paradigm from measuring the amplitude of a classical wave to monitoring the correlations of quantum states, offering a sensing modality inherently robust against classical intensity noise and providing a complementary window into the gravitational universe.}

{The search for relic GW backgrounds offers a new window into the micro-physics of the early universe. Crucially, the standard inflationary paradigm dictates that primordial gravitons are not produced in simple classical or coherent states. Instead, through parametric particle creation from the vacuum, they emerge as highly squeezed quantum states~\cite{Grishchuk:1974ny,Albrecht:1992kf}. This implies a massive average graviton occupation number per mode ($N_k = \sinh^2(r_k)$). Since, these gravitons interact so weakly with matter fields, this extreme squeezing parameter $r_k$ is potentially preserved and is considered as a unique signature in the redshifted GW background present at the current epoch~\cite{Ricciardone:2016ddg,Wang:2024gko,Bass:2026wzo,KAGRA:2021kbb}.}

Importantly, while alternative early-universe scenarios --- such as bouncing cosmologies with an initial matter-dominated contraction phase --- can be engineered to yield an almost identical scale-invariant primordial power spectrum, their underlying quantum state features are distinct~\cite{Chandran:2024utf}. Recent work has demonstrated that while standard two-point correlation functions (which underpin classical power spectra) are often degenerate between inflation and bounce scenarios, quantum information-theoretic measures can cleanly break this degeneracy. Specifically, the Dynamical Fidelity Susceptibility (DFS) of the primordial perturbations tracks the full quantum state space density matrix overlap rather than just field amplitudes, proving that the underlying quantum state phase space geometry retains a unique signature of the cosmic history~\cite{Chandran:2023ogt,Chandran:2024utf}.

Translating this theoretical necessity into a physical observable presents a formidable challenge to modern GW metrology. Traditional space- and ground-based interferometers (such as LISA or LIGO) function primarily as first-order ($g^{(1)}$) field-amplitude detectors. They map gravitational metric fluctuations onto a classical optical phase shift $\Delta \phi \propto \langle \hat{h}_{ij} \rangle$. While highly efficient for continuous-wave or stochastic classical strains, these $g^{(1)}$ frameworks are blind to the non-classical quantum statistics --- such as the squeezing phase $\theta_k$ or state purity --- of the graviton background. To isolate these signatures and distinguish between competitive early-universe paradigms like inflation and cosmic bounces, one must measure higher-order moments of the gravitational field.

In this work, we bridge this gap by introducing a quantum optics framework based on second-order intensity correlations ($g^{(2)}$) utilizing Hong-Ou-Mandel (HOM) interference. Because the HOM coincidence counting rate is driven by the fourth-order moment of the EM field ($\langle \hat{a}^\dagger \hat{a}^\dagger \hat{a} \hat{a} \rangle$) coupled via inelastic photon-graviton scattering, the resulting observable \emph{does not} register a classical time-delay. Instead, as we prove via a rigorous open-quantum-system approach, it maps the anomalous variance and quantum coherence of the squeezed graviton background directly onto a potentially measurable photon decoherence signature. This setup allows us to access the Quantum Fisher Information (QFI) of the spacetime metric, establishing a novel metrological paradigm capable of laboratory execution and specifically designed to measure the quantum nature of gravity.

This manuscript is organized as follows. In Sec.~\ref{sec:ph-gr-dynamics}, we quantize linearized metric perturbations and examine the matter-gravity interaction Hamiltonian, demonstrating that photon-graviton scattering induces kinematic, Raman-like photon frequency shifts. In Sec.~\ref{sec:evolution-of-ph-states}, we model the graviton background using general non-classical states ---specifically, squeezed number states --- to track the bipartite density matrix evolution. By tracing out the gravitational environment, we derive the reduced optical dynamics, proving that squeezed graviton states drive a visibility-degrading decoherence signature, whereas coherent states precisely recover classical interferometric phase shifts and General Relativistic antenna patterns. In Sec.~\ref{sec:towards-qauntum-detection}, we motivate the metrological transition from first-order field amplitudes to second-order intensity correlations ($g^{(2)}$). Building on this foundation, Sec.~\ref{sec:HOM} establishes a Hong-Ou-Mandel interference scheme that encodes gravitational phase shifts into photon distinguishability. In Sec.~\ref{sec:SNR}, we analyze the signal-to-noise ratio (SNR), experimental feasibility, and candidate detector geometries. Finally, Sec.~\ref{sec:conclusion} highlights the isotropic sensitivity advantages of a 3D pyramidal constellation and offers concluding remarks. Throughout this work, we adopt natural units $c = \hbar = 1$, where $\kappa = \sqrt{16\pi G}$ denotes the gravitational coupling constant.

\color{black}

\section{Dynamics of the Photon-Graviton interaction}\label{sec:ph-gr-dynamics}
To establish the feasibility of HOM-based detection, we derive the effective photon dynamics induced by its interaction with the graviton field. We work in the interaction picture and trace out the graviton degrees of freedom to obtain the reduced EM density matrix. 

The total Hamiltonian is:
%
%%%%%%%%%%%%%%%%
\begin{equation}
    H = H_{\mathrm{ph}} + H_{\mathrm{GW}} + H_{\mathrm{int}} .
\end{equation}
%%%%%%%%%%%%%%%%
%
%
where $H_{\mathrm{ph}}$ ($H_{\mathrm{GW}}$) is the free EM (graviton) Hamiltonian given by:
%
%%%%%%%%%%%%%%%%
\begin{equation}
H_{\mathrm{ph}} = \sum_{\mathbf{k},\lambda}  \omega_{\mathbf{k}}\, \hat{a}^{\dagger}_{\mathbf{k},\lambda} \hat{a}_{\mathbf{k},\lambda} \, , \, 
H_{\mathrm{GW}} = \sum_{\mathbf{q},\sigma} \omega_{\mathbf{q}}\, \hat{b}^{\dagger}_{\mathbf{q},\sigma} \hat{b}_{\mathbf{q},\sigma}~\, ,
\end{equation}
%%%%%%%%%%%%%%%%
%
where, $a_{\mathbf{k},\lambda}$ annihilates a photon of wave vector $\mathbf{k}$, polarization $\lambda$, and frequency $\omega_{\mathbf{k}} = |\mathbf{k}|$ and the vacuum energy is excluded. Similarly, $b_{\mathbf{q},\sigma}$ annihilates a graviton of wave vector $\mathbf{q}$ with polarization $\sigma$.

To explicitly define the field expansions, we must fix the gauge. We adopt the \emph{Transverse-Traceless (TT) gauge} for the gravitational field ($h_{0\mu}=0$, $h^i_i=0$, $\partial^j h_{ij}=0$) and the \emph{Coulomb gauge} for the EM field ($A^0=0$, $\nabla \cdot \mathbf{A} = 0$). In the TT gauge, the quantized metric perturbation is:
\begin{align}
    h_{ij}(\mathbf{x},t) = \sum_{\mathbf{q},\sigma}\frac{\kappa}{\sqrt{2\omega_\mathbf{q}V}} \Big[ b_{\mathbf{q},\sigma} e^{i(\mathbf{q}\cdot\mathbf{x}-\omega_{\mathbf{q}} t)} \nn \\
    + b^{\dagger}_{\mathbf{q},\sigma} e^{-i(\mathbf{q}\cdot\mathbf{x}-\omega_{\mathbf{q}} t)} \Big] \varepsilon^{(\mathbf{q},\sigma)}_{ij}~.
    \label{eqn:hij-def}
\end{align}
where $V$ is the volume associated with box normalization and $b_{\mathbf{q},\sigma}$ annihilates a graviton with polarization $\sigma$.  Similarly, in the Coulomb gauge, the vector potential is:
\begin{align}
    A^{i}(\mathbf{x},t)= \sum_{\mathbf{k},\lambda}\frac{1}{\sqrt{2\omega_\mathbf{k}V}} \Big[ a_{\mathbf{k},\lambda} e^{i(\mathbf{k}\cdot\mathbf{x}-\omega_{\mathbf{k}} t)} \nn \\
    + a^{\dagger}_{\mathbf{k},\lambda} e^{-i(\mathbf{k}\cdot\mathbf{x}-\omega_{\mathbf{k}} t)} \Big] \varepsilon^{(\mathbf{k},\lambda)}_{i}~.
\end{align}

\begin{figure}[h]
\includegraphics[width=8.5cm]{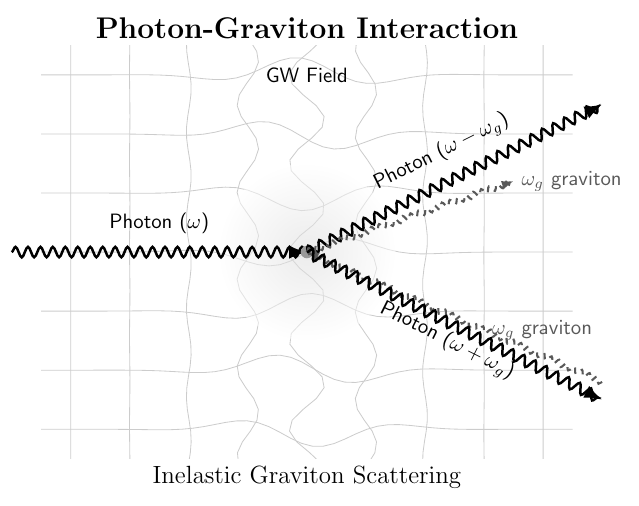}
    \caption{\label{fig1} Schematic diagram of the inelastic graviton scattering.}
\end{figure}

In the linearized gravity regime, the interaction between the EM stress-energy tensor $T^{ij}_{\text{EM}}$ and the metric perturbation $h_{ij}$ (via minimal coupling) is~\cite{Gupta:1954abc,Feynman:1996kb}:
\begin{equation}
H_{\mathrm{int}} = \frac{1}{2} \int d^3x\; h_{ij}(\mathbf{x},t)\, T^{ij}_{\mathrm{EM}}(\mathbf{x},t)~,
\end{equation}
where $T^{ij}_{\mathrm{EM}} = \l(F^{i\mu} F_{\mu}^{j} - \frac{1}{4} \delta^{ij} F_{\mu \nu}F^{\mu\nu}\r)$. 
%For explicit derivation, see Appendix~\ref{app:ph-gr-int-hamiltonian}.
%
Expanding the fields in terms of creation ($\hat{a}^\dagger, \hat{b}^\dagger$) and annihilation ($a, b$) operators, the interaction Hamiltonian is:
\begin{widetext}
\begin{align}
    H_{\mathrm{int}} = \frac{\kappa}{2}\sum_{\mathbf{k},\mathbf{k}^\prime,\mathbf{q}} \sum_{\sigma,\lambda,\lambda^\prime} \frac{\sqrt{\omega_\mathbf{k}\,\omega_{\mathbf{k}^\prime}}}{2(2\pi)^{3/2} \sqrt{\omega_{\mathbf{q}}}} \, \mathcal{G}^{\sigma,\lambda,\lambda^\prime}_{\mathbf{k},\mathbf{k}^\prime,\mathbf{q}} \times \bigg[ & a^\dagger_{\mathbf{k},\lambda} a_{\mathbf{k}^\prime,\lambda^\prime} b_{\mathbf{q},\sigma} \, e^{-i(\omega_{\mathbf{k}}-\omega_{\mathbf{k}^\prime} -\omega_{\mathbf{q}})t} \, \delta^3(\mathbf{k}-\mathbf{k}^\prime-\mathbf{q}) \nonumber \\
    & \quad + a^\dagger_{\mathbf{k},\lambda} a_{\mathbf{k}^\prime,\lambda^\prime} b^\dagger_{\mathbf{q},\sigma} \, e^{-i(\omega_{\mathbf{k}}-\omega_{\mathbf{k}^\prime} +\omega_{\mathbf{q}})t} \, \delta^3(\mathbf{k}-\mathbf{k}^\prime+\mathbf{q}) \bigg]~,
    \label{eqn:int-hamiltonian-1}
\end{align}
\end{widetext}
where, 
{\small 
$$ \, {\mathcal{G}}^{(\sigma,\lambda,\lambda^\prime)}_{\hat{\mathbf{k}},\hat{\mathbf{k}^\prime},\hat{\mathbf{q}}} =\l[ \varepsilon^{*i}_{(\hat{\mathbf{k}},\lambda)} \varepsilon^{j}_{(\hat{\mathbf{k}}^\prime,\lambda^\prime)} 
+ (\hat{\mathbf{k}}\times \varepsilon_{(\hat{\mathbf{k}},\lambda)})^{*i} (\hat{\mathbf{k}}^\prime\times\varepsilon_{(\hat{\mathbf{k}}^\prime,\lambda^\prime)})^j\r] \varepsilon_{ij}^{(\hat{\mathbf{q}},\sigma)}
$$
}
\hspace*{-8pt} is the geometric polarization factor. The momentum-conserving delta functions has an energy resonance condition $\omega_{\mathbf{k}} - \omega_{\mathbf{k}^\prime} = \pm \omega_{\mathbf{q}}$.
As schematically shown in Fig.~\ref{fig1}, this describes an \textit{inelastic scattering process} where the photon gains or loses energy $\omega_{\mathbf{q}}$ by absorbing or emitting a graviton. This mechanism is formally analogous to the Stokes (emission) and anti-Stokes (absorption) shifts in Raman spectroscopy, where the GW acts as a pump field modulating the photon energy~\cite{Schutzhold:2025vti}. 
{Unlike traditional molecular Raman scattering, where the target possesses internal, discrete electronic or vibrational quantum states, the graviton field alters the photon via continuous space-time translations, making the resulting frequency shift entirely kinematic and directional ($\Omega_q = \omega_q(1 - \hat{k}\cdot\hat{q})$).}
As we show in our proposal using quantum interference this energy exchange (graviton absorption or emission) will cause frequency shift of photons leading to an additional phase for the photons in the interferometric experiment. 

For GWs of astrophysical interest, $\omega_{\mathbf{q}} \ll \omega_{\mathbf{k}}$. Consequently, the scattering is quasi-elastic: the photon direction is largely preserved ($\hat{\mathbf{k}} \approx \hat{\mathbf{k}}^\prime$), and the frequency shift is small. Using the approximation $\omega_{\mathbf{k}} - \omega_{\mathbf{k}^\prime} \approx \omega_{\mathbf{k}} - |\mathbf{k} \mp \mathbf{q}| \approx \pm \mathbf{q} \cdot \hat{\mathbf{k}}$, the phase arguments simplify to:
\begin{equation}
    (\omega_{\mathbf{k}} - \omega_{\mathbf{k}^\prime} \mp \omega_{\mathbf{q}})t \approx \mp \omega_{\mathbf{q}}(1 - \hat{\mathbf{k}} \cdot \hat{\mathbf{q}})t \equiv \mp \Omega_{\mathbf{q}} t~.
\end{equation}
The effective interaction Hamiltonian governing the phase evolution is then:
\begin{multline}
H_{\text{int}} \approx \frac{\kappa }{2}\sum_{\mathbf{k},\mathbf{q}} \sum_{\sigma,\lambda} \frac{\omega_\mathbf{k}}{(2\pi)^3 2 \sqrt{\omega_{\mathbf{q}}}} \, {\mathcal{G}}^{\sigma,\lambda}_{\hat{\mathbf{k}},\hat{\mathbf{q}}} \times\\
    \left[ \hat{n}_{\mathbf{k},\lambda} \left(\hat{b}_{\mathbf{q},\sigma} e^{i\Omega_{\mathbf{q}} t} + \hat{b}^{\dagger}_{\mathbf{q},\sigma} e^{-i\Omega_{\mathbf{q}} t} \right) \right]
    \label{eqn:Hint}
\end{multline}
where $\Omega_\mathbf{ q} =\omega_{\mathbf{q}}(1-\hat{\mathbf{k}}\cdot \hat{\mathbf{q}})$. 
Equation~\eqref{eqn:Hint} is obtained by summing over $\mathbf{k}^\prime$ where delta function $\delta(\mathbf{k}-\mathbf{k}^\prime\pm \mathbf{q})$ is involved for each terms and then using
 $\omega_{\mathbf{k}}-\omega_{\mathbf{k}^\prime} \approx \mp \omega_{\mathbf{q}} \hat{\mathbf{k}}\cdot \hat{\mathbf{q}}$.
This interaction reveals a \textit{graviton-photon energy exchange}~\cite{Saulson:1997vi,Ma:2014pqa}. {Just as in Raman spectroscopy, where photons undergo inelastic scattering with molecular vibrations (Stokes and anti-Stokes lines), here the photon gains or loses energy by absorbing or emitting a graviton. For GW detection, where $\omega_{\mathbf{k}}\gg\omega_{\mathbf{q}}$, this manifests as frequency shift, which as we show below can be  treated as an accumulative phase $\phi_{\mathbf{k}}(t)$. 
While the Hamiltonian explicitly contains energy-exchange terms, the macroscopic time delay induced by a coherent GW state emerges primarily from elastic forward scattering\footnote{While the interaction Hamiltonian is evaluated in the TT gauge for computational convenience, the resulting two-photon coincidence probability measured at the detector ports represents a
coordinate-independent, gauge-invariant physical observable.}. The inelastic energy-exchange signatures only become dominant when probing non-classical (e.g., squeezed) graviton states, as discussed soon.}

While Eq.~\eqref{eqn:Hint} describes a fundamentally microscopic, inelastic energy-exchange process between single photons and gravitons, it establishes the foundation for macroscopic GW detection. As we will demonstrate in the next section, when the background gravitational field is modeled as a highly populated coherent state --- the quantum analog of a classical GW --- the expectation value of the interaction phase accumulated by the photon field recovers the classical geometric time delay. Crucially, the geometric polarization factor $\mathcal{G}_{\hat{\mathbf{k}},\hat{\mathbf{q}}}^{(\sigma,\lambda)}$ defined above acts as the mathematical bridge; its tensor contraction over photon polarizations directly yields the standard quadrupolar antenna patterns (e.g., $h_{xx} - h_{yy}$) utilized in macroscopic interferometry.

%\noindent \underline{\emph{Emergence of the Classical Phase:}} 
\section{Evolution of photon states under graviton energy exchange}\label{sec:evolution-of-ph-states}
%

%\color{blue}
To study the effect of graviton-photon interaction, as depicted in Fig.~\ref{fig1}, we assume the initial state to be a product state $\hat{\rho}(0) = \hat{\rho}_{\text{ph}}(0)\otimes\hat{\rho}_{\text{gw}}(0)$. By considering the photon initial state as a $n$ photon Fock state, i.e., $\hat{\rho}_{\mathrm{gw}}(0)=\rho_{nn^\prime}\ket{n}\bra{n}$, the graviton state as a generic quantum state given by squeezed number state -- $\ket{l,\gamma,\beta} = \hat{S}(\gamma)\hat{D}(\beta)\ket{l}$~\cite{mvs:prd32}, and evaluating the time-evolution operator $U_I(t)$, we obtain the following density matrix:
\begin{equation}
\label{eqn:evolution-1}
\!\!\!\!\hat{\rho}(t) = \sum%_{n,n^\prime,m,m^\prime} 
\rho_{nn^\prime}\ket{n}\bra{n^\prime} \otimes \hat{D}(\widetilde{\alpha}_n) \ket{l,\gamma,\beta} \bra{l,\gamma,\beta} \hat{D}^\dagger(\widetilde{\alpha}_{n^\prime})   
\end{equation}
where the displacement operator $D(\widetilde{\alpha}_{n})$ denotes the unitary time evolution operator with $\widetilde{\alpha}_{n}:={\widetilde{\alpha}}^{\sigma}_{\mathbf{q}} = \sum_{\mathbf{k},\lambda,\lambda^\prime} \omega_\mathbf{k}\left(2(2\pi)^3 \sqrt{2\omega_{\mathbf{q}}}\right)^{-1} \, \hat{n}_{\mathbf{k},\lambda,\lambda^\prime} \, \alpha^{\sigma,\lambda,\lambda^\prime}_{\mathbf{k},\mathbf{q}}(t)$. To obtain the photon evolution, we evaluate the reduced density matrix (details in Appendix \eqref{app:red-photon-dyn}) as:
\begin{widetext}
\begin{equation}
    \!\!\!\!\hat{\rho}_{\mathrm{ph}}(t)= \sum_{n,n^\prime} \rho_{nn^\prime} \ket{n}\bra{n^\prime} e^{i\mathrm{Im}(\widetilde{\alpha}_n\widetilde{\alpha}^{*}_{n^\prime})} e^{2i\mathrm{Im}(\zeta_{nn^\prime} \beta^*)}e^{-\frac{1}{2}|\zeta_{nn^\prime}|^2} L_l (|\zeta_{nn^\prime}|^2 )~.
\label{eqn:evolution-final}
\end{equation}
\end{widetext}
where,  
%\textcolor{red}{for the forward scattering process, $k\approx k^\prime$, the phase factor $\mathrm{Im}(\widetilde{\alpha}_n\widetilde{\alpha}^{*}_{n^\prime})$ will vanish} and
$L_{l}(|\zeta_{nn^\prime}|^2)$ denotes the Laguerre polynomial resulting from the action of the displacement operator on the number state, and $\zeta_{nn^\prime}$ is defined as (see Appendix \eqref{app:red-photon-dyn}) $\zeta_{nn^\prime} = (\widetilde{\alpha}_n-\widetilde{\alpha}_{n^\prime}) \cosh{r} + (\widetilde{\alpha}_n-\widetilde{\alpha}_{n^\prime})^{*} e^{i\varphi} \sinh{r}$ with $\{r,\varphi\}$ being the magnitude of squeezing and $\varphi$ is the phase angle of the complex squeezing parameter $z = r e^{i\varphi}$. The above equation is central to our analysis, and we emphasize the following key aspects:
First, the term  $2i\Im{\sum_{\mathbf{q}}\zeta^{\sigma}_{\mathbf{q},nn^\prime}\beta^*_{\mathbf{q}}}$ is purely imaginary and hence leads to the additional phase in photons due to its interaction with gravitons. 
Consequently, it is clear from Eq.~\eqref{eqn:evolution-final} that the photon state $\ket{n}$ evolves in time as, $\ket{n}\to e^{2i \Im{\sum_\mathbf{q}{\zeta}^{\sigma}_{\mathbf{q},n} \beta_{\mathbf{q}}^*}}\ket{n}$. 

Second, the remaining terms, including the exponential decay factor, $e^{-\frac{1}{2}|{\zeta}_{nn^\prime}|^2}$, represent gravitational decoherence. These decay terms arises because the interaction creates entanglement between the photons and the graviton environment. If the graviton  \emph{measures} which path the photon took (by picking up a distinct state ${\zeta}_{nn^\prime}$), the photon's quantum coherence is suppressed. In the \textit{weak-field limit}, the coupling constant (contained in ${\zeta}_{nn^\prime}$) is extremely small. Hence, the information about the path is negligible ($|{\zeta}_{nn^\prime}|^2 \ll 1$), meaning the decay factor is approximately unity, and the evolution remains unitary for typical astrophysical GW detections.

However, isolating the decoherence terms becomes crucial when probing the quantum nature of gravity. Specifically, interactions with non-classical graviton field states, such as Fock or squeezed states, can imprint distinct decoherence signatures on the interferometric signal.

\subsection{Squeezed States for Gravitons}
\label{sec:squeezed_gravitons}

To analyze the impact of non-classical gravitational backgrounds, we consider a vacuum-squeezed graviton state $\hat{\rho}_{\mathrm{gw}} = \hat{S}(\gamma)\ket{0}\bra{0}\hat{S}^\dagger(\gamma)$, where 
$$\hat{S}(\gamma) = \exp\left[\frac{1}{2}\left(\gamma^* \hat{b}^2 - \gamma \hat{b}^{\dagger 2}\right)\right]$$ 
is the squeezing operator characterized by the complex parameter $\gamma = r e^{i\theta}$. Formally, this corresponds to taking the unexcited Fock limit ($m \to 0$) in the general squeezed-number state framework. Assuming the same initial optical state as before, the unitary displacement operator acts on the bipartite system to yield the evolved state:
\begin{align}
    \hat{\rho}(t) = \sum_{n,n^\prime} \rho_{nn^\prime} \ket{n}\bra{n^\prime} \otimes \hat{D}(\widetilde{\alpha}_n) \hat{S}(\gamma)\ket{0}\bra{0}\hat{S}^\dagger(\gamma) \hat{D}^\dagger(\widetilde{\alpha}_{n^\prime}) \, .
\end{align}
Tracing out the environmental graviton degrees of freedom gives the reduced photon density matrix:
{\small{
\begin{align}
    \hat{\rho}_{\mathrm{ph}}(t) = \sum_{n,n^\prime} \rho_{nn^\prime} \ket{n}\bra{n^\prime} {\underbrace{\exp\left( i \Im \left[ \widetilde{\alpha}_n \widetilde{\alpha}^{*}_{n^\prime}  \right] \right)}_{\text{Unitary Phase (Signal)}}} \underbrace{\exp\left( - \frac{1}{2} |\zeta_{nn^\prime}|^2 \right)}_{\text{Decoherence (Noise)}} ~,
    \label{eqn:rho_evolved_squeezed}
\end{align}
}}
\noindent 
where the effective displacement parameter $\zeta_{nn^\prime}$ encapsulates the quantum variance of the squeezed background:
\begin{equation}
    \zeta_{nn^\prime} = (\widetilde{\alpha}_{n}-\widetilde{\alpha}_{n^\prime})\cosh(r) + (\widetilde{\alpha}_{n}-\widetilde{\alpha}_{n^\prime})^* e^{i\theta}\sinh(r) \, .
    \label{eqn:zeta_def}
\end{equation}
Here, $r$ represents the squeezing amplitude, $\theta$ denotes the squeezing phase angle, and $\widetilde{\alpha}_n$ is the photon-number-dependent displacement eigenvalue. 

Equation~\eqref{eqn:rho_evolved_squeezed} is the first key result of this work. It  demonstrates that non-classical features in the gravitational field alter the optical state dynamics. While a classical or coherent gravitational background ($\beta \neq 0$) induces a purely unitary phase accumulation, quantum squeezing ($r > 0$) exponentially amplifies the effective displacement distance $|\zeta_{nn^\prime}|$. Physically, this occurs because squeezing stretches quantum field fluctuations along a preferred phase-space quadrature, strengthening the interaction between the optical modes and the background metric perturbations.

{Please note since the gravity coupling is tiny~\cite{Unruh1980}, as discussed in   Appendix~\eqref{app:interf_visibility}, even with large enhancement the detectability can be masked by other noises, including stochastic gravitational wave background~\cite{KAGRA:2021kbb}. For instance, a Gaussian graviton background (like squeezed states) acts on the photons as a classical stochastic field with the same quadrature variance \cite{Carney:2023nzz,Miyauchi:2026cnt}, so the visibility loss cannot be distinguished from thermal noise without other independent diagnostics such as detector cross-correlation, as in QUEST/GQuEST experiment~\cite{quest,gquest}. To validate non-classicality would require $|\zeta|\gtrsim0.1$ or non-Gaussian features (the Laguerre factor in Eq.~(10)). %\textcolor{red}{Our results should therefore be read as a framework for quantifying this decoherence rather than as a claim of observability in the present setup.}}

\begin{figure}
\includegraphics[width=\linewidth]{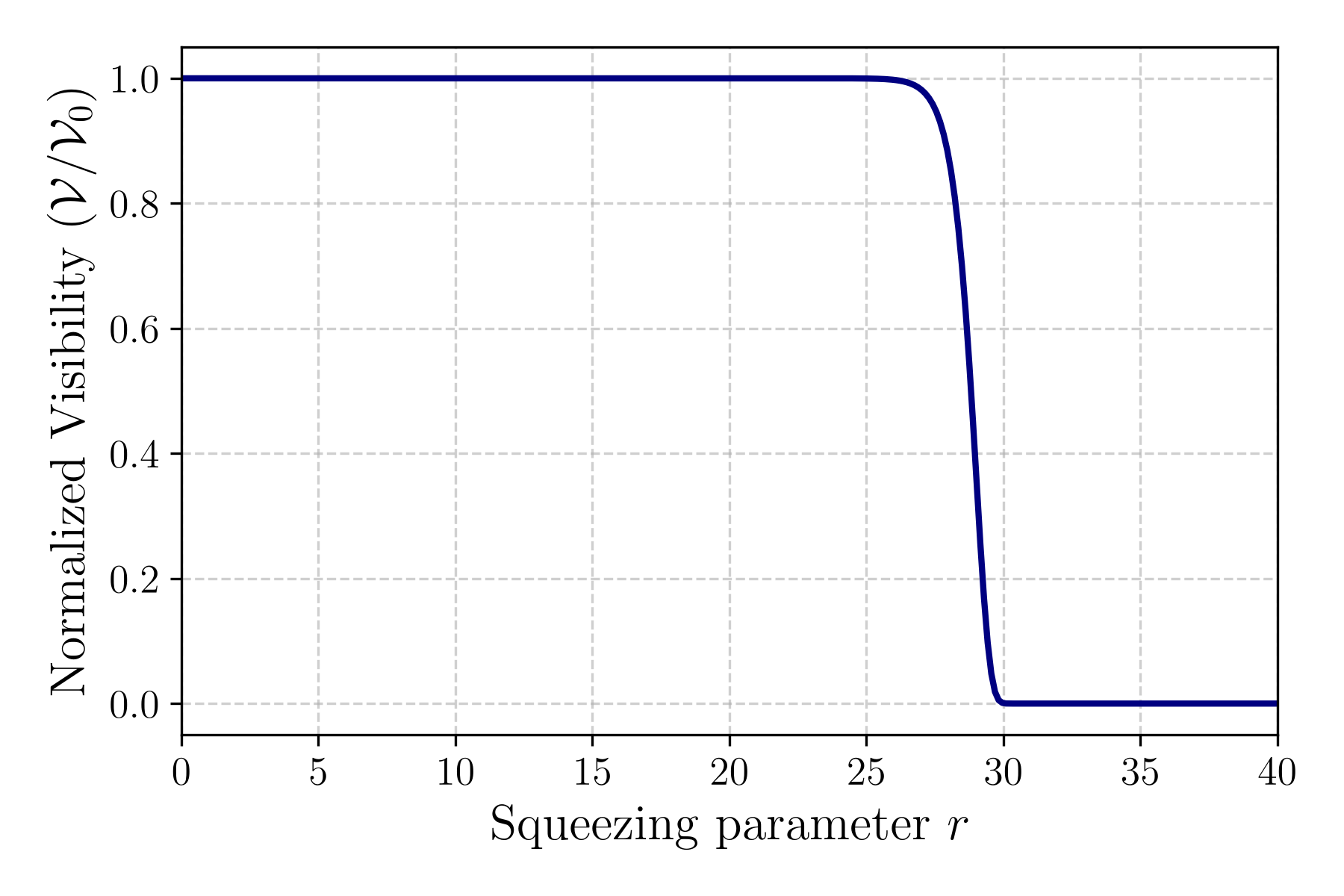}
\caption{Interferometric visibility as a function of squeezing amplitude for deci-Hertz frequency GW ($\sim 10\,\mathrm{Hz}$). For the plot, we have chosen the photon frequency to be 400 THz, the squeezing angle to be $\pi/4$, the arm length to be $10^7~\mathrm{m}$, {the photon flux to be $10^{12}$ photons/s, and the integration time to be 1 day (see Appendix~\ref{App:zeta} for details).}}
    \label{fig:visibility-squeezing}
\end{figure}

\color{black}
Figure~\ref{fig:visibility-squeezing} illustrates the resulting behavior of the normalized \textit{interferometric visibility} $\mathcal{V}/\mathcal{V}_0$ as a function of the squeezing parameter $r$ (see Appendix~\ref{app:interf_visibility} for detailed derivations). In the weak-squeezing regime ($r \ll 1$), the decay factor satisfies $|\zeta_{nn^\prime}|^2 \ll 1$, rendering gravitational decoherence negligible and preserving maximum optical fringe contrast ($\mathcal{V}/\mathcal{V}_0 \approx 1$). However, as $r$ increases, the hyperbolic factors $\cosh(r)$ and $\sinh(r)$ in Eq.~\eqref{eqn:zeta_def} trigger a sharp, non-linear suppression of off-diagonal density matrix elements, causing the visibility to fall off rapidly toward zero. This contrast degradation serves as a direct macroscopic signature of quantum entanglement with the background metric, wherein the squeezed graviton field effectively extracts \emph{which-path} information from the traversing optical photon pair.

\color{black}

\subsection{Coherent states of Gravitons and emergence of classical phase}
\label{sec:Coherentstates}
To connect our quantum scattering formalism with standard GW literature~\cite{Saulson:2017jlf}, we consider the limit of a coherent macroscopic GW. We approximate the graviton field state $\ket{\beta}$ by a coherent amplitude~\cite{Kanno:2025how} sharply peaked at the GW frequency $\omega_{\mathrm{gw}}$, such that $\kappa\,\beta^{(\sigma)}_{\mathbf{q}}/\sqrt{2\omega_{\mathbf{q}}V} \approx h^{(\sigma)}_0 \,\delta_{\omega_{\mathbf{q}},\omega_{\mathrm{gw}}}$. In the single-photon sector relevant to our interference scheme ($\ket{n_{\mathbf{k}}=1}$), the evolution is characterized by the phase acquired by the creation operator, $a^\dagger_{\mathbf{k}} \to a^\dagger_{\mathbf{k}} e^{i\phi_\mathbf{k}(t)}$, where:
%%%%%%%%%%%%%%%%
\begin{align}
\label{eqn:phase-time-delay}
    \phi_{\mathbf{k}}(t) 
    & = \frac{1}{2}\Im{\sum_{\lambda}\frac{\omega_\mathbf{k}\,{\mathcal{G}}^{\sigma,\lambda}_{\hat{\mathbf{k}},\mathbf{q}_\mathrm{gw}}}{\Omega_{\mathrm{gw}}}(1-e^{-i\Omega_\mathrm{gw} t}) h_0^{(\sigma)} }~,
\end{align}
%%%%%%%%%%%%%%%%
%
and $\Omega_{\mathrm{gw}} = \omega_{\mathrm{gw}}(1 - \hat{\mathbf{k}} \cdot \hat{\mathbf{q}})$. We identify the term $(1-e^{-i\Omega_{\mathrm{gw}} t})/\Omega_{\mathrm{gw}}$ as the time integral of the interaction phase factor $-i\int_0^t dt' e^{-i\Omega_{\mathrm{gw}} t'}$. Appendix \ref{app:coherent-state} for details. This allows us to recast the accumulated phase into a form directly comparable to classical interferometry~\cite{Finn:2008np,Saulson:2017jlf}:
\begin{equation}
\phi_{\mathbf{k}}(t) = -\omega_\mathbf{k} \left( \frac{1}{2} \int_0^t dt' h_{\mathrm{eff}}(t') \right)~.
\label{eqn:effective-phase-time-delay}
\end{equation}
Here, $h_{\mathrm{eff}}(t) = \sum_{\lambda} {\mathcal{G}}^{\sigma,\lambda}_{\hat{\mathbf{k}},\hat{\mathbf{q}}}\,e^{-i\Omega_{\mathbf{q}}t} h_0^{(\sigma)}$ encapsulates the geometric projection of the metric perturbation onto the photon polarization basis, recovering the standard antenna pattern functions (see Appendix \ref{app:antenna-pattern-geo-factor} for details) derived in general relativity~\cite{Finn:2008np}.

\hspace*{7pt} Equation~\eqref{eqn:effective-phase-time-delay} is the second key result of this work regarding we would discuss the following points. First, it demonstrates that the accumulated quantum phase shift arising from graviton-photon scattering is mathematically identical to the phase shift induced by variations in optical path length in the classical description. In observatories like LIGO-VIRGO-KAGRA~\cite{LIGOScientific:2014pky,VIRGO:2014yos,Aso:2013eba}, this phase is operationally interpreted as a time delay $\Delta \tau = \phi_{\mathbf{k}}/\omega_{\mathbf{k}}$. Our derivation reveals that this \emph{time delay} can be explained as a cumulative record of inelastic energy exchanges between the photons and gravitons. While consistent in the classical limit, this field-theoretic perspective is essential for describing interferometers where the quantum state of light (e.g., entanglement or squeezing) plays a non-trivial role in the detection statistics~\cite{Schnabel:2010rha,Dimopoulos:2008sv,Badurina:2019hst}. 

{Second, to physically bridge the microscopic photon-graviton scattering dynamics ($H_{\mathrm{int}}$) to the macroscopic time-delays of these large-scale observatories, requires rigorous mapping that recovers the classical GW strain from the expectation value of the quantum phase. Specifically, the classical metric tensor projections (e.g., $h_{xx} - h_{yy}$) that define the standard antenna patterns should emerge directly from the geometric tensor contraction encapsulated by $\sum_{\lambda}\mathcal{G}_{\hat{\mathbf{k}},\hat{\mathbf{q}}}^{(\sigma,\lambda)}$.  In Appendix \ref{app:antenna-pattern-geo-factor}, we have explicitly derived the same. Lastly, since the detector's sensitivity is entirely governed by these directional geometric projections, optimizing the spatial architecture is critical. We will discuss more about the configurations in Sec.~\ref{sec:MicroMacrobridge}.}

Having established that the interaction between gravitons and photons induces a frequency shift, effectively manifesting as a phase accumulation or time delay in the photon propagation, a detection scheme sensitive to such frequency-dependent phase shifts is required to observe this effect. In the rest of this work, we propose using Hong-Ou-Mandel (HOM) interference~\cite{Hong:1987}, a two-photon quantum phenomenon, as the probe.

HOM has certain advantages over classical Michelson interferometry. In a standard Michelson interferometer, the GW signal is transduced into optical intensity modulation, which is susceptible to classical laser power fluctuations and requires complex stabilization loops. In contrast, our proposed scheme encodes the gravitational interaction into the quantum distinguishability of the photon pair.
The observable here is the photon coincidence rate (a second-order correlation function, $g^{(2)}$), not the field intensity ($g^{(1)}$). Since the HOM dip relies on the destructive quantum interference of amplitudes rather than the stability of macroscopic fringes, the measurement is robust against classical intensity noise. The GW acts as a \emph{distinguisher} ---inducing a spectral shift that partially destroys the quantum interference --- allowing us to probe the interaction through photon statistics rather than phase-locked fringes.

\section{Towards the Quantum Detection of Gravitons}\label{sec:towards-qauntum-detection}
%
%\color{blue}

Current and planned GW detectors based on photon interference (such as LIGO and LISA) capture the phase difference between photons traveling in different arms due to their interaction with GWs (or gravitons). To measure this phase difference, the amplitude correlations of the photon field are used. This correlation is referred to as the first-order coherence, $g^{(1)}$, which is defined by \cite{Glauber:1963tqt,Loudon:tqt,Fox:2006qo},
\begin{equation}
    g^{(1)}(\mathbf{r}_1,\mathbf{r}_2,\tau) = \frac{\langle E^{(-)}(\mathbf{r}_1,t) E^{(+)}(\mathbf{r}_2,t+\tau)\rangle}{\sqrt{\langle |E(\mathbf{r}_1,t)|^2 \rangle \langle |E(\mathbf{r}_2,t+\tau)|^2 \rangle}} 
\end{equation}
where $E^{\pm}$ denote the positive and negative frequency mode functions, and $|E|^2 = |E^{(-)}E^{(+)}|$. This quantity captures the amount of coherence between the photon fields. For Michelson or Mach-Zehnder interferometers, the visibility is defined as the modulus of the first-order coherence, with the maximum being unity for perfectly coherent fields. 

To explore the quantum properties of light, such as bunching and anti-bunching, it is necessary to define the second-order coherence, $g^{(2)}$, which measures correlations in intensity rather than amplitude. The second-order coherence is defined as
\cite{Glauber:1963tqt,Loudon:tqt,Fox:2006qo},
\begin{align}
    &g^{(2)}(\mathbf{r}_1,\mathbf{r}_2,\tau) = \nonumber\\
    &\;\;   \frac{\langle E^{(-)}(\mathbf{r}_1,t)E^{(+)}(\mathbf{r}_1,t) E^{(-)}(\mathbf{r}_2,t+\tau)E^{(+)}(\mathbf{r}_2,t+\tau)\rangle}{\langle |E(\mathbf{r}_1,t)|^2 \rangle \langle |E(\mathbf{r}_2,t+\tau)|^2 \rangle}~.
\end{align}
For stationary states, this reduces to $g^{(2)}(\tau) = \langle \hat{n}_1(t) \hat{n}_2(t+\tau) \rangle / (\langle \hat{n}_1(t)\rangle \langle \hat{n}_2(t+\tau)\rangle )$, with $\hat{n}$ being the photon number operator. While $g^{(1)}$ probes the classical wave nature of light, $g^{(2)}$ probes the statistical distribution of energy packets by measuring how the arrival times of these packets correlate with one another \cite{Loudon:tqt,Fox:2006qo}. Since the quantum nature of photons directly modulates their bunching statistics, probing GW-induced phase differences via second-order coherence offers a distinctly quantum-mechanical sensitivity that classical first-order interference lacks.

Crucially, measuring second-order coherence introduces the principle of photon indistinguishability as a fundamental metrological resource. In HOM setup, the coincidence counts are dictated by the overlap of photon wave packets \cite{Hong:1987}. Any differential phase shift or time delay induced by graviton-photon scattering alters this indistinguishability, translating directly into a measurable variation in the $g^{(2)}$ statistics. This mechanism is particularly advantageous for extended multi-baseline or constellation-based architectures \cite{Crowder:2005nr,Chen:2006zra,Sato:2006gk,Liu:2020xcv}, where $g^{(2)}$ interferometry can provide a noise-resilient channel. By relying on intensity correlations across separate detectors rather than classical field amplitudes, such a setup intrinsically isolates localized quantum-gravitational signatures from classical environmental decoherence.

\color{black}

%\noindent \underline{\emph{GW-induced HOM Interference:}}~
\section{GW-induced HOM Interference}\label{sec:HOM}
In the standard HOM experiment \cite{Hong:1987}, the indistinguishability of photons at a beam splitter dictates their bunching behavior. Our proposal leverages the fact that the GW-induced phase shift $\phi_{\mathbf{k}}(t)$ (Eq.~\eqref{eqn:phase-time-delay}) acts as a distinguishing marker. This \emph{gravitational distinguishability} modifies the quantum interference pattern, converting the graviton-photon interaction into a measurable change in coincidence counts. 

\begin{figure}[h]
\includegraphics[width=9cm]{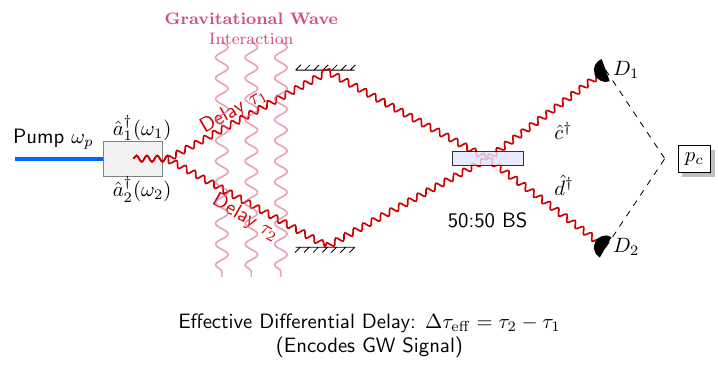}
    \caption{\label{fig2} Schematic diagram of effective differential time delay in the HOM-type experiment.}
\end{figure}

To go about this, we first formally describe the HOM interference pattern in the presence of GW-induced time delays. We consider the general case of frequency-entangled photons generated, for instance, by spontaneous parametric down-conversion (SPDC)\cite{Pan:2011xkw} pumped by a monochromatic beam of frequency $\omega_p$. The signal and idler photons satisfy energy conservation. %$\omega_s + \omega_i = \omega_p$.
As shown in Fig.~\ref{fig2}, the quantum state of the photon pair entering the interferometer is described by the joint spectral amplitude (JSA), $f(\omega_1, \omega_2)$:
\begin{equation}
    \ket{\Psi}_{\mathrm{in}} = \int \d\omega_1 \, \d\omega_2 \, f(\omega_1, \omega_2) \, \hat{a}_1^\dagger(\omega_1) \hat{a}_2^\dagger(\omega_2) \ket{0}_{1,2}~,
\end{equation}
where $\hat{a}_1^\dagger$ and $\hat{a}_2^\dagger$ create photons in the spatial modes $a$ and $b$ (the two interferometer arms), and the state is normalized such that $\int d\omega_1 d\omega_2 |f(\omega_1, \omega_2)|^2 = 1$. As the photons traverse the arms, the graviton-photon interaction induces effective time delays $\tau_1$ and $\tau_2$ in the respective arms (see Eq.~\eqref{eqn:effective-phase-time-delay}). This imposes a frequency-dependent phase shift on the creation operators: $\hat{a}_1^\dagger(\omega) \to \hat{a}_1^\dagger(\omega) e^{-i\omega \tau_{1}}$ and $\hat{a}_2^\dagger(\omega) \to \hat{a}_2^\dagger(\omega) e^{-i\omega \tau_2}$. The state arriving at the beam splitter is then:
{\small{
\begin{equation*}
\ket{\Psi'}_{\mathrm{in}} = \int \d\omega_1  \d\omega_2  f(\omega_1, \omega_2) e^{-i(\omega_1 \tau_1 + \omega_2 \tau_2)} \, \hat{a}_1^\dagger(\omega_1) \hat{a}_2^\dagger(\omega_2) \ket{0}_{1,2}.
\end{equation*}
}}
As shown in Fig.~\ref{fig2}, the 50:50 beam splitter mixes the modes $\hat{a}_1$ and $\hat{a}_2$ into output modes $\hat{c}$ and $\hat{d}$ via the transformations $\hat{a}_1^\dagger \to (\hat{c}^\dagger + \hat{d}^\dagger)/\sqrt{2}$ and $\hat{a}_2^\dagger \to (\hat{c}^\dagger - \hat{d}^\dagger)/\sqrt{2}$. Substituting these into the state vector, the output state $\ket{\Psi}_{\mathrm{out}}$ becomes:
{\small{
\begin{align}
\ket{\Psi}_{\mathrm{out}} &= \frac{1}{2} \int \d\omega_1 \d\omega_2  f(\omega_1, \omega_2) e^{-i(\omega_1 \tau_1 + \omega_2 \tau_2)} 
\Big( \hat{c}^\dagger(\omega_1) \hat{c}^\dagger(\omega_2) - 
\nonumber \\
& \hat{d}^\dagger(\omega_1) \hat{d}^\dagger(\omega_2) - \hat{c}^\dagger(\omega_1) \hat{d}^\dagger(\omega_2) + \hat{c}^\dagger(\omega_2) \hat{d}^\dagger(\omega_1) \Big) \ket{0}
\end{align}
}}
The observable of interest is the probability of a coincidence detection, $p_c$, where one photon is detected in port $D_1$ and one in port $D_2$. This corresponds to the projection onto the subspace spanned by $\hat{c}^\dagger \hat{d}^\dagger$. The probability $p_c = \bra{\Psi}_{\mathrm{out}} \hat{\Pi}_{\mathrm{cd}} \ket{\Psi}_{\mathrm{out}}$, where $\hat{\Pi}_{\mathrm{cd}}$ is the projection operator for mode c and d that describe the detection of these modes, is calculated as:
{\small{
\begin{equation}
\!\! p_{c} = \frac{1}{2} - \frac{1}{4} \int \!\! \d\omega_1 \d\omega_2 \left[ f^{*}(\omega_2, \omega_1) f(\omega_1, \omega_2) e^{-i \Delta \omega \Delta \tau_{\mathrm{eff}}} + \mathrm{c.c.} \right]
    \label{eqn:cc-general}
\end{equation}
}}
where $\Delta \omega = \omega_2 - \omega_1$ and $\Delta \tau_{\mathrm{eff}} = \tau_2 - \tau_1$ is the effective differential time delay induced by the GW. Equation~\eqref{eqn:cc-general} describes the famous HOM dip. When $\Delta \tau_{\mathrm{eff}} = 0$, the interference term is maximized, and for a symmetric JSA ($f(\omega_1, \omega_2) = f(\omega_2, \omega_1)$), the coincidence probability drops to zero ($p_c \to 0$). The GW signal $\Delta \tau_{\mathrm{eff}}(t)$ modulates this probability, encoding the graviton interaction into the coincidence rate.

\subsection{Working point of the HOM interferometer}

\subsubsection{Optimizing the working point}

To detect the minute variations in $\Delta \tau_{\mathrm{eff}}(t)$ caused by gravitons, we must optimize the working point of the interferometer. We assume a Gaussian spectral distribution for the photons with bandwidth $\sigma$. For unentangled identical photons (or entangled photons with symmetric Gaussian joint spectra), the coincidence probability {as a function of time delay} $\Delta \tau$ is given by:
\begin{equation}
p_c(\Delta \tau) = \left( 1 - e^{-\sigma^2 \Delta\tau^2} \right)/2.
\end{equation}
The total path length difference is $\Delta \tau(t) = \Delta \tau_0 + \Delta \tau_{\mathrm{eff}}(t)$, where $\Delta \tau_0$ is a static offset (the operating point) and $\Delta \tau_{\mathrm{eff}}(t) \propto h(t)$ is the time-varying GW signal. Expanding the coincidence probability around $\Delta \tau_0$ to first order in the strain yields:
%where $\Delta \tau=\Delta \tau_0+\Delta \tau_{\mathrm{eff}}$ is the total path length difference, with operating point at $\Delta \tau_0$ and GW signal $\Delta \tau_{\mathrm{eff}}$. For GW signal, $\Delta \tau_{\mathrm{eff}} \propto h_0$ and expanding the coincidence probability around this operating point:
%
\begin{align}\label{eqn:ccp-prob}
    p_{\mathrm{net}} &\approx p_c(\Delta \tau_0) + \left. {\d_\tau p_c}\right|_{\tau_0} \Delta \tau_{\mathrm{eff}} \\
    &= \frac{1}{2} \left[1 - e^{-\sigma^2 \Delta \tau_0^2} \right] + \sigma^2 \Delta \tau_0 \, e^{-\sigma^2 \Delta \tau_0^2} \Delta \tau_{\mathrm{eff}}(t) + \mathcal{O}(h_0^2). \nonumber
\end{align}
\begin{figure}[h]
\includegraphics[width=8.5cm]{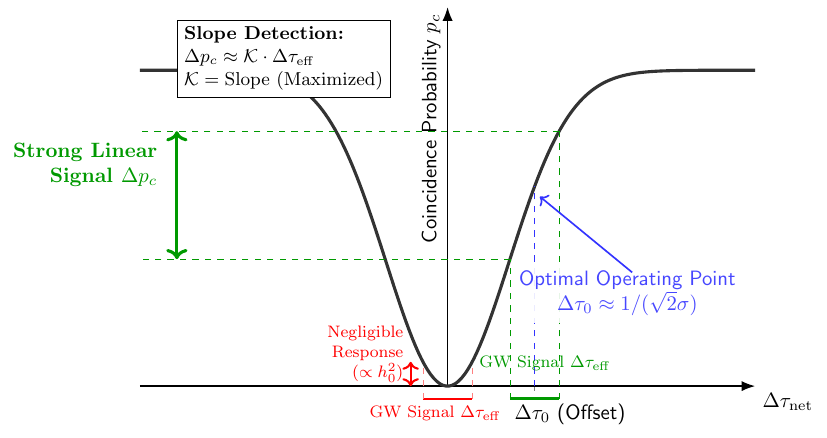}
    %\caption{\label{fig3} Illustration of the slope detection vs.   quadratic detection.
    \caption{\label{fig3} Illustration of slope vs. quadratic detection.}
\end{figure}
The signal manifests as a time-varying fluctuation in the coincidence probability, $\Delta p_c(t)$, which is linearly proportional to the effective delay:
\begin{equation}
\Delta p_c(t) \approx \mathcal{K}(\sigma, \Delta \tau_0)  \Delta \tau_{\mathrm{eff}}(t)~.
\end{equation}
The sensitivity factor $\mathcal{K} = \sigma^2 \Delta \tau_0 \, e^{-\sigma^2 \Delta \tau_0^2}$ is maximized when the operating point is set to the point of steepest slope, $\Delta \tau_0 = 1/(\sqrt{2}\sigma)$, as illustrated in Fig.~\ref{fig3}. We will see soon how combining this approach with Fisher information allows us to pinpoint the optimal working point for the detector.
%where, as shown in Fig.~\ref{fig3}, the sensitivity factor $\mathcal{K} = \sigma^2 \Delta \tau_0 \, e^{-\sigma^2 \Delta \tau_0^2}$ is {maximized when $\Delta \tau_0 = 1/(\sqrt{2}\sigma)$}.
The observable quantity is the instantaneous photon coincidence rate $R_c(t) = \Gamma p_{\mathrm{net}}(t)$, where $\Gamma$ is the incident photon flux. The GW signal is extracted from the time-dependent modulation of this rate:
%\begin{equation}
$\delta R_c(t) = \Gamma \, \mathcal{K} \, \Delta \tau_{\mathrm{eff}}(t)$.
%\end{equation}
%
This demonstrates that by biasing the interferometer to the slope of the HOM dip, the phase shifts induced by graviton-photon scattering can be read out directly via photon counting statistics.

%The total number of coincidence counts $N_c$ measured over a time $T$ is $N_c = \Gamma \int_0^T p_{\mathrm{net}}(t) \, dt$, where $\Gamma$ is the photon flux. The GW signal is extracted from the fluctuations in these counts:
%
%\begin{equation}
%$
%    \delta N_c = \Gamma \,\mathcal{K} \int_0^T \Delta \tau_{\mathrm{eff}}(t) \, dt~.
%$
%\end{equation}
%
%This result demonstrates that by monitoring the variations in photon coincidence rates at the slope of the HOM dip, one can directly measure the phase shifts induced by the graviton-photon scattering.

It is important to note that if the operating point is set to the bottom of the dip ($\Delta \tau_0 = 0$), the first-order sensitivity vanishes ($\mathcal{K}=0$). In this case, $p_c \approx \frac{1}{2}\sigma^2 (\Delta \tau_{\mathrm{eff}})^2$, resulting in a signal quadratic in the strain $h$. This is analogous to the \emph{dark port} problem in classical Michelson interferometers~\cite{Saulson:2017jlf}, where detection at the exact dark fringe yields zero sensitivity to small displacements; a local oscillator or offset is required to linearize the signal. 
%Note that, for the operating point set as the dip coincidence probability, ($\Delta \tau_0=0$), $p_c \approx \sigma^2 \Delta \tau_{\mathrm{eff}}^2/2 + \mathcal{O}(\Delta \tau_{\mathrm{eff}}^4)~$ -- a quadratic dependence on the strain $h_0$, which vanishes to first order. This "dark port" problem is analogous to detecting at the dark fringe in a classical Michelson interferometer~\cite{Saulson:2017jlf}; the sensitivity to small displacements is zero at the extremum.

%\color{blue}
\subsubsection{Fisher Information and Cram\'er-Rao Bound}
We have seen that, for the linear response, we have to choose the operating point away from the center of the dip. However, to rigorously quantify the ultimate sensitivity limit, we must analyze the classical Fisher information at this operating point. The Fisher information establishes the maximum amount of information that can be extracted about an unknown parameter from a given probability distribution \cite{Fisher1922,Lyon:sciadv.aap9416}. It is defined as \cite{Fisher1922}:
\begin{equation}
\mathcal{F}_{\Delta\tau} = \sum_{i}\frac{1}{p(i|\Delta\tau)} \left(\frac{\partial p(i|\Delta\tau)}{\partial \Delta\tau} \right)^2~,
\label{eqn:Fisher-Info}
\end{equation}
where $\Delta\tau$ is the unknown parameter and $p(i|\Delta\tau)$ is the probability for event $i$ conditioned on the value of $\Delta\tau$.

To model a realistic experimental scenario, we incorporate a photon loss probability $\gamma$ for each input arm, as well as a mismatch parameter $m$ that accounts for distinguishability between the input photons (e.g., due to imperfect polarization or spatial mode overlap) \cite{Lyon:sciadv.aap9416}. This mismatch reduces the HOM visibility, modifying the coincidence probability to $p_c(\Delta \tau) = (1-me^{-\sigma^2\Delta \tau^2})/2$. Since photons are lost independently with probability $\gamma$, the net probability of detecting zero photons is simply $\gamma^2$. Following Ref. \cite{Lyon:sciadv.aap9416}, the probabilities for detecting zero ($p_0$), one ($p_1$), or two ($p_2$) photons can be compactly written as:
\begin{align}
    \begin{pmatrix}
        p_0\\
        p_1\\
        p_2
    \end{pmatrix}
    =
    \begin{pmatrix}
        \gamma^2 & \gamma^2\\
        2\gamma(1-\gamma) & 1-\gamma^2\\
        (1-\gamma)^2 & 0
    \end{pmatrix}
    \begin{pmatrix}
        p_c\\
        1-p_c
    \end{pmatrix}~.
\end{align}
Using the definition in Eq.~\eqref{eqn:Fisher-Info}, and noting that the probabilities sum to unity such that $\partial_{\Delta\tau} p_2 = -\partial_{\Delta\tau} p_1 = (1-\gamma)^2\partial_{\Delta\tau} p_c $ (since $p_0$ is independent of $\Delta\tau$), the Fisher information reduces to:
\begin{equation}
\mathcal{F}_{\Delta\tau} = \frac{4m^2(1-\gamma)^2(1+\gamma)\sigma^4\Delta\tau^2}{(e^{\sigma^2\Delta\tau^2}-m)\left[m(1-\gamma)+e^{\sigma^2\Delta\tau^2}(1+3\gamma)\right]}~.
\label{eqn:Fisher-Info-2}
\end{equation}
In Fig.~\ref{fig:Fisher-Info}, we plot the Fisher information alongside the coincidence probability to illustrate the optimal operating point for maximizing detector sensitivity. For this illustration, we assume typical experimental parameters of $m=0.9$ and $\gamma=10^{-4}$, which results in the Fisher information exhibiting two symmetric peaks around $\Delta \tau=0$. Notably, in the ideal case of perfect indistinguishability ($m=1$), the Fisher information has maximum value when very close to the center of the dip, $\Delta \tau=0$ (although it is undefined exactly at $\Delta \tau=0$). However, the introduction of any distinguishability ($m<1$) prevents the coincidence probability from reaching true zero, causing the Fisher information to drop identically to zero at $\Delta \tau=0$. This abrupt splitting into double peaks represents a stark qualitative transition in the detector's optimal operating regime (a similar feature is also seen in Ref. \cite{Lyon:sciadv.aap9416}).
\begin{figure}
    \includegraphics[width=\linewidth]{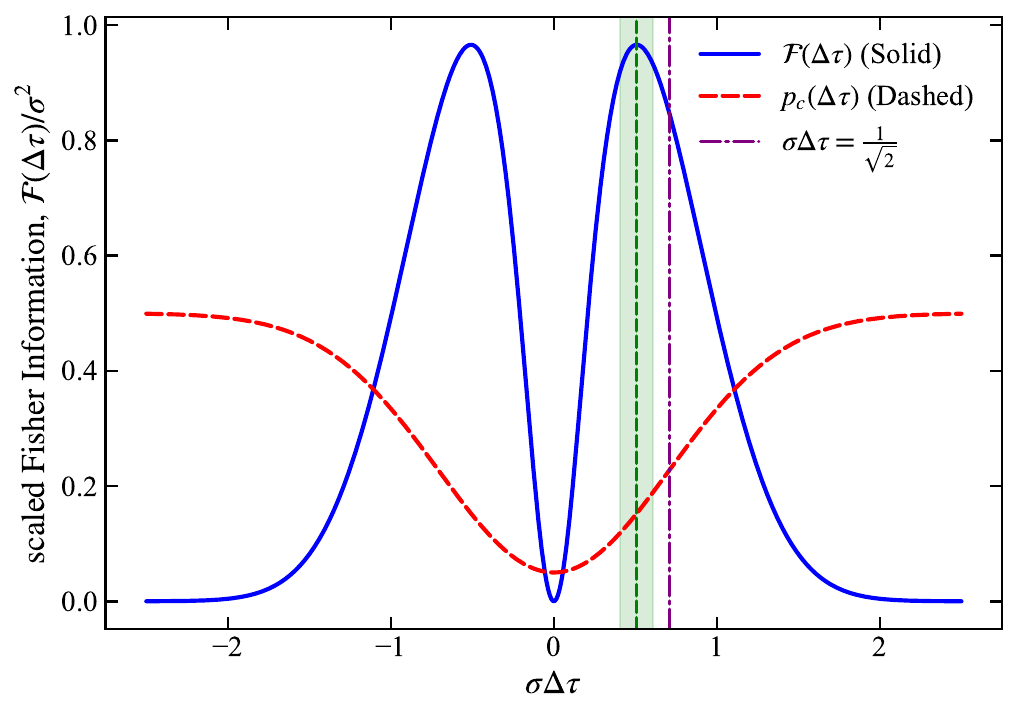}
    \caption{The Fisher information exhibit double peaks symmetric about $\Delta \tau=0$ when photons have some amount of distinguishability. The optimal working point can be chosen by tuning the setup for a preferred distinguishability of the photons (here $m=0.9$) such that both the Fisher information and the linear response is maximized. We have used $\gamma=10^{-4}$.}
    \label{fig:Fisher-Info}
\end{figure}

\noindent\textit{Cram\'er-Rao Bound:~} The Cram\'er-Rao bound establishes the fundamental lower limit on the estimator variance \cite{Rao1992,cramer1999mathematical}. Since the GW signal is encoded as minute variations in the coincidence count at the operating point $\Delta \tau = \Delta \tau_0$, this bound dictates the maximum achievable measurement precision. For $N$ independent measurements, the variance in the delay parameter is bounded by \cite{cramer1999mathematical}
\begin{equation}
    \text{Var}(\Delta\tau) \geq \frac{1}{N\mathcal{F}_{\Delta \tau}} \, .
\end{equation}

Using the above Cram\'er--Rao bound, the maximum attainable SNR can be estimated as $\mathcal{S}/\mathcal{N}\simeq\Delta\tau_{\rm rms}/\sqrt{\mathrm{Var}(\Delta\tau)}=\Delta\tau_{\rm rms}\sqrt{N\mathcal{F}_{\Delta\tau}}$, where $N=\Gamma T$ is the number of photon pairs and, for a monochromatic GW, $\Delta\tau_{\rm rms}=\mathcal{C}\,\mathscr{T}\,L\,h_0/(\sqrt{2}\,c)$ (see Eq.~\eqref{eqn:time-delay}, with $\mathscr{T}=|\mathrm{sinc}(\omega_{\rm eff}L/2c)|$). Near the optimal operating point ($\sigma\Delta\tau_0\simeq0.5$--$0.7$, $m\approx0.9$) one finds $\mathcal{F}_{\Delta\tau}\approx\sigma^2$ [from Eq.~\eqref{eqn:Fisher-Info-2}], which gives 
\begin{equation} 
\frac{\mathcal{S}}{\mathcal{N}}\approx\frac{\mathcal{C}\,\mathscr{T}\,L\,h_0\,\sigma}{\sqrt{2}\,c}\sqrt{\Gamma T}\, . 
\label{eqn:SNR-Fisher}
\end{equation} 
\color{black}
% 
%A similar expression, up to $\mathcal{O}(1)$ factors, is obtained below (Eq.~\eqref{eqn:SNR-t}) from a direct shot-noise calculation.
While this provides a theoretical limit to the detector's sensitivity, physical noise sources --- such as shot noise --- will inevitably elevate the experimental noise floor. {We provide a detailed estimation of this bound and the corresponding signal-to-noise ratio (SNR) in Section \ref{sec:SNR} (see Eq.~\eqref{eqn:SNR-final}).}

\noindent\textit{Fisher Information Matrix (FIM):} While our preceding analysis optimizes the working point of the interferometer using single-parameter Fisher information, this framework can naturally be extended to multiparameter estimation. Specifically, employing a maximum likelihood estimator (MLE) in conjunction with the Fisher Information Matrix enables the simultaneous extraction of multiple GW parameters --- such as wave amplitudes, polarization states, and high-frequency noise characteristics --- directly from the modulated coincidence counts.

\section{SNR and Feasibility}
\label{sec:SNR}

To evaluate the observational feasibility of this scheme, we consider a realistic broadband photon source characterized by a Gaussian spectral width $\sigma \approx 10^{13}~\mathrm{rad\,s^{-1}}$ (bandwidth $\sim 1.5~\mathrm{THz}$). At the optimal operating point, the sensitivity scaling factor evaluates to $\mathcal{K} \approx \sigma \approx 10^{13}~\mathrm{s^{-1}}$.  For a characteristic GW strain $h_0 \sim 10^{-19}$, the effective differential time delay induced across the optical paths ($L$) is $\Delta \tau_{\mathrm{eff}} \approx (h_0/c) L \approx (L \times 10^{-28})~\mathrm{s}$. 

Considering a space-borne configuration with an arm length $L \approx 10^{4}~\mathrm{km}$ ($10^7~\mathrm{m}$), the resulting delay is on the order of $\Delta \tau_{\mathrm{eff}} \sim 10^{-21}~\mathrm{s}$. This translates into a fractional modulation of the photon coincidence probability given by $\Delta p_{c} \approx \mathcal{K} \Delta \tau_{\mathrm{eff}} \sim 10^{-8}$. Although small, this signal can be definitively resolved given adequate photon flux and integration duration. Conservatively, assuming a photon pair-generation rate of $\Gamma \sim 10^{16}~\mathrm{photons\,s^{-1}}$ (corresponding to $\approx 3~\mathrm{mW}$ at optical wavelengths)~\cite{Zhong:2018jgi,Erhard:2019bke}, a signal-to-noise ratio (SNR) exceeding unity can be achieved within realistic measurement windows, demonstrating that this sensitivity regime is accessible to near-future space-based quantum technologies.

{A rigorous quantitative estimate of the SNR is established by assuming Poissonian photon-counting shot noise as the dominant noise contribution. In a HOM coincidence measurement, the primary readout is the integrated coincidence count. The expected signal-to-noise ratio is projected as:
\begin{equation}
    \frac{\mathcal{S}}{\mathcal{N}} = \frac{N_{\mathrm{gw}}}{\Delta N_{\mathrm{noise}}} \, ,
    \label{eqn:SNR}
\end{equation}
where $N_{\mathrm{gw}}$ represents the fluctuation in coincidence counts induced by the passing GW, and $\Delta N_{\mathrm{noise}}$ denotes the statistical fluctuation of the noise floor. Under Poissonian counting statistics, the noise fluctuation scales as $\Delta N_{\mathrm{noise}} = \sqrt{N_0}$, where $N_0$ is the baseline coincidence count determined by the static operating delay $\Delta \tau_0$.}

The signal-induced variation in coincidence counts $N_{\mathrm{gw}}$ is obtained from the coincidence probability in Eq.~\eqref{eqn:ccp-prob}. For a monochromatic GW, $\Delta\tau_{\rm eff}(t)=\mathcal{C}\, \mathscr{T} \,(L/c)\,h_0\cos(\Omega_\mathbf{q} t+\phi_0)$, so the coincidence rate is modulated at the GW frequency, $\delta R(t)=\Gamma m\mathcal{K}\,\mathcal{C}\, \mathscr{T} \,(L/c)\,h_0\cos(\Omega t+\phi_0)$. Matched filtering this rate over a coherent observation time $T\gg 1/\Omega_{\mathbf{q}}$ gives an effective signal count,
\begin{equation}
\!\!\! N_{\mathrm{gw}} \approx \frac{\Gamma m\mathcal{K} \mathcal{C}\, \mathscr{T} \, (L/c)\, h_0\, T}{\sqrt{2}} \, ,
\end{equation}
where the factor $1/\sqrt{2}$ is the rms of $\cos(\Omega t+\varphi)$, $\Gamma$ is the signal-idler pair production rate at the source, $m$ is the spatial/polarization mode mismatch parameter, $\mathcal{K}$ is the dip sensitivity factor, $\mathcal{C}$ is a geometric projection factor, $L$ is the nominal baseline arm length, $c$ is the speed of light, and $T$ is the total integration time. The transfer function $\mathscr{T} =|\mathrm{sinc}(\omega_{\rm eff}L/2c)|$, with $\omega_{\rm eff}=\omega_{\rm gw}(1-\hat{\mathbf q}\cdot\hat{\mathbf k})$ (Eq.~\eqref{eqn:time-delay} in Appendix~D), satisfies $ \mathscr{T} \approx 1$ in the long-wavelength limit $\omega_{\rm gw}L/c\ll 1$. Note that this is a condition on the arm length and not on the integration time: a time-integrated count $\int_0^T h\,\mathrm{d}t$ would average to zero for $\omega_{\rm gw}T\gg 1$, so the signal must be read out at $\omega_{\rm gw}$. The baseline coincidence count corresponding to the offset $\Delta \tau_0$ is $N_0 = \frac{1}{2}\Gamma T (1- me^{-\sigma^2\Delta \tau_0^2})$, and the Poissonian noise is $\Delta N_{\rm noise}=\sqrt{N_0}$. Substituting these quantities into Eq.~\eqref{eqn:SNR} yields the explicit signal-to-noise ratio:
\begin{align}
    \frac{\mathcal{S}}{\mathcal{N}} & \approx \frac{\Gamma m\mathcal{K} \mathcal{C}\,  \mathscr{T} \, (L/c) h_0 T/\sqrt{2}}{\sqrt{\Gamma T (1-me^{-\sigma^2 \Delta \tau_0^2})/2}} \nonumber \\
    & = \frac{\sqrt{\Gamma T}\, m\mathcal{K} \mathcal{C}\, \mathscr{T} \, (L/c) h_0}{\sqrt{1-me^{-\sigma^2 \Delta \tau_0^2}}} \, .
    \label{eqn:SNR-final}
\end{align}
Evaluating this expression near the optimal working point {($\mathcal{K}=\sigma e^{-1/2}/\sqrt{2}\approx 4\times 10^{12}\,\mathrm{s^{-1}}$)} with representative baseline parameters: $\mathcal{C} \sim 1$, $\mathscr{T} \approx 1$, $L=10^{7}\,\mathrm{m}$ ($L/c\approx 3\times 10^{-2}\, \mathrm{s}$), $h_0 \sim 10^{-19}$, $m=0.9$, and a pair generation rate of $\Gamma \sim 10^{12}\,\mathrm{s^{-1}}$ -- gives the simplified scaling:
\begin{equation}
    \frac{\mathcal{S}}{\mathcal{N}} \approx 2\times 10^{-2} \sqrt{T} \, ,
    \label{eqn:SNR-t}
\end{equation}
which should be regarded as an order-of-magnitude estimate. For $\omega_{\rm gw}L/c\gtrsim 1$ (e.g.\ $\mathscr{T} \approx 0.8$ at $10\,\mathrm{Hz}$ for $L=10^7\,\mathrm{m}$) the result is reduced by the factor $\mathscr{T}$.

\color{black}
Consequently, for integration periods on the order of a few days ($T \sim 10^5\text{--}10^6 \,\mathrm{s}$), the SNR comfortably exceeds unity (see Fig. \ref{fig:HOM-SNR}), enabling unambiguous signal extraction.
Under these parameter choices, the corresponding Cram\'er-Rao lower bound on the measurement precision of the effective GW time delay $\Delta\tau_{\mathrm{gw}}$ evaluates to $\mathrm{SD}({\Delta\tau_{\mathrm{gw}}}) \approx 10^{-24}\,\mathrm{s}$.
\begin{figure}
    \includegraphics[width=\linewidth]{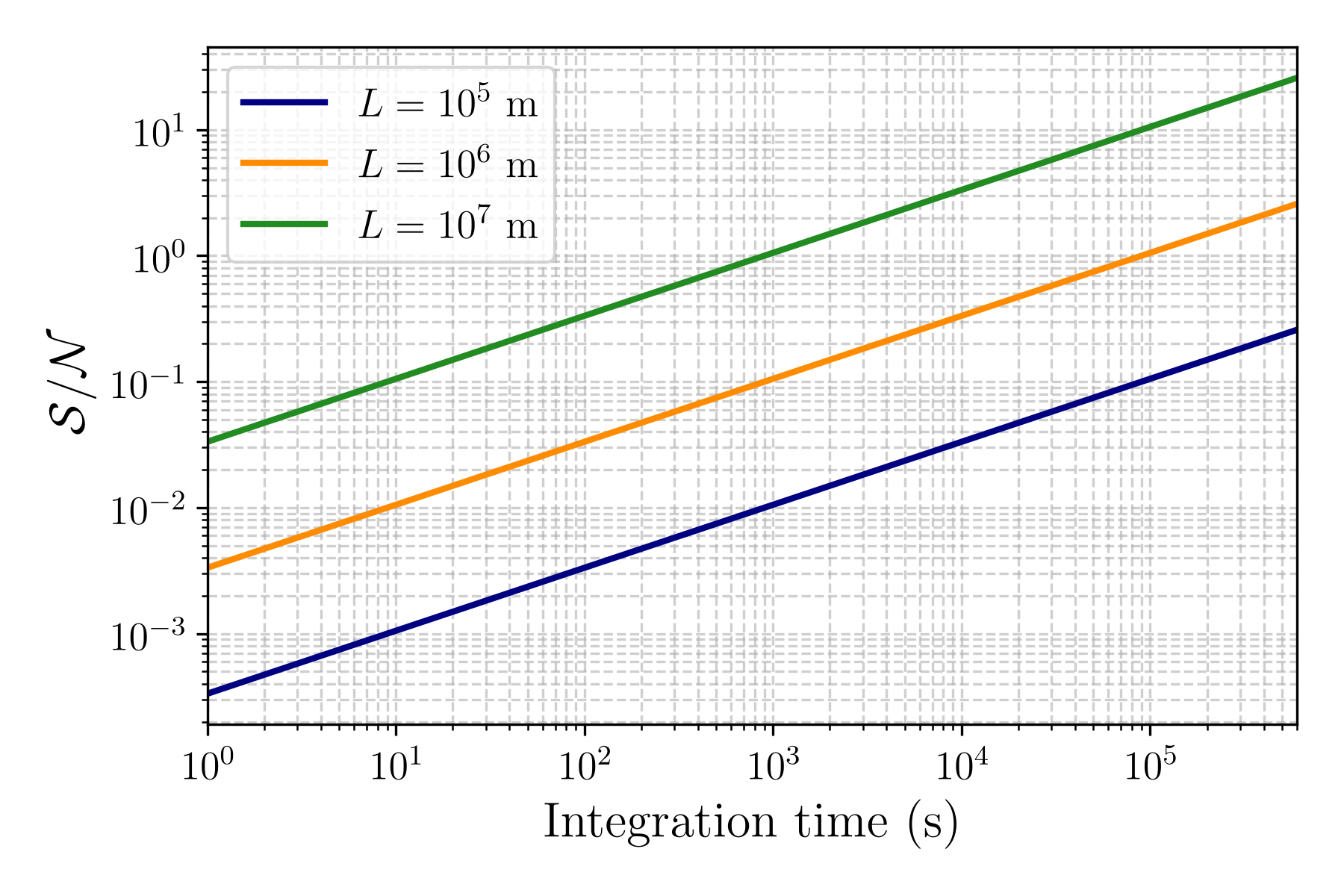}
    \caption{Plot of ${\mathcal{S}}/{\mathcal{N}}$  in Eq. \eqref{eqn:SNR-t} as a function of integration time $T$ for different arm lengths.}
    \label{fig:HOM-SNR}
\end{figure}

To place this quantum measurement capability in direct comparison with standard GW detection metrics, we express the SNR for periodic sources in terms of the strain power spectral density (PSD) $S_n$~\cite{Maggiore:2007ulw}:
\begin{equation}
    \frac{\mathcal{S}}{\mathcal{N}} = h_0 \sqrt{\frac{T}{S_n}} \, ,
    \label{eqn:SNR-Maggiore}
\end{equation}
Equating Eq.~\eqref{eqn:SNR-Maggiore} with the expression for $\mathcal{S}/\mathcal{N}$ yields the equivalent strain power spectral density of the HOM interferometer:
\begin{equation}
    S_n \approx \frac{c^2}{\Gamma \mathcal{K}^2 \mathcal{C}^2 L^2} \, .
\end{equation}
A detailed comparison between this strain noise power spectral density and that of classical $g^{(1)}$ interferometry is provided in Appendix~\ref{app:comparison-classical-interf}.

\subsection{Geometrical Configurations and Angular Response}
\label{sec:MicroMacrobridge}

Having established in Sec.~\eqref{sec:Coherentstates} that the photon-graviton interaction in a coherent graviton state reproduces the macroscopic gravitational time delay via geometric projection factors, we now evaluate how the spatial layout of the optical trajectories directs the overall detector sensitivity. Since, the net differential phase shift is governed by these directional geometric projections of the metric tensor, optimizing the spatial architecture of the optical arms is critical for maximizing both signal response and sky coverage. Realizing the macroscopic baselines ($L \sim 10^4~\mathrm{km}$) required for detecting minute differential delays, while maintaining optical coherence across the entangled photon channels, inherently necessitates a space-borne architecture. Although, standard planar geometries (e.g., LISA) are sufficient for classical intensity interferometry, second-order quantum coincidence measurements ($g^{(2)}$) benefit significantly from isotropic sensitivity to the GW background. To this end, we investigate two specific topological configurations: a baseline 2D planar setup suitable for single-plane platforms, and a 3D pyramidal geometry that can be engineered to eliminate directional blind spots.

Using the scattering phase derived in Eq.~\eqref{eqn:effective-phase-time-delay}, the effective delay for a photon propagating along a optical path defined by the unit vector $\hat{\mathbf{k}}$ over a distance $L$ is given by the spatial projection of the metric perturbation $h_{ij}$ along the photon trajectory~\cite{Finn:2008np}:
\begin{equation}
    c \Delta t \approx L + \frac{1}{2} \hat{k}^i \hat{k}^j \int_{t_0}^{t_0+L/c} \mathrm{d}t \, h_{ij}(t, \mathbf{x}(t))~.
    \label{eqn:general_delay}
\end{equation}
Equation~\eqref{eqn:general_delay} highlights the fundamentally directional nature of the interaction: the tensor contraction $\hat{k}^i \hat{k}^j h_{ij}$ determines which components of the gravitational strain tensor couple to the optical phase. We propose two distinct geometric topologies to exploit this response:

\begin{figure}[!th]
    \centering
    \includegraphics[width=0.425\linewidth]{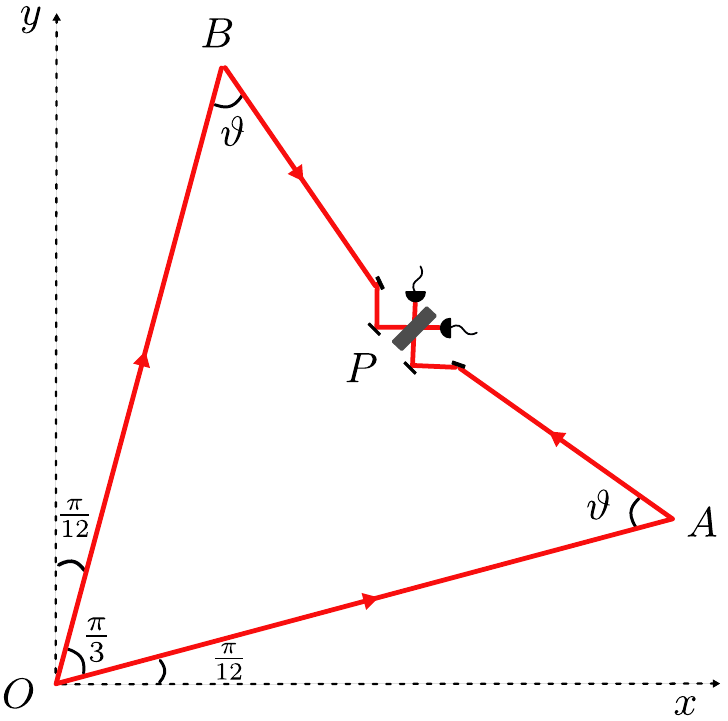}
    \hskip10pt
    \includegraphics[width=0.525\linewidth]{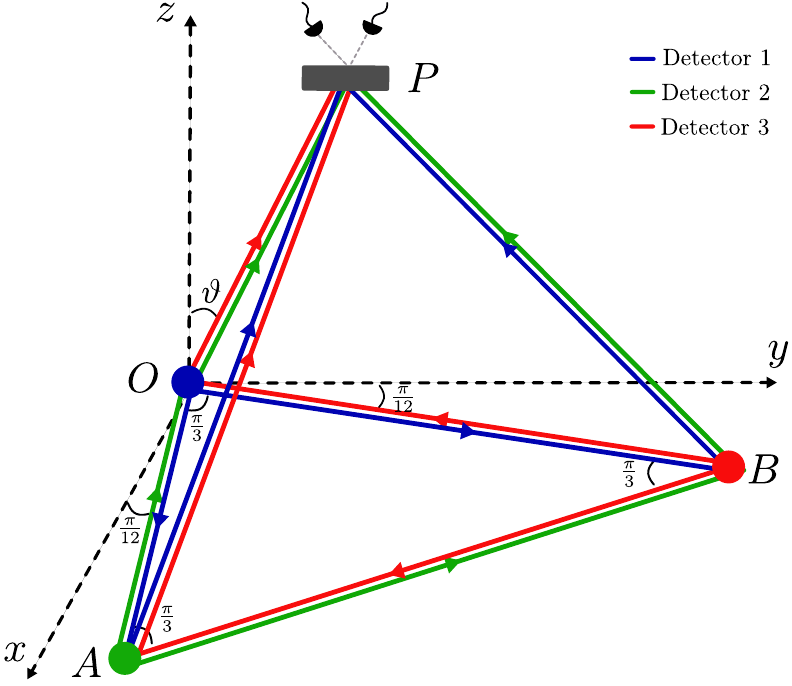}
    \caption{Possible geometrical configurations for HOM-based GW detectors: \textbf{Left:} 2D planar geometry, and \textbf{Right:} 3D pyramidal configuration featuring non-coplanar optical paths.}
    \label{fig:2D3D-geometry}
\end{figure}

\noindent \textbf{2D Planar Setup:} Consider first a planar configuration suitable for terrestrial or single-plane space missions, as illustrated in the left panel of Fig.~\ref{fig:2D3D-geometry}. Since, all optical segments lie within a two-dimensional plane (defined here as the $xy$-plane, where $k_z = 0$), the photon wavevectors possess no out-of-plane directional components. Consequently, in the long-wavelength limit ($\omega_{\mathrm{gw}} L / c \ll 1$), the differential time delay between orthogonal optical arms simplifies to:
\begin{equation}
    c \Delta \tau_{\mathrm{eff}} \approx \mathcal{C}_{\mathrm{2D}} L (h_{xx} - h_{yy}) \, ,
\end{equation}
where $\mathcal{C}_{\mathrm{2D}}$ is a geometric projection factor of order unity. This setup recovers the standard quadrupolar antenna pattern characteristic of existing planar observatories (such as LIGO and LISA), yielding maximum sensitivity to the $+$ polarization mode along the normal axis, but exhibiting vanishing response (directional blind spots) for waves incident along the arm bisectors or matching the $\times$ polarization frame (for example, see Ref. \cite{Sathyaprakash:2009xs}).

\textbf{3D Pyramidal Setup:} To overcome these directional limitations in space-based missions, we propose a three-dimensional pyramidal geometry (depicted in the right panel of Fig.~\ref{fig:2D3D-geometry}) consisting of three non-coplanar interferometers sharing a central common vertex $P$~\cite{Chen:2006zra,Sato:2006gk,Liu:2020xcv,Jin:2024mma}. By elevating the central vertex out of the baseline plane, the optical paths acquire non-zero vertical directional projections ($\hat{k}_z \neq 0$). This non-coplanar geometry directly couples the optical propagation to the off-diagonal metric components ($h_{xz}, h_{yz}$) in the detector frame. For instance, the explicit differential response for one component interferometer (Detector 1, or $D_1$) is given by:
\begin{align}
    \frac{c\,\Delta \tau_{\rm eff}}{L} \Bigg|_{\rm D_1}
    =\;& \frac{1}{4}\left( h_{xx} + h_{yy} \right)
    + \frac{3\sqrt{3}}{4} \left( h_{xx} - h_{yy}\right) \nn \\
    &\quad + \sqrt{\frac{3}{2}} \left(h_{yz} - h_{xz}\right)
    + \frac{1}{2} h_{xy}~,
    \label{eqn:3D-eff-time-delay-approx}
\end{align}
where $L$ denotes the characteristic arm length.

The explicit appearance of the off-diagonal shear components $h_{xz}$ and $h_{yz}$ in Eq.~\eqref{eqn:3D-eff-time-delay-approx} represents a crucial architectural advantage over planar detectors. Because the three constituent interferometers in the 3D network sample distinct linear combinations of all spatial strain components, their individual null directions (blind spots) are mutually complementary. As a result, this non-coplanar sensitivity significantly smooths the sky-averaged response, providing a quasi-isotropic antenna pattern (see Appendix~\ref{app:config-details} for detailed geometric derivations), eliminating directional blind spots, and substantially improving source localization and polarization-deconvolution capabilities compared to single-plane detectors.

\begin{figure}[!ht]
    \centering
    \includegraphics[width=0.7\linewidth]{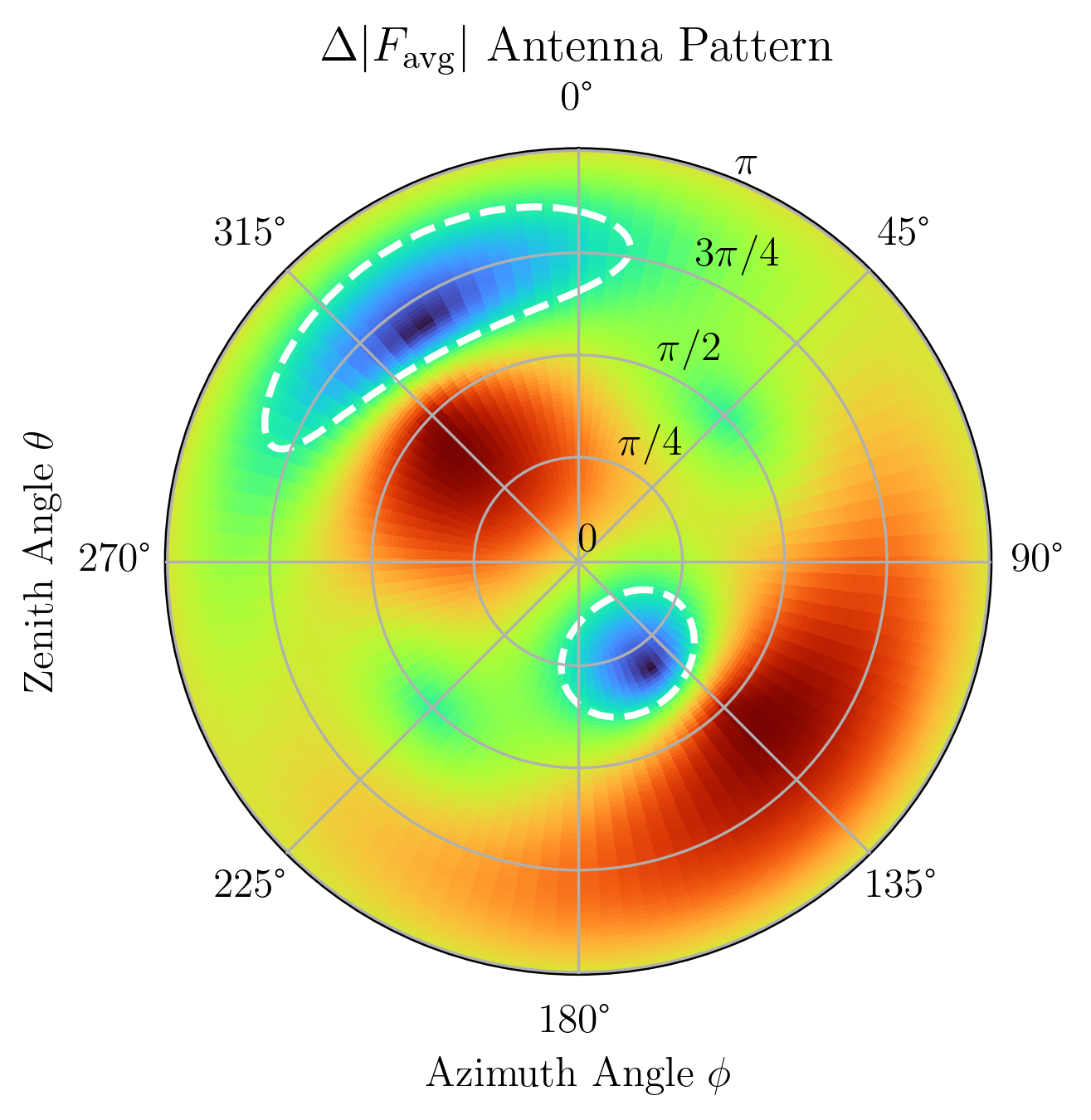}
    \caption{The difference of the antenna pattern functions ($\Delta |F_{\mathrm{avg}}| = |F_{D_1}| - |F_{\mathrm{planar}}|$) between 3D and planar configurations (corresponding to LIGO).}
    \label{fig:comparison-with-planar}
\end{figure}

\section{3D Sensitivity and Experimental Outlook}\label{sec:conclusion}
The specific coupling to off-diagonal metric components ($h_{xz}, h_{yz}$) shown in Eq.~\eqref{eqn:3D-eff-time-delay-approx} represents a significant deviation from the response of planar interferometers. In standard configurations like LIGO or LISA, the detector response is maximized for GWs propagating perpendicular to the detector plane, while vanishing for waves incident along the arm bisectors (blind spots). 
 
As illustrated in Fig.~\ref{fig:comparison-with-planar}, even an individual interferometer in our proposed 3D pyramidal configuration has a better angular sensitivity compared to planar setups (LIGO or LISA), except in the specified blue regions. Each of the individual detectors also possesses a distinct antenna pattern. Crucially, the orientation of these patterns is mutually complementary (details in Appendix \ref{app:config-details}). When operating as a network, the combined response --- averaged over three detectors --- approximates a monopole-like sensitivity (Fig.~\ref{fig:3D-antenna-avg}). This quasi-isotropic coverage ensures that there are \emph{no blind spots on the sky}, enabling continuous monitoring of transient sources regardless of their sky location. Furthermore, the simultaneous measurement of distinct linear combinations of the strain tensor components ($h_{ij}$) by the three detectors breaks the degeneracy typically found in single-detector observations, significantly enhancing the capability for source localization and polarization resolution.
%
%\begin{comment}
\begin{figure}[!h]
    \centering
    \includegraphics[width=0.7\linewidth]{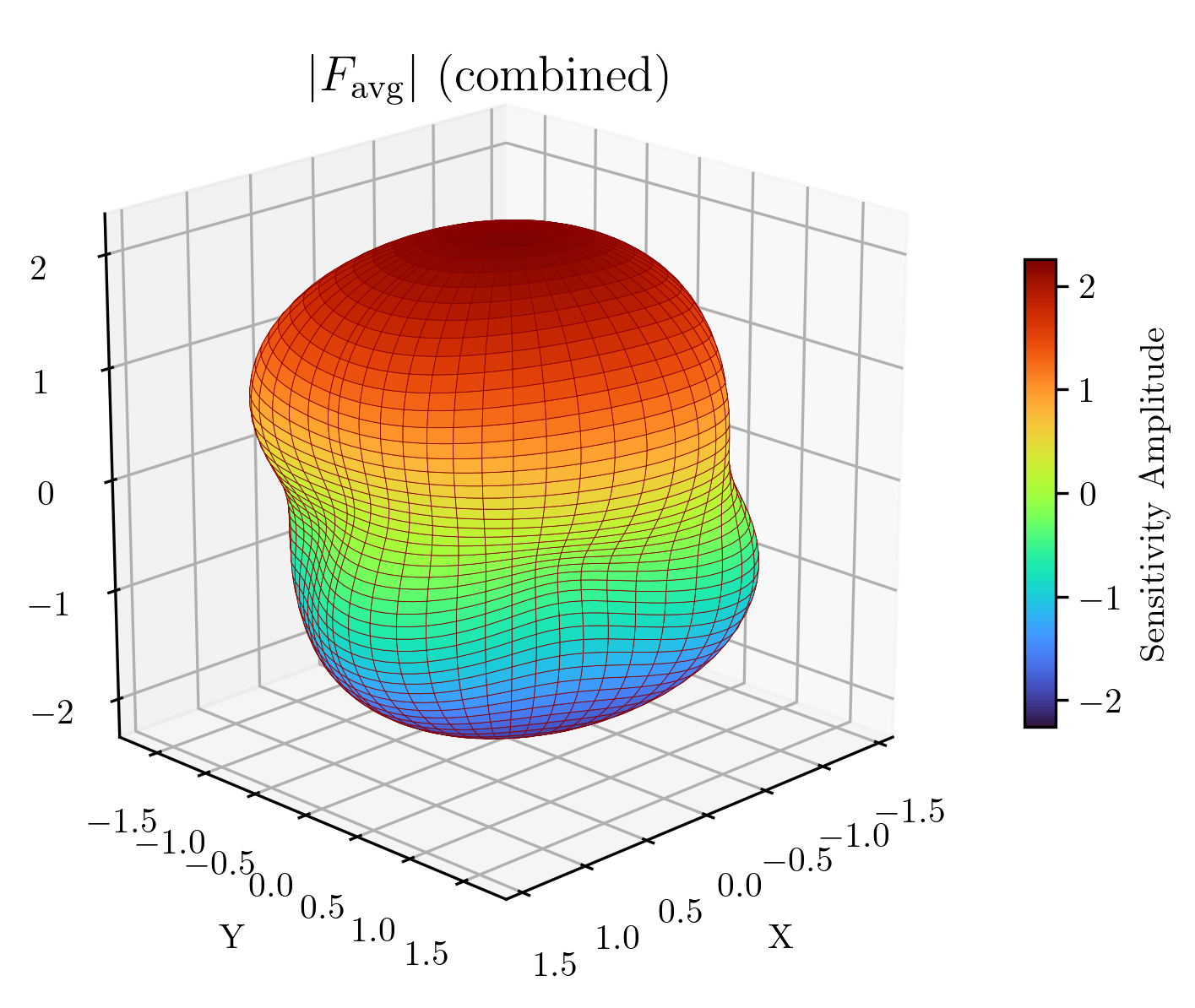}
    \caption{\label{fig:3D-antenna-avg} The combined angular sensitivity of the 3D network (see Fig. \ref{fig:3D-antenna-each} for each detector).}
    %, calculated as the root-mean-square response of the three component interferometersThe resulting pattern is nearly isotropic, demonstrating full-sky coverage without significant blind spots.}
\end{figure}
%\end{comment}
%

As mentioned earlier, this architecture is naturally suited for a space-based mission. In the vacuum of space, long baselines can be achieved without scattering losses. Overcoming the fundamental shot-noise limitations of this $g^{(2)}$ counting scheme will require substantial technological milestones, specifically in the development of ultra-bright, narrow-band SPDC photon sources and exquisite drag-free control technologies, adapted from LISA, to stabilize spacecraft formations to the requisite precision. 
In such an environment, the quantum advantages of this counting experiment --- specifically its robustness against laser intensity noise compared to classical fringe locking --- could offer a complementary observational window for the gravitational universe. As discussed in the introduction, metrics like the DFS retain a unique signature of cosmic history~\cite{Chandran:2023ogt,Chandran:2024utf}. While previous studies established the theoretical necessity of evaluating state-space density matrix overlaps, our proposed HOM $g^{(2)}$ interferometer provides the concrete physical detection architecture required for future laboratory execution.

Finally, it is instructive to distinguish the present framework from the seminal analysis of quantum mechanical noise in interferometers by Caves~\cite{Caves:1981hw,Martynov:2017ufx,LIGOScientific:2020luc}. Caves' formalism focuses on the fundamental limits of measurement precision, demonstrating that the shot-noise limit in a classical interferometer arises from vacuum fluctuations entering the unused port, and can be surpassed using squeezed states. However, in Caves' analysis, the GW is treated as a classical tidal force that modifies the boundary conditions (mirror positions) of the EM field modes. In contrast, our work addresses the microscopic origin of the signal itself. By treating the interaction as a scattering process between the quantized EM and gravitational fields, we derive the accumulated phase not as a geometric boundary effect, but as a consequence of inelastic energy exchange. {While Caves' analysis pertains to intensity measurements (first-order coherence, $g^{(1)}$), our proposed detection scheme relies on HOM interference (second-order coherence, $g^{(2)}$). This shifts the measurement basis from field amplitude quadratures to photon distinguishability, offering a complementary modality that may be less susceptible to specific classes of low-frequency laser noise. For instance, our analysis can be extended for a single-graviton exchange. While this is currently beyond the detector capabilities, the formalism can be applied in Fock representation (taking $r \to 0$ and $\beta \to 0$ limit in Eq.~\eqref{eqn:evolution-1}).}

{Crucially, our analysis reveals that for highly non-classical gravitational backgrounds --- such as squeezed states predicted by inflation~\cite{Chandran:2023ogt,Chandran:2024utf}---the photon-graviton interaction induces significant decoherence. This effect scales non-linearly with the squeezing parameter, distinct from the response to a coherent state (see Appendix \ref{app:coherent-state} and \ref{app:interf_visibility}). This suggests that the loss of interferometric visibility itself could serve as a probe of the quantum nature of gravity. Consequently, evaluating the impact of squeezed graviton states is essential for characterizing the observational potential and fundamental noise limits of third-generation detectors.}

\vspace{0.5cm}
\noindent\emph{\underline{Acknowledgment:}} The authors are grateful to S. Bhaumik, I. Chakraborty, A. Chowdhury, P. G. Christopher, J. Fernandes, S. Ghosh, N. Jaiswal, D. Kothawala, K. Lochan, S. Mondal, A. Pai, and T. Parvez for their valuable discussions and feedback on the earlier draft. The authors thank the anonymous referees for the feedback on the manuscript which have significantly improved the presentation of the work.
The work is supported by ANRF-Advanced
Research Grant (ANRF/ARG/2025/001514/PS).

\appendix

%%%%%%%%%%%%%%%%%%%%%%%%%%%%%
\section{Photon dynamics for squeezed number state} \label{app:ph-gr-density-evolution}
%%%%%%%%%%%%%%%%%%%%%%%%%%%%%
%
We consider the density matrix of the total system, initialized as a product state of the photon and graviton sectors:
%
%%%%%%%%%%%%%%%%
\begin{equation}
    \hat{\rho}(0) = \hat{\rho}_{\mathrm{ph}}(0)\otimes\hat{\rho}_{\mathrm{gw}}(0)~.
\end{equation}
%%%%%%%%%%%%%%%%
%
The photon state is general, $\hat{\rho}_{\mathrm{ph}} = \sum_{n,n^\prime} \rho_{nn^\prime} \ket{n}\bra{n^\prime}$, where $\ket{n}$ represents a Fock state of the relevant optical mode. The unitary time evolution in the interaction picture is given by the time-ordered exponential: 
\begin{equation}
    U_I(t) = \mathcal{T} \exp\left( -i \int_0^t dt'\, H_{\mathrm{int}}(t') \right)~.
\end{equation}
Using the Magnus expansion up to second order, this can be written as:
\begin{align}
    U_I(t) = & \exp\bigg( -i \int_0^t dt' H_{\mathrm{int}}(t') \nonumber \\ & \quad-\frac{1}{2} \int_{0}^t dt' \int_0^{t'} dt'' \left[H_{\mathrm{int}}(t'), H_{\mathrm{int}}(t'')\right] \bigg)~.
\end{align}
The first-order term is linear in graviton operators ($\hat{b}, \hat{b}^\dagger$) and drives the displacement of the gravitational field:
\begin{widetext}
\begin{align}
    -i \int_0^t dt' H_{\mathrm{int}}(t') &=  \sum_{\mathbf{k},\mathbf{q}} \sum_{\sigma,\lambda} \frac{\omega_\mathbf{k}}{4\sqrt{2\omega_{\mathbf{q}}V}} \, \hat{n}_{\mathbf{k},\lambda} \left[ \hat{b}^{\dagger}_{\mathbf{q},\sigma} \alpha^{(\sigma,\lambda)}_{\mathbf{k},\mathbf{q}}(t) - \hat{b}_{\mathbf{q},\sigma} \left(\alpha^{(\sigma,\lambda)}_{\mathbf{k},\mathbf{q}}(t)\right)^* \right]~,
\end{align}
\end{widetext}
where we have defined the time-dependent coupling coefficient:
\begin{equation}
    \alpha^{(\sigma,\lambda)}_{\mathbf{k},\mathbf{q}}(t) = \kappa \,{\mathcal{G}}^{(\sigma,\lambda)}_{\hat{\mathbf{k}},\hat{\mathbf{q}}} \left( \frac{1 - e^{-i\Omega_{\mathbf{q}} t}}{\Omega_{\mathbf{q}}} \right) = i \kappa\, {\mathcal{G}}^{(\sigma,\lambda)}_{\hat{\mathbf{k}},\hat{\mathbf{q}}} \int_0^t dt' e^{-i\Omega_{\mathbf{q}}t'}~,
\end{equation}
where ${\mathcal{G}}^{(\sigma,\lambda)}_{\hat{\mathbf{k}},\hat{\mathbf{q}}}$ is already defined in the main text. The second-order term involves the commutator $[H_{\mathrm{int}}(t'), H_{\mathrm{int}}(t'')]$, which reduces to the c-number commutator $[\hat{b}_{\mathbf{q}}, \hat{b}^\dagger_{\mathbf{q}'}] = \delta_{\mathbf{q}\mathbf{q}'}$. This term contributes only a global phase factor $\Phi(t)$ (a c-number in the graviton subspace, though diagonal in the photon number basis), and all higher-order commutators vanish. The total evolution operator can thus be factorized as:
\begin{equation}
    U_{I}(t) = e^{i\Phi(t)} \prod_{\mathbf{q},\sigma} \hat{D}_{\mathbf{q}}\left({\widetilde{\alpha}}^{\sigma}_{\mathbf{q}} \right)~,
\end{equation}
where $\hat{D}_{\mathbf{q}}(\gamma) = \exp(\gamma \hat{b}^\dagger_{\mathbf{q}} - \gamma^* \hat{b}_{\mathbf{q}})$ is the displacement operator acting on the graviton field, and the displacement amplitude depends on the photon number operator:
\begin{equation}
    {\widetilde{\alpha}}^{(\sigma)}_{\mathbf{q}} = \sum_{\mathbf{k},\lambda} \frac{\omega_\mathbf{k}}{4\sqrt{2\omega_{\mathbf{q}}V}} \, \hat{n}_{\mathbf{k},\lambda} \, \alpha^{(\sigma,\lambda)}_{\mathbf{k},\mathbf{q}}(t)~.
    \label{eqn:alphatilde-def}
\end{equation}

\subsection{Reduced Photon Dynamics} \label{app:red-photon-dyn}
%\color{blue}

We now evolve the total density matrix $\hat{\rho}(t) = U_I(t) [\hat{\rho}_{\mathrm{ph}} \otimes \hat{\rho}_{\mathrm{gw}}] U_I^\dagger(t)$. We use a generic quantum state called squeezed number state \cite{mvs:prd32}, $\ket{k,\gamma,\beta} = \hat{S}(\gamma) \hat{D}(\beta) \ket{l}$ for graviton, so $\hat{\rho}_{\mathrm{gw}} = \ket{l,\gamma,\beta}\bra{l,\gamma,\beta}$. The photon state is expanded in the number basis $\ket{n}$. Since the operator $\hat{\widetilde{\alpha}}$ is diagonal in the photon number basis, let $\widetilde{\alpha}_n$ denote its eigenvalue for the state $\ket{n}$. The evolved state is:
\begin{align}
    \!\!\!\!\hat{\rho}(t) &= \sum_{n,n^\prime} \rho_{nn^\prime} \ket{n}\bra{n^\prime} \otimes \nonumber \\
    & \qquad \hat{D}(\widetilde{\alpha}_n) \hat{S}(\gamma) \hat{D}(\beta)\ket{l} \bra{l} \hat{D}^{\dagger}(\beta) \hat{S}^{\dagger}(\gamma)  \hat{D}^\dagger(\widetilde{\alpha}_{n^\prime})~.
\end{align}
Here, $\gamma=re^{i\varphi}$ is the squeezing parameter with squeezing amplitude $r$.
%first to second equality we have used the property $\hat{D}(\widetilde{\alpha}_n)\hat{S}(\gamma) = \hat{S}(\gamma) \hat{D}(\zeta_n)$ where $\zeta_n = \widetilde{\alpha}_n \cosh{r} + \widetilde{\alpha}_n^{*} e^{i\theta} \sinh{r}$, with $r$ being the squeezing amplitude, $\gamma=r e^{i\theta}$ and from second to third equality we have used the displacement property $\hat{D}(\alpha)\ket{\beta} = e^{i\Im(\alpha\beta^*)} \ket{\alpha+\beta}$ (up to global phases). The term $\Theta_{nn'}$ collects the phase factors arising from the displacement algebra.
To obtain the reduced photon density matrix, we trace out the gravitational degrees of freedom:
\begin{widetext}
\begin{subequations}
\begin{align}
    \hat{\rho}_{\mathrm{ph}}(t) &= \mathrm{Tr}_{\mathrm{gw}} [\hat{\rho}(t)] \\
    &= \sum_{n,n^\prime,m} \rho_{nn^\prime} \ket{n}\bra{n^\prime} \times \bra{m}\hat{D}(\widetilde{\alpha}_n) \hat{S}(\gamma) \hat{D}(\beta)\ket{l} \bra{l} \hat{D}^{\dagger}(\beta) \hat{S}^{\dagger}(\gamma)  \hat{D}^\dagger(\widetilde{\alpha}_{n^\prime}) \ket{m} \\
    %&= \sum_{n,n^\prime,m} \rho_{nn^\prime} \ket{n}\bra{n^\prime} \times  \bra{k} \hat{D}^{\dagger}(\beta) \hat{S}^{\dagger}(\gamma)  \hat{D}^\dagger(\widetilde{\alpha}_{n^\prime}) \ket{m} \bra{m}\hat{D}(\widetilde{\alpha}_n) \hat{S}(\gamma) \hat{D}(\beta)\ket{k} \\
    &= \sum_{n,n^\prime} \rho_{nn^\prime} \ket{n}\bra{n^\prime} \times  \bra{l} \hat{D}^{\dagger}(\beta) \hat{S}^{\dagger}(\gamma)  \hat{D}^\dagger(\widetilde{\alpha}_{n^\prime})\hat{D}(\widetilde{\alpha}_n) \hat{S}(\gamma) \hat{D}(\beta)\ket{l} \\
    &= \sum_{n,n^\prime} \rho_{nn^\prime} \ket{n}\bra{n^\prime} \times  e^{\frac{1}{2}(\widetilde{\alpha}_n \widetilde{\alpha}_{n^\prime}^* - \widetilde{\alpha}_{n^\prime} \widetilde{\alpha}_n^*)} \bra{l} \hat{D}^{\dagger}(\beta) \hat{S}^{\dagger}(\gamma) \hat{D}(\widetilde{\alpha}_n-\widetilde{\alpha}_{n^\prime}) \hat{S}(\gamma) \hat{D}(\beta)\ket{l} \\
    &= \sum_{n,n^\prime} \rho_{nn^\prime} \ket{n}\bra{n^\prime} \times  \bra{l} \hat{D}^{\dagger}(\beta) \hat{D}(\zeta_{nn^\prime})  \hat{D}(\beta)\ket{l} \times e^{i\mathrm{Im}\l(\widetilde{\alpha}_{n} \widetilde{\alpha}^{*}_{n^{\prime}}\r)}
    \label{eqn:rho-photon-interm}
\end{align}
\end{subequations}
\end{widetext}
where, we used the property of the displacement operator and the squeezing operators. The displacement operators in the last line, $\hat{D}^{\dagger}(\beta) \hat{D}(\zeta_{nn^\prime})  \hat{D}(\beta)$, will reduce to $e^{\mathrm{Im}(\zeta_{nn^\prime} \beta^*)}\hat{D}(\zeta_{nn^\prime})$ using the property of the displacement operator. 
%\color{blue}
The parameter $\zeta_{nn^\prime}$ is given by,
\begin{widetext}
\begin{align}
\zeta_{nn^\prime} &= (\widetilde{\alpha}_n-\widetilde{\alpha}_{n^\prime}) \cosh{r} + (\widetilde{\alpha}_n-\widetilde{\alpha}_{n^\prime})^{*} e^{i\varphi} \sinh{r}    \nonumber \\
& = \frac{1}{4\sqrt{2\omega_{\mathbf{q}}V}} \sum_{\mathbf{k}\mathbf{k}^\prime,\lambda\lambda^\prime} \Big[\left(\omega_\mathbf{k} \, \hat{n}_{\mathbf{k},\lambda} \, \alpha^{(\sigma,\lambda)}_{\mathbf{k},\mathbf{q}}(t)- \omega_{\mathbf{k}^\prime} \, \hat{n}_{\mathbf{k}^\prime,\lambda^\prime} \, \alpha^{(\sigma,\lambda^\prime)}_{\mathbf{k}^\prime,\mathbf{q}}(t)\right)\cosh r \nonumber \\
& \qquad \qquad \qquad \qquad\quad + \left(\omega_\mathbf{k} \, \hat{n}_{\mathbf{k},\lambda} \, \alpha^{(\sigma,\lambda)}_{\mathbf{k},\mathbf{q}}(t)- \omega_{\mathbf{k}^\prime} \, \hat{n}_{\mathbf{k}^\prime,\lambda^\prime} \, \alpha^{(\sigma,\lambda^\prime)}_{\mathbf{k}^\prime,\mathbf{q}}(t)\right)^* e^{i\varphi}\sinh r \Big]
\label{eqn:zeta-param}
\end{align}
\end{widetext}
\color{black}
Using the representation of the coherent number state in terms of the Fock state from Ref. \cite{mvs:prd32}, the overlap of the Fock state with the coherent number state in the last step can be obtained as, 
\begin{align}
    \braket{l}{l,\zeta_{nn^\prime}} &= e^{-\frac{1}{2}|\zeta_{nn^\prime}|^2} \times \nonumber \\
    &\qquad \sum_{m=0}^{\infty} \left(\frac{l!}{m!} \right)^{1/2} L_{l}^{m-l}(|\zeta_{nn^\prime}|^2) \,\zeta_{nn^\prime}^{m-l} \braket{l}{m}~, \nonumber \\
    &= e^{-\frac{1}{2}|\zeta_{nn^\prime}|^2}  L_{l}\l(|\zeta_{nn^\prime}|^2\r)
    \label{eqn:squeezed-num-overlap}
\end{align}
where, $L_{l}$ is the Laguerre polynomial. The above equation gives the expectation value of the displacement operator in a specific excited state. Then the final reduced photon density matrix is given by,
\begin{widetext}
\begin{equation}
    \!\!\!\!\hat{\rho}_{\mathrm{ph}}(t)= \sum_{n,n^\prime} \rho_{nn^\prime} \ket{n}\bra{n^\prime} e^{i\mathrm{Im}\l(\widetilde{\alpha}_{n} \widetilde{\alpha}^{*}_{n^{\prime}}\r)} e^{2i\mathrm{Im}(\zeta_{nn^\prime} \beta^*)} e^{-\frac{1}{2}|\zeta_{nn^\prime}|^2} L_l (|\zeta_{nn^\prime}|^2 )~.
    \label{eqn:rho-evolved-general}     
\end{equation}
\end{widetext}
Here, $e^{i\mathrm{Im}\l(\widetilde{\alpha}_{n} \widetilde{\alpha}^{*}_{n^{\prime}}\r)}$ and $e^{2i\mathrm{Im}(\zeta_{nn^\prime} \beta^*)}$ are the unitary phase term. All the remaining terms such as $e^{-\frac{1}{2}|\zeta_{nn^\prime}|^2} L_l (|\zeta_{nn^\prime}|^2 )$ contribute to decoherence terms. We will take different limits to show what the noise signatures will look like for different choices of graviton states. An important result on coherent state of gravitons that is close to the classical GWs can be obtained by taking the limit, $l \to 0$.

\subsection{Interferometric visibility} \label{app:interf_visibility}
The reduced density matrix for the photons after interaction with the GW is given by:
\begin{equation}
    \hat{\rho}_{\mathrm{ph}}(t) = \sum_{n,n^\prime} e^{i \phi_{n,n^\prime}(t)} e^{-\frac{1}{2}|\zeta_{nn^\prime}|^2} \rho_{nn^\prime} \ket{n} \bra{n^\prime}~.
\end{equation}
The coincidence probability is evaluated by taking the expectation value of the joint detection projection operators at each arm of the HOM interferometer:
\begin{align}
    p_{cc} &= \sum_{m} \bra{m} \hat{\rho}_{\mathrm{ph}}(t) \mathbb{P}_{c}\otimes\mathbb{P}_d \ket{m} \nonumber \\
    & = \sum_{mn^\prime} e^{i \phi_{mn^\prime}(t)} e^{-\frac{1}{2}|\zeta_{mn^\prime}|^2} \rho_{mn^\prime} \bra{n^\prime} \mathbb{P}_{c}\otimes\mathbb{P}_d \ket{m}~.
\end{align}
The visibility is defined as:
\begin{equation}
    \mathcal{V} = \frac{p_{cc}^{\mathrm{max}}-p_{cc}^{\mathrm{min}}}{p_{cc}^{\mathrm{max}}}~.
\end{equation}
The maximum coincidence probability corresponds to the perfectly distinguishable case, denoted as $p_{cc}^{\mathrm{max}} = p_{0}$. The minimum coincidence occurs at zero time delay, where the photons are indistinguishable, yielding $p_{cc}^{\mathrm{min}} = p_{0} - p_{\mathrm{int}}e^{-\frac{1}{2}|\zeta|^2}$. Here, $p_{\mathrm{int}}$ represents the ideal interference amplitude at minimum coincidence without the decoherence factor. The visibility can then be written as:
\begin{equation}
    \mathcal{V} = \mathcal{V}_0 e^{-\frac{1}{2}|\zeta_{nn^\prime}|^2}~,
\end{equation}
where $\mathcal{V}_0$ is the intrinsic visibility of the interferometer in the absence of decoherence from the environment.

\vskip12pt
%\color{blue}
%
\subsection{Magnitude of $\zeta_{nn'}$} 
\label{App:zeta}

From Eqs.~\eqref{eqn:zeta-param} and  \eqref{eqn:alphatilde-def}, we see that 
$$\tilde\alpha_n-\tilde\alpha_{n'}\implies(n-n')\,\mathscr{A} \, $$ 
with $|\mathscr{A}|=\omega_k|g|\,h_q\,|\sin(\Omega t/2)|/(2\Omega)$, $h_\mathbf{q}=\kappa/\sqrt{2\omega_{\mathbf{q}}V}$ the single-graviton strain amplitude. The mode volume $V\sim L^3$ is chosen for the current calculations, where $L$ is the arm length of the detector. For estimating the visibility, the following choices have been made:
\begin{align}
    &\nu_{\mathrm{ph}} = 400\,\mathrm{THz};\; L=10^7\,\mathrm{m}; \;\nonumber\\ 
    &n_{\mathrm{ph}} \equiv \Gamma T \simeq (10^{12}\,\mathrm{s}^{-1})\times(10^5\,\mathrm{s});
\end{align}
The above choices lead to an estimate for $\widetilde{\alpha}$ as $\widetilde{\alpha} \simeq 5\times10^{-13}$. This value can be substituted directly into the visibility equation and plotted, as shown in Fig. \ref{fig:visibility-squeezing}. To obtain a rough estimate of the critical point can be obtained from the definition of $\zeta$, $\zeta \sim 5\times10^{-13} e^{r}$, where the squeezing angle is not taken into account. Since the visibility decays as $e^{-|\zeta|^2/2}$ and assuming the decay starts from $\zeta=0.5$, the squeezing parameter should be $e^{r}=10^{12}$ or $r\sim28$, which is evident from Fig. \ref{fig:visibility-squeezing}.
\color{black}

\section{Photon dynamics for coherent state of graviton} \label{app:coherent-state}

For the choice of coherent state of gravitons, which aligns very close to the classical GWs, the above Eq.~\eqref{eqn:rho-evolved-general} can be further simplified ($k\to 0$) into:
\begin{widetext}
\begin{align}
    \hat{\rho}_{\mathrm{ph}}(t) = \sum_{n,n^\prime} \rho_{nn^\prime} \ket{n}\bra{n^\prime} \underbrace{\exp\left( 2   i \Im \left[ (\widetilde{\alpha}_n - \widetilde{\alpha}_{n^\prime}) \beta^* + \frac{1}{2} \widetilde{\alpha}_{n}\widetilde{\alpha}^{*}_{n^\prime}\right] \right)}_{\text{Unitary Phase (Signal)}} \underbrace{\exp\left( - \frac{1}{2} |\widetilde{\alpha}_n - \widetilde{\alpha}_{n^\prime}|^2 \right)}_{\text{Decoherence (Noise)}}~.
    \label{eqn:rho_evolved}
\end{align}
\end{widetext}
The second exponential represents decoherence due to entanglement with the graviton vacuum (spontaneous emission of gravitons), which is negligible for laboratory parameters. The first exponential contains the \emph{GW signal}.

For the two-photon state traversing the interferometer arms (the HOM input state $\ket{1,1}\bra{1,1}$), the relative phase accumulated between the distinct optical paths is determined by the term linear in $\beta$.
%For a single photon traversing the arm (HOM input state $\ket{1}\bra{1}$ vs vacuum $\ket{0}\bra{0}$), the relative phase accumulated is determined by the term linear in $\beta$. 

To identify $\beta$, consider the graviton coherent state defined by $\hat{b}_{\mathbf{q}}\ket{\beta} = \beta_{\mathbf{q}} \ket{\beta}$. To map this quantum state to the classical field, the classical GW strain must correspond to the expectation value of $\hat{h}_{ij}$ in this coherent state,
\begin{equation}
    \bra{\beta} \hat{h}_{ij} \ket{\beta} = h_0 \varepsilon_{ij}^{(\sigma)}\,\cos(\mathbf{q}_{\mathrm{gw}}\cdot \mathbf{x}-\omega_\mathrm{gw} t)~.
\end{equation}
Using the definition for $\hat{h}_{ij}$ in Eq.~\eqref{eqn:hij-def}, this expectation value can be evaluated to:
\begin{align}
    \bra{\beta}\hat{h}_{ij} \ket{\beta} &= \sum_{q,\sigma} \frac{\kappa \,\varepsilon_{ij}^{(\sigma)}}{\sqrt{2\omega_{\mathbf{q}} V}} \Big( \beta_{\mathbf{q}}^{(\sigma)} e^{i(\mathbf{q}\cdot \mathbf{x}-\omega_\mathbf{q} t)} \nonumber \\
    & \qquad \qquad + \beta_{\mathbf{q}}^{(\sigma)*} e^{-i(\mathbf{q}\cdot \mathbf{x}-\omega_\mathbf{q} t)}\Big)~,
    \label{eqn:coherent-state-expect-value}
\end{align}
where we have assumed a normalized coherent state, $\braket{\beta}{\beta} = 1$. Assuming a continuous wave of frequency $\omega_{\mathrm{gw}}$, we set $\beta_{\mathbf{q}} = \beta_{\mathbf{q}_{\mathrm{gw}}} \delta_{\mathbf{q},\mathbf{q}_{\mathrm{gw}}}$. By equating this to the classical strain, we arrive at the \textit{identification}:
\begin{equation}
\beta_{\mathbf{q}} \approx \frac{\sqrt{2\omega_{\mathbf{q}}V}}{2\kappa} \, h_0 \, \delta_{\mathbf{q}, \mathbf{q}_{\mathrm{gw}}} \, . 
\end{equation}
Consequently, the phase $\phi_{\mathbf{k}}(t)$ acquired by the creation operator $\hat{a}^\dagger_{\mathbf{k}}$ is:
\begin{align}
    \phi_{\mathbf{k}}(t) &= 2 \Im \left( \widetilde{\alpha}_{n=1} \beta^* \right) \nonumber \\
    &= \Im \left[ \sum_{\lambda} \frac{\omega_\mathbf{k}}{2} \, {\mathcal{G}}^{(\sigma,\lambda)}_{\hat{\mathbf{k}},\hat{\mathbf{q}}} \left( i \int_0^t dt' e^{-i\Omega_{\mathbf{q}}t'} \right) h_0^{(\sigma)} \right] \nonumber \\
    &= -\omega_\mathbf{k} \left[ \frac{1}{2} \hat{k}^i \hat{k}^j h_0 \varepsilon_{ij} \int_0^t dt' \cos(\Omega_{\mathbf{q}}t') \right]~,
    \label{eqn:photon-phase}
\end{align}
where we used $\sum_{\lambda} {\mathcal{G}}^{(\sigma,\lambda)}_{\hat{\mathbf{k}},\hat{\mathbf{q}}} = -2 \hat{\mathbf{k}}^i\hat{\mathbf{k}}^j\varepsilon^{(\hat{\mathbf{q}},\sigma)}_{ij}$ (see Appendix.~\ref{app:antenna-pattern-geo-factor}) and the phase $2i\mathrm{Im}(\widetilde{\alpha}_{n}\widetilde{\alpha}^{*}_{n^\prime})$ will vanish. This matches the standard expression for the GW-induced time delay $\Delta t = \frac{1}{2} \int h_{ij} n^i n^j dt$ derived in geometric optics~\cite{Finn:2008np}. Thus, our quantum field-theoretical derivation correctly recovers the classical phase shift in the limit of a coherent GW state, while providing the full density matrix evolution necessary for analyzing quantum interference.

%\color{blue}

\subsection{Mapping geometric factor to antenna pattern} \label{app:antenna-pattern-geo-factor}
The geometric factor defined in the main text is given by,
\begin{widetext}
\begin{equation}
    {\mathcal{G}}^{\sigma,\lambda,\lambda^\prime}_{\hat{\mathbf{k}},\hat{\mathbf{k}^\prime},\hat{\mathbf{q}}} =\l( \varepsilon^{*i}_{(\hat{\mathbf{k}},\lambda)} \varepsilon^{j}_{(\hat{\mathbf{k}}^\prime,\lambda^\prime)}+(\hat{\mathbf{k}}\times\varepsilon_{(\hat{\mathbf{k}},\lambda)})^{*i} (\hat{\mathbf{k}}^\prime\times\varepsilon_{(\hat{\mathbf{k}}^\prime,\lambda^\prime)})^j\r) \varepsilon_{ij}^{(\hat{\mathbf{q}},\sigma)}~.
\end{equation}
\end{widetext}
After performing the Dirac delta function integral and using $\mathbf{k}\approx\mathbf{k}^\prime$, this simplies to (in index notation),
\begin{equation}
    {\mathcal{G}}^{(\sigma,\lambda)}_{\hat{\mathbf{k}},\hat{\mathbf{q}}} =\l( \varepsilon^{*i}_{(\hat{\mathbf{k}},\lambda)} \varepsilon^{j}_{(\hat{\mathbf{k}},\lambda)}+\epsilon^{ilm}\epsilon^{jnp} \,\hat{\mathbf{k}}_{l} \hat{\mathbf{k}}_{n} \varepsilon_{m}^{(\hat{\mathbf{k}},\lambda)*} \varepsilon_{p}^{(\hat{\mathbf{k}},\lambda)}\r) \varepsilon_{ij}^{(\hat{\mathbf{q}},\sigma)}~.
\end{equation}
Using the completeness relation for the photon polarization vector, $\sum_{\lambda=1,2} \varepsilon^{i}_{\lambda} \varepsilon^{*j}_{\lambda}=\delta^{ij}-\hat{\mathbf{k}}^{i}\hat{\mathbf{k}}^{j}$, then the above equation becomes,
\begin{align}
    \sum_{\lambda=1,2}&{\mathcal{G}}^{(\sigma,\lambda)}_{\hat{\mathbf{k}},\hat{\mathbf{q}}} \nn \\ &=\Big( (\delta^{ij} - \hat{\mathbf{k}}^{i}\hat{\mathbf{k}}^{j}) + \epsilon^{ilm}\epsilon^{jnp} \,\hat{\mathbf{k}}_{l} \hat{\mathbf{k}}_{n} (\delta_{mp} - \hat{\mathbf{k}}_{m} \hat{\mathbf{k}}_{p})\Big)\times \varepsilon_{ij}^{(\hat{\mathbf{q}},\sigma)}~.
\end{align}
Noting that $\delta^{ij}\varepsilon^{(\hat{\mathbf{q}},\sigma)}_{ij} = 0$, since $\varepsilon^{(\hat{\mathbf{q}},\sigma)}_{ij}$ is transverse traceless, the geometric factor simplifies to,
\begin{equation}
    \sum_{\lambda=1,2} {\mathcal{G}}_{\hat{\mathbf{k}},\hat{\mathbf{q}}} = -2 \,\hat{\mathbf{k}}^{i} \hat{\mathbf{k}}^{j}\varepsilon^{(\hat{\mathbf{q}},\sigma)}_{ij}~,
\end{equation}
which corresponds to the antenna pattern function of the GW detector.

\section{Comparison with Classical Interferometry} \label{app:comparison-classical-interf}

While the derived power spectral density $S_n \propto \Gamma^{-1}$ exhibits the fundamental shot-noise scaling, it is crucial to properly contextualize this against standard first-order ($g^{(1)}$) interferometry. A classical macroscopic Michelson interferometer operates with continuous-wave laser powers yielding a photon flux $N$ many orders of magnitude higher than the SPDC pair-production rate $\Gamma$ achievable in quantum setups. Consequently, for the detection of classical, coherent GW strains, a traditional Michelson natively possesses a superior raw strain sensitivity. 

\vskip12pt
%\color{blue}
To compare with Michelson interferometer (shot-noise-limited) with the same arm length $L$, optical frequency $\omega_{\mathrm{ph}}$, we use the standard shot noise SNR for Michelson for monochromatic signal (see Ref. \cite{Maggiore:2007ulw}) with no cavities or power recycling and mid-fringe readout (with $G=\mathcal{C}=1$):
\begin{equation}
\left(\frac{\mathcal{S}}{\mathcal{N}}\right)_{\mathrm{Mich}} = \left(\frac{P_0 T}{2\hbar\omega_{\mathrm{ph}}}\right)^{1/2}\frac{2\omega_{\mathrm{ph}}L}{c}h_0~,
\end{equation}
where, $P_0$ is the output power of the laser. Then the ratio of the SNR in both cases can be written as,
\begin{equation}
\frac{(\mathcal{S}/\mathcal{N})_{\mathrm{HOM}}}{(\mathcal{S}/\mathcal{N})_{\mathrm{Mich}}} = \frac{{\sigma}\sqrt{\Gamma\hbar}}{{\sqrt{P_0\omega_{\mathrm{ph}}}}}~.
\end{equation}
\vskip12pt

Then for the output power of Michelson, $P_0=$1\,$\mu$W, 1\,mW and 1\,W, the ratio can be evaluated as --- $10^{-3},\,5\times10^{-5},\,10^{-6}$ respectively. The HOM scheme is therefore less sensitive in raw strain than a shot-noise-limited Michelson, because of the lower SPDC pair rate and the THz-scale spectral width $\sigma\ll\omega_{\rm ph}$. %Any increase in spectral width, for eg., $\sigma=10^{14}\,\mathrm{s}^{-1}$, state-of-the-art, would increase the ratio by one more order.
Its advantage is the insensitivity to optical-phase (fringe) locking and to laser intensity noise.%to first order%; differential-delay noise enters as in a Michelson

\vskip12pt
Thus, while HOM interferometry is subject to stricter shot-noise bounds due to lower photon flux, it provides a possible route to quantum nature of gravitational waves.

\color{black}

\section{Details of the Proposed Interferometer Configuration} \label{app:config-details}
%%%%%%%%%%%%%%%%%%%%%%%%%%%%%%%%%%%%%%%%%%%%%%%%%%%%%%%%%%%%%%%%%%%%%%%%%%%%%%

In the previous section, we established that the photon-graviton scattering induces a phase shift on the EM field mode. For a coherent GW background, this phase shift can be mapped to an effective variation in the optical path length (or time delay) experienced by the photon. Following the formalism of Ref.~\cite{Finn:2008np}, the effective time delay $\Delta t$ for a photon traversing a path of length $L$ with initial position $\mathbf{x}_0$ at time $t_0$ is given by:
\begin{equation}
    c \Delta t = L + \frac{1}{2} \hat{k}^i \hat{k}^j \int_{t_0}^{t_0+L/c} dt \, h_{ij}(t, \mathbf{x}_{\gamma}(t))~,
    \label{eqn:time-delay-general}
\end{equation}
where $\hat{k}^i$ is the unit vector along the photon trajectory, $\mathbf{x}_{\gamma}(t) = \mathbf{x}_0 + c \hat{\mathbf{k}}(t - t_0)$ is the unperturbed photon path, and $h_{ij}$ is the GW metric perturbation. For a plane wave propagating along direction $\hat{\mathbf{q}}$ with frequency $\omega_{\mathrm{gw}}$, this integrates to:
\begin{align}
    c \Delta t &= L + \frac{1}{2} \hat{k}^i \hat{k}^j (h_0 \varepsilon_{ij}) \times \nn \\
    &\qquad\int_{t_0}^{t_0+L/c} dt \, \cos\left[ \omega_{\mathrm{gw}} \left( t - \hat{\mathbf{q}} \cdot (\mathbf{x}_0 + c \hat{\mathbf{k}}(t-t_0))/c \right) \right] \nonumber \\
    &= L + \frac{L}{2} h_0 \mathscr{F}(k, \varepsilon) \, \mathrm{sinc}\left( \frac{\omega_{\mathrm{eff}} L}{2c} \right) \cos\left( \omega_{\mathrm{gw}} t_{\mathrm{mid}} - \Phi_{\mathrm{gw}} \right)~,
    \label{eqn:time-delay}
\end{align}
where $\mathscr{F}(k, \varepsilon) = \hat{k}^i \hat{k}^j \varepsilon_{ij}$ is the geometric projection factor. We define the effective frequency seen by a photon in a segment with direction $\hat{\mathbf{k}}$ as $\omega_{\mathrm{eff}} = \omega_{\mathrm{gw}}(1 - \hat{\mathbf{q}} \cdot \hat{\mathbf{k}})$.

%%%%%%%%%%%%%%%%%%%%%%%%%%%%%
\subsection{2D Planar setup} \label{app:config-2D}
%%%%%%%%%%%%%%%%%%%%%%%%%%%%%

We consider the planar configuration depicted in the main text (Fig. 5). The setup is symmetric about the $y$-axis. The source is located at the origin $O$. The photons travel to intermediate mirrors at $A$ and $B$, and reflect to a final beam splitter at $P$. The unit vectors for the four optical segments are:
%
%%%%%%%%%%%%%%%%
\begin{subequations}
\begin{align}
    \hat{k}^{i}_{\small{\rm OA}} &= \Big(\cos \frac{\pi}{12},\, \sin\frac{\pi}{12},\, 0\Big) \\
    \hat{k}^{i}_{\small{\rm AP}} &= \Bigg(\cos \left(\frac{13\pi}{12} -\vartheta\right),\, \sin \left(\frac{13\pi}{12} -\vartheta\right),\, 0 \Bigg) \nonumber \\
    & = \Bigg(-\cos \left(\frac{\pi}{12} -\vartheta\right),\, -\sin \left(\frac{\pi}{12} -\vartheta\right),\, 0 \Bigg) \\
    \hat{k}^{i}_{\small{\rm OB}} &= \Big(\sin \frac{\pi}{12},\, \cos\frac{\pi}{12},\, 0 \Big) \\
    \hat{k}^{i}_{\small{\rm BP}} &= \Bigg(\cos \left(\frac{7\pi}{12} -\vartheta\right),\, -\sin \left(\frac{7\pi}{12} -\vartheta\right),\, 0 \Bigg) \nonumber \\
    &= \Bigg(-\sin \left(\frac{\pi}{12} -\vartheta\right),\, -\cos \left(\frac{\pi}{12} -\vartheta\right),\, 0 \Bigg) 
\end{align}
\end{subequations}
%%%%%%%%%%%%%%%%
%
For arm 1 $(OAP)$, the time taken by a photon to travel the first leg $(OA)$ is:
%
%%%%%%%%%%%%%%%%
\begin{widetext}
\begin{align}
    c\Delta t_{\small{\rm OA}} & = L + \frac{1}{2} \hat{k}^{l}_{\small{\rm OA}} \hat{k}^{m}_{\small{\rm OA}}\,\overset{\rm o}{h}_{l m} \int_{t_0}^{t_0+ L/c} \cos[\omega_{\rm gw} (t-(t-t_0)q_{i}\hat{k}^{i}_{\small{\rm OA}})] \nonumber \\ 
    & = L + \f{L}{2}\left( h_{xx} \cos^2 \frac{\pi}{12} + h_{yy} \sin^2 \frac{\pi}{12} + h_{xy} \sin\frac{\pi}{6} \right) \cos \left({\omega_{\rm gw} t_0 + \omega_{\small{\rm OA}} L/2}\right) \,\sinc\l({\omega_{\rm OA }L /2}\r)
\end{align}
\end{widetext}
%%%%%%%%%%%%%%%%
%
where, $\omega_{\rm OA}=\omega_{\rm gw}\left( 1-\cos{[\frac{\pi}{12} - \phi] \sin{\theta}}\right)$. % 
Similarly, using Eq.~\eqref{eqn:time-delay} the remaining time delays can be obtained. 
%
%%%%%%%%%%%%%%%%
\begin{widetext}
\begin{subequations}
\begin{align}
    c\Delta t_{\small{\rm OB}} & = L \l[ 1+ \frac{1}{2}\left( h_{xx} \sin^2 \frac{\pi}{12} + h_{yy} \cos^2 \frac{\pi}{12} + h_{xy} \sin\frac{\pi}{6} \right) \cos \left({\omega_{\rm gw} t_0 + \omega_{\small{\rm OB}} L/2}\right) \sinc{(\omega_{\rm OB }L /2}) \r] \\ 
    c\Delta t_{\small{\rm AP}} & = \ell_{\rm AP} \Bigg[1+ \frac{1}{2} \Bigg( h_{xx} \cos^2 \l(\frac{\pi}{12}-\vartheta\r) + h_{yy} \sin^2 \l(\frac{\pi}{12}-\vartheta\r) + h_{xy} \sin \l( \frac{\pi}{6} -2 \vartheta\r) \Bigg)  \nn \\ 
    &  \qquad\qquad \qquad \times \, \sinc{\l(\omega_{\rm AP} \,\ell_{\rm AP}/2\r)}\, \cos \left({\omega_{\rm gw} t_0 + \omega_{\small{\rm AP}} \left(L+\ell_{\rm AP}/2\right)}\right)  \Bigg]\\    
    c\Delta t_{\small{\rm BP}} & = \ell_{\rm BP} \Bigg[1+ \frac{1}{2} \Bigg( h_{xx} \sin^2 \l(\frac{\pi}{12}-\vartheta\r) + h_{yy} \cos^2 \l(\frac{\pi}{12}-\vartheta\r) + h_{xy} \sin \l( \frac{\pi}{6} -2 \vartheta\r) \Bigg)  \nn \\ 
    &  \qquad\qquad \qquad \times \, \sinc{\l(\omega_{\rm BP} \,\ell_{\rm BP}/2\r)}\, \cos \left({\omega_{\rm gw} t_0 + \omega_{\small{\rm BP}} \left(L+\ell_{\rm BP}/2\right) }\right)  \Bigg]
\end{align}
\end{subequations}
\end{widetext}
%%%%%%%%%%%%%%%%
%
The length of the arms $AP$ and $BP$ is $\ell_{\rm AP}=\ell_{\rm BP}=L/(2\sin(\pi/6 + \vartheta))$ used in the above equations. In the long wavelength approximations, $\omega_{\rm gw} L/c \ll1$ the sinc function can be taken as 1 and $\omega_{\rm arm} \approx \omega_{\rm gw}$. The differential time delay between the two photons in the two different arms, is $\Delta t_{\rm net} = \Delta t_{\rm OA} + \Delta t_{\rm AP} - \Delta t_{\rm OB} - \Delta t_{\rm BP} $, and simplifying gives, $\Delta t_{\rm OA}$ and $\Delta t_{\rm AP} $ as,
%
%%%%%%%%%%%%%%%%
{\small{
\begin{subequations}
\begin{align}
    c\Delta t_{\small{\rm OA}} &- c\Delta t_{\small{\rm OB}}  \approx L \l[ \frac{\sqrt{3}}{4}( h_{xx} - h_{yy}) \cos \left({\omega_{\rm gw} (t_0 + L/2)}\right)  \r]~, \nn \\ 
    c\Delta t_{\small{\rm AP}} &- c\Delta t_{\small{\rm BP}}  \approx \ell_{\rm AP} \Bigg[\frac{1}{2} \Bigg( (h_{xx} - h_{yy}) \cos \l(\frac{\pi}{6}-2\vartheta\r) \Bigg)  \times \nn \\
    &\qquad \qquad \qquad \qquad\cos \left({\omega_{\rm gw} \left(t_0 + L+\ell_{\rm AP}/2\right)}\right)  \Bigg],  \nn 
\end{align}
\end{subequations}
}}
%%%%%%%%%%%%%%%%
%
and then the net time delay is given by,
%
%%%%%%%%%%%%%%%%
\begin{widetext}
\begin{align}
    c\Delta t_{\rm net} & \approx (h_{xx} - h_{yy})
    \Bigg[ \frac{\sqrt{3} L}{4}  \cos \left({\omega_{\rm gw} (t_0 + L/2)}\right) + \frac{\ell_{\rm AP}}{2} \Bigg(\cos \l(\frac{\pi}{6}-2\vartheta\r) \Bigg) \cos \left({\omega_{\rm gw} \left(t_0 + L+\ell_{\rm AP}/2\right)}\right)  \Bigg]~.
    \label{eqn:2D-diff-time-delay-app}
\end{align}
\end{widetext}
%%%%%%%%%%%%%%%%
%

%%%%%%%%%%%%%%%%%%%%%%%%%%%%%
\subsection{3D pyramidal setup} \label{app:config-3D}
%%%%%%%%%%%%%%%%%%%%%%%%%%%%%

We extend the analysis to the 3D pyramidal configuration proposed in the main text. We focus on "Detector 1," formed by the arms OAP and OBP.  The time required for the photon to travel OAP, $\Delta _{\mathrm{OAP}}$to $\mathcal{O}(h^2)$ can be found by further splitting it into intervals $\Delta t_{\mathrm{OA}}$ and $\Delta t_{\mathrm{AP}}$. Keeping $O$ as the origin, the photon propagation directional vectors for each interval $\mathrm{OA}$ and $\mathrm{AP}$ are: 
%
%%%%%%%%%%%%%%%%
\begin{align*}
\hat{k}^{i}_{OA} & = \left(\cos \frac{\pi}{12}, \, \sin \frac{\pi}{12},\, 0 \right)~, \\
\hat{k}^{i}_{AP} & = \left(\sin\frac{\pi}{12} \sin{\vartheta} - \cos{\frac{\pi}{12}},\, \cos\frac{\pi}{12} \sin{\vartheta}-\sin \frac{\pi}{12}, \, \cos{\vartheta}\right)~.
\end{align*}
%%%%%%%%%%%%%%%%
%
Then the corresponding time delay for $\vartheta=\pi/6$ is given by,
\begin{widetext}
\begin{align}
    c\Delta t_{OA} & = L + \frac{L}{2}\left(h_{xx} \cos^2\frac{\pi}{12} + h_{yy} \sin^2 \frac{\pi}{12} + 2 h_{xy} \cos\frac{\pi}{12} \sin\frac{\pi}{12}\right) \sinc(\omega_{OA} L/2) \cos{\omega_{\rm gw} t_0 + \omega_{OA} L/2}\\ 
    c\Delta t_{AP} & = L^\prime + \frac{L^\prime}{2}\Bigg(h_{xx} \left(\sin\frac{\pi}{12}\sin\vartheta - \cos \frac{\pi}{12} \right)^2+ h_{yy} \left( \cos\frac{\pi}{12} \sin\vartheta -\sin\frac{\pi}{12}\right)^2 +  h_{zz} \cos^2\vartheta \nonumber \\ 
    & \quad + 2 h_{xy} \left(\sin\frac{\pi}{12}\sin\vartheta - \cos \frac{\pi}{12} \right) \left( \cos\frac{\pi}{12} \sin\vartheta -\sin\frac{\pi}{12}\right)  + 2 h_{xz} \left(\sin\frac{\pi}{12}\sin\vartheta - \cos \frac{\pi}{12} \right)\cos \vartheta \nonumber \\   
    & \quad + 2 h_{yz} \left(\cos\frac{\pi}{12}\sin\vartheta - \sin \frac{\pi}{12} \right)\cos \vartheta\Bigg) \sinc(\omega_{AP} L^\prime/2) \cos{\omega_{\rm gw} t_0 + \omega_{AP} L^\prime/2}
    \label{eqn:time-delay-3D}
\end{align}
\end{widetext}
where, $\omega_{OA} = \omega_{\rm gw} (1-\cos{[\pi/12 -\phi]}\sin\theta)$ and $\omega_{AP} = \omega_{\rm gw}(1+ \cos[\pi/12-\phi] \sin \theta - \cos\theta\cos\vartheta - \sin\vartheta\sin\theta\sin[\pi/12+\phi])$. Similarly, the time of flight of photon for the arm $OBP$ can be found.\\

%\textcolor{blue}{eqn for $\Delta t_{OB}$ and $\Delta t_{BP}$}. \\

To maximize the response of the detector, we can use the long wavelength approximation $\omega_{\rm gw} L/c \ll 1$ and the time delays can be simplified. Assuming that $L=L^\prime$, and taking $\vartheta = \pi/6$ for symmetric configuration, the net time difference, $\Delta t_{\rm net} = \Delta t_{OAP} -\Delta t_{OBP} = (\Delta t_{OA} - \Delta t_{OB}) + (\Delta t_{AP} - \Delta t_{BP})$ can be reduced to a simpler form. 
\begin{align}
    \Delta t_{OA} - \Delta t_{OB} & = \frac{\sqrt{3}L}{2c} \left(h_{xx}  - h_{yy} \right) \\
    \Delta t_{AP} - \Delta t_{BP} & = \frac{L}{c}\Bigg[ \frac{1}{4}\left( h_{xx} + h_{yy} \right) + \frac{\sqrt{3}}{4} \left( h_{xx} - h_{yy}\right) \nn \\
    &\qquad\quad+ \sqrt{\frac{3}{2}} \left(h_{yz} - h_{xz}\right) - \frac{1}{2} h_{xy}\Bigg]
\end{align}
Hence the net time delay due to the passage of GW in the HOM interferometer is given by,
\begin{align}
    \frac{c\Delta t_{\rm net}}{L} = & \frac{1}{4}\left( h_{xx} + h_{yy} \right) + \frac{3\sqrt{3}}{4} \left( h_{xx} - h_{yy}\right) + \frac{1}{2} h_{xy} \nn\\
    & + \sqrt{\frac{3}{2}} \left(h_{yz} - h_{xz}\right)
\end{align}
Here, the components of the metric $h_{ij}$ are in the detector frame.

\subsection{Antenna pattern function for 3D (Pyramidal) configuration}

The GW strain signal $h(t)$ is defined as the fractional time delay, $h(t) \equiv c \Delta t_{\mathrm{net}} / L_{\mathrm{tot}}$. This scalar signal can be decomposed into the two polarization modes of the GW, $h_+$ and $h_\times$:
\begin{equation}
    h(t) = F_+(\theta, \phi, \psi) h_+(t) + F_\times(\theta, \phi, \psi) h_\times(t)~,
\end{equation}
where $F_{+,\times}$ are the antenna pattern functions, $(\theta, \phi)$ define the propagation direction of the wave, and $\psi$ is the polarization angle.

To derive $F_{+,\times}$, we transform the metric perturbation from the wave frame (where it is transverse-traceless) to the detector frame. The metric in the wave frame is:
\begin{equation}
    h'_{\mu\nu} = \begin{pmatrix}
        h_+ & h_\times & 0 \\
        h_\times & -h_+ & 0 \\
        0 & 0 & 0
    \end{pmatrix}_{\text{TT}}~.
\end{equation}
The transformation to the detector frame is given by $h_{ij} = \mathcal{R}_{ik} \mathcal{R}_{jl} h'_{kl}$, where $\mathcal{R}$ is the rotation matrix aligning the wave frame with the detector frame:
\begin{widetext}
\begin{align}
    \mathcal{R}(\theta, \phi, \psi) = \begin{pmatrix}
        \cos\psi & \sin\psi & 0 \\
        -\sin\psi & \cos\psi & 0 \\
        0 & 0 & 1
    \end{pmatrix}
    \begin{pmatrix}
        \cos\theta & 0 & -\sin\theta \\
        0 & 1 & 0 \\
        \sin\theta & 0 & \cos\theta
    \end{pmatrix}
    \begin{pmatrix}
        \cos\phi & \sin\phi & 0 \\
        -\sin\phi & \cos\phi & 0 \\
        0 & 0 & 1
    \end{pmatrix}~.
\end{align}
\end{widetext}
\begin{figure*}[!ht]
    \centering
    \includegraphics[width=\linewidth]{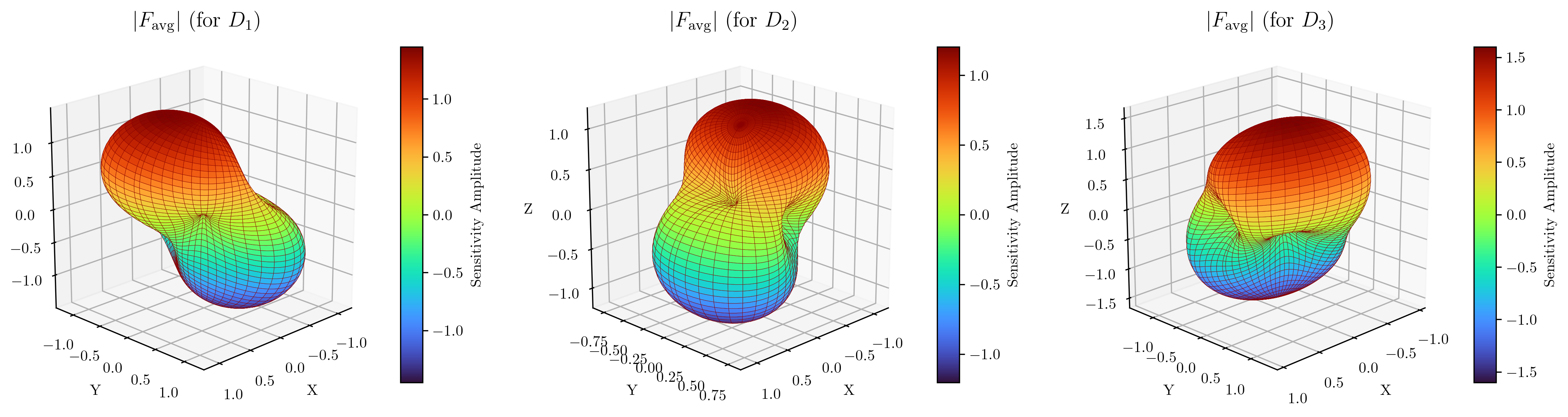}
    \caption{Directional sensitivity (antenna patterns) for the three individual detectors in the pyramidal
    %tetrahedral 
    configuration ($\vartheta = \pi/6$). Unlike planar detectors which share common null directions, the 3D geometry ensures that the maxima of one detector cover the minima of the others.}
    \label{fig:3D-antenna-each}
\end{figure*}
\begin{figure*}[!ht]
    \centering
    \includegraphics[width=\linewidth]{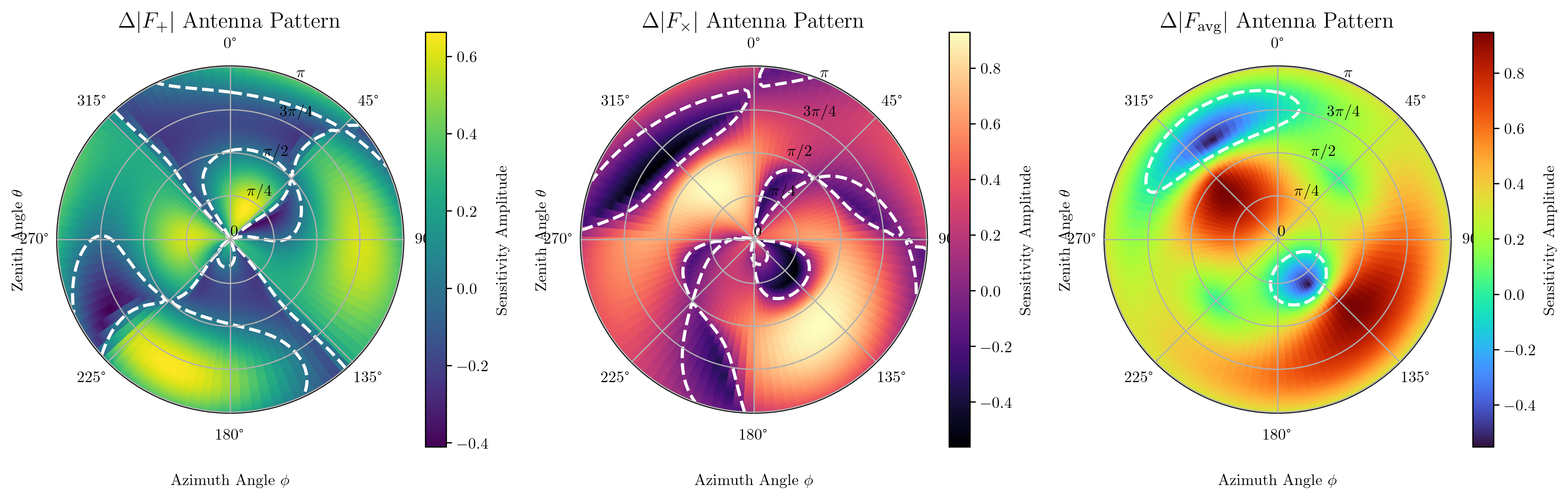}
    \caption{Comparison using polar sky map of 3D configuration with planar (for eg: LIGO and LISA) by taking the difference of antenna pattern functions, $\Delta F = |F_{D_1}| - |F_{\mathrm{planar}}|$. The average angular sensitivity for 3D configuration is better than the planar configuration (LIGO or LISA) which is only better in the region inside the dashed contour.}
    \label{fig:comparison-with-planar-indv}
\end{figure*}
Substituting the transformed components $h_{ij}$ into Eq.~\eqref{eqn:time-delay-3D} and extracting the coefficients of $h_+$ and $h_\times$, we obtain the antenna pattern functions for Detector 1 as:
\begin{widetext}
\begin{align}
    F_{+}\Big|_{{D_1}} = \frac{1}{16} & \bigg[ 2 \sqrt{6} \cos(2 \psi) \sin(2 \theta) (\cos\phi + \sin\phi) + \cos(2 \theta) \cos(2 \psi) (\cos\phi + \sin\phi) \big( (1 + 3\sqrt{3}) \cos\phi + (1 - 3\sqrt{3}) \sin\phi\big) \nonumber \\
    & + \cos(2 \psi) \big(-1 + 9 \sqrt{3} \cos(2 \phi) + 3 \sin(2 \phi)\big) + 4 \sqrt{6} \sin\theta (\cos\phi - \sin\phi) \sin(2 \psi) \nonumber \\
    & + 4 \cos\theta \sin(2 \psi) \big(\cos(2 \phi) - 3 \sqrt{3} \sin(2 \phi)\big)  \bigg]~,
\end{align}
\begin{align}
    F_{\times}\Big|_{D_1} = \frac{1}{16} \bigg[ &
    -4\sqrt{6}\cos(2\psi)\sin\theta(\cos\phi-\sin\phi) + \cos(2\theta)\sin(2\psi)(\cos\phi+\sin\phi) 
    \big( (1 + 3\sqrt{3})\cos\phi + (1 - 3\sqrt{3})\sin\phi\big) \nonumber \\
    & + \sin(2\psi)\big(-1+9\sqrt{3}\cos(2\phi)+3\sin(2\phi)\big) + 2\sqrt{6}\sin(2\theta)(\cos\phi+\sin\phi)\sin(2\psi) \nonumber \\
    & - 4\cos\theta\cos(2\psi)\big(\cos(2\phi)-3\sqrt{3}\sin(2\phi)\big) \bigg]~.
\end{align}
\end{widetext}
The corresponding antenna patterns for Detectors 2 and 3 are obtained by $\pm 120^\circ$ rotations in the azimuth $\phi$. The averaged response of the full network is discussed in the main text. The individual antenna responses are given in Fig.~\ref{fig:3D-antenna-each} and the comparison with planar setup (LIGO) is shown in Fig.~\ref{fig:comparison-with-planar-indv}

\newpage

%%%%%%%%%%%%%%%%%%%%%%%%%%%%%%%%%%%%%%%%%%%%%%%%%%%%%%%%%%%%%%%%%%%%%%%%%%%%%%
\bibliographystyle{apsrev4-1}
%\bibliographystyle{unsrt} % or apsrev4-2 for APS stylet
%\bibliography{References}

\begin{thebibliography}{71}%
\makeatletter
\providecommand \@ifxundefined [1]{%
 \@ifx{#1\undefined}
}%
\providecommand \@ifnum [1]{%
 \ifnum #1\expandafter \@firstoftwo
 \else \expandafter \@secondoftwo
 \fi
}%
\providecommand \@ifx [1]{%
 \ifx #1\expandafter \@firstoftwo
 \else \expandafter \@secondoftwo
 \fi
}%
\providecommand \natexlab [1]{#1}%
\providecommand \enquote  [1]{``#1''}%
\providecommand \bibnamefont  [1]{#1}%
\providecommand \bibfnamefont [1]{#1}%
\providecommand \citenamefont [1]{#1}%
\providecommand \href@noop [0]{\@secondoftwo}%
\providecommand \href [0]{\begingroup \@sanitize@url \@href}%
\providecommand \@href[1]{\@@startlink{#1}\@@href}%
\providecommand \@@href[1]{\endgroup#1\@@endlink}%
\providecommand \@sanitize@url [0]{\catcode `\\12\catcode `\$12\catcode `\&12\catcode `\#12\catcode `\^12\catcode `\_12\catcode `\%12\relax}%
\providecommand \@@startlink[1]{}%
\providecommand \@@endlink[0]{}%
\providecommand \url  [0]{\begingroup\@sanitize@url \@url }%
\providecommand \@url [1]{\endgroup\@href {#1}{\urlprefix }}%
\providecommand \urlprefix  [0]{URL }%
\providecommand \Eprint [0]{\href }%
\providecommand \doibase [0]{http://dx.doi.org/}%
\providecommand \selectlanguage [0]{\@gobble}%
\providecommand \bibinfo  [0]{\@secondoftwo}%
\providecommand \bibfield  [0]{\@secondoftwo}%
\providecommand \translation [1]{[#1]}%
\providecommand \BibitemOpen [0]{}%
\providecommand \bibitemStop [0]{}%
\providecommand \bibitemNoStop [0]{.\EOS\space}%
\providecommand \EOS [0]{\spacefactor3000\relax}%
\providecommand \BibitemShut  [1]{\csname bibitem#1\endcsname}%
\let\auto@bib@innerbib\@empty
%</preamble>
\bibitem [{\citenamefont {Milonni}(1994)}]{Milonni:1994xx}%
  \BibitemOpen
  \bibfield  {author} {\bibinfo {author} {\bibfnamefont {P.~W.}\ \bibnamefont {Milonni}},\ }\href@noop {} {\emph {\bibinfo {title} {{The Quantum vacuum: An Introduction to quantum electrodynamics}}}}\ (\bibinfo  {publisher} {Academic Press},\ \bibinfo {year} {1994})\BibitemShut {NoStop}%
\bibitem [{\citenamefont {Cohen-Tannoudji}\ \emph {et~al.}(1987)\citenamefont {Cohen-Tannoudji}, \citenamefont {Dupont-Roc},\ and\ \citenamefont {Grunberg}}]{Cohen-Tannoudji:1987oxb}%
  \BibitemOpen
  \bibfield  {author} {\bibinfo {author} {\bibfnamefont {C.}~\bibnamefont {Cohen-Tannoudji}}, \bibinfo {author} {\bibfnamefont {J.}~\bibnamefont {Dupont-Roc}}, \ and\ \bibinfo {author} {\bibfnamefont {G.}~\bibnamefont {Grunberg}},\ }\href@noop {} {\emph {\bibinfo {title} {{Photons and Atoms: Introduction to quantum electrodynamics}}}}\ (\bibinfo {year} {1987})\BibitemShut {NoStop}%
\bibitem [{\citenamefont {Peskin}\ and\ \citenamefont {Schroeder}(1995)}]{Peskin:1995ev}%
  \BibitemOpen
  \bibfield  {author} {\bibinfo {author} {\bibfnamefont {M.~E.}\ \bibnamefont {Peskin}}\ and\ \bibinfo {author} {\bibfnamefont {D.~V.}\ \bibnamefont {Schroeder}},\ }\href {\doibase 10.1201/9780429503559} {\emph {\bibinfo {title} {{An Introduction to quantum field theory}}}}\ (\bibinfo  {publisher} {Addison-Wesley},\ \bibinfo {address} {Reading, USA},\ \bibinfo {year} {1995})\BibitemShut {NoStop}%
\bibitem [{\citenamefont {Mandel}\ and\ \citenamefont {Wolf}(1995)}]{Mandel:1995seg}%
  \BibitemOpen
  \bibfield  {author} {\bibinfo {author} {\bibfnamefont {L.}~\bibnamefont {Mandel}}\ and\ \bibinfo {author} {\bibfnamefont {E.}~\bibnamefont {Wolf}},\ }\href {\doibase 10.1017/CBO9781139644105} {\emph {\bibinfo {title} {{Optical Coherence and Quantum Optics}}}}\ (\bibinfo  {publisher} {Cambridge Univ. Pr.},\ \bibinfo {year} {1995})\BibitemShut {NoStop}%
\bibitem [{\citenamefont {Saulson}(2017)}]{Saulson:2017jlf}%
  \BibitemOpen
  \bibfield  {author} {\bibinfo {author} {\bibfnamefont {P.~R.}\ \bibnamefont {Saulson}},\ }\href {\doibase 10.1142/10116} {\emph {\bibinfo {title} {{Fundamentals of Interferometric Gravitational Wave Detectors}}}},\ \bibinfo {edition} {2nd}\ ed.\ (\bibinfo  {publisher} {World Scientific},\ \bibinfo {year} {2017})\BibitemShut {NoStop}%
\bibitem [{\citenamefont {Abbott}\ \emph {et~al.}(2016)\citenamefont {Abbott} \emph {et~al.}}]{LIGOScientific:2016ao}%
  \BibitemOpen
  \bibfield  {author} {\bibinfo {author} {\bibfnamefont {B.~P.}\ \bibnamefont {Abbott}} \emph {et~al.} (\bibinfo {collaboration} {LIGO Scientific, Virgo}),\ }\href {\doibase 10.1103/PhysRevLett.116.061102} {\bibfield  {journal} {\bibinfo  {journal} {Phys. Rev. Lett.}\ }\textbf {\bibinfo {volume} {116}},\ \bibinfo {pages} {061102} (\bibinfo {year} {2016})},\ \Eprint {http://arxiv.org/abs/1602.03837} {arXiv:1602.03837 [gr-qc]} \BibitemShut {NoStop}%
\bibitem [{\citenamefont {Gefen}\ \emph {et~al.}(2024)\citenamefont {Gefen}, \citenamefont {Tarafder}, \citenamefont {Adhikari},\ and\ \citenamefont {Chen}}]{Gefen:2022ztc}%
  \BibitemOpen
  \bibfield  {author} {\bibinfo {author} {\bibfnamefont {T.}~\bibnamefont {Gefen}}, \bibinfo {author} {\bibfnamefont {R.}~\bibnamefont {Tarafder}}, \bibinfo {author} {\bibfnamefont {R.~X.}\ \bibnamefont {Adhikari}}, \ and\ \bibinfo {author} {\bibfnamefont {Y.}~\bibnamefont {Chen}},\ }\href {\doibase 10.1103/PhysRevLett.132.020801} {\bibfield  {journal} {\bibinfo  {journal} {Phys. Rev. Lett.}\ }\textbf {\bibinfo {volume} {132}},\ \bibinfo {pages} {020801} (\bibinfo {year} {2024})},\ \Eprint {http://arxiv.org/abs/2209.02998} {arXiv:2209.02998 [quant-ph]} \BibitemShut {NoStop}%
\bibitem [{\citenamefont {Yu}\ \emph {et~al.}(2022)\citenamefont {Yu}, \citenamefont {Martynov}, \citenamefont {Adhikari},\ and\ \citenamefont {Chen}}]{Yu:2022upc}%
  \BibitemOpen
  \bibfield  {author} {\bibinfo {author} {\bibfnamefont {H.}~\bibnamefont {Yu}}, \bibinfo {author} {\bibfnamefont {D.}~\bibnamefont {Martynov}}, \bibinfo {author} {\bibfnamefont {R.~X.}\ \bibnamefont {Adhikari}}, \ and\ \bibinfo {author} {\bibfnamefont {Y.}~\bibnamefont {Chen}},\ }\href {\doibase 10.1103/PhysRevD.106.063017} {\bibfield  {journal} {\bibinfo  {journal} {Phys. Rev. D}\ }\textbf {\bibinfo {volume} {106}},\ \bibinfo {pages} {063017} (\bibinfo {year} {2022})},\ \Eprint {http://arxiv.org/abs/2205.14197} {arXiv:2205.14197 [gr-qc]} \BibitemShut {NoStop}%
\bibitem [{\citenamefont {Finn}(2009)}]{Finn:2008np}%
  \BibitemOpen
  \bibfield  {author} {\bibinfo {author} {\bibfnamefont {L.~S.}\ \bibnamefont {Finn}},\ }\href {\doibase 10.1103/PhysRevD.79.022002} {\bibfield  {journal} {\bibinfo  {journal} {Phys. Rev. D}\ }\textbf {\bibinfo {volume} {79}},\ \bibinfo {pages} {022002} (\bibinfo {year} {2009})},\ \Eprint {http://arxiv.org/abs/0810.4529} {arXiv:0810.4529 [gr-qc]} \BibitemShut {NoStop}%
\bibitem [{\citenamefont {Rakhmanov}(2005)}]{Rakhmanov:2004eh}%
  \BibitemOpen
  \bibfield  {author} {\bibinfo {author} {\bibfnamefont {M.}~\bibnamefont {Rakhmanov}},\ }\href {\doibase 10.1103/PhysRevD.71.084003} {\bibfield  {journal} {\bibinfo  {journal} {Phys. Rev. D}\ }\textbf {\bibinfo {volume} {71}},\ \bibinfo {pages} {084003} (\bibinfo {year} {2005})},\ \Eprint {http://arxiv.org/abs/gr-qc/0406009} {arXiv:gr-qc/0406009} \BibitemShut {NoStop}%
\bibitem [{\citenamefont {Aggarwal}\ \emph {et~al.}(2021)\citenamefont {Aggarwal} \emph {et~al.}}]{Aggarwal:2020olq}%
  \BibitemOpen
  \bibfield  {author} {\bibinfo {author} {\bibfnamefont {N.}~\bibnamefont {Aggarwal}} \emph {et~al.},\ }\href {\doibase 10.1007/s41114-021-00032-5} {\bibfield  {journal} {\bibinfo  {journal} {Living Rev. Rel.}\ }\textbf {\bibinfo {volume} {24}},\ \bibinfo {pages} {4} (\bibinfo {year} {2021})},\ \Eprint {http://arxiv.org/abs/2011.12414} {arXiv:2011.12414 [gr-qc]} \BibitemShut {NoStop}%
\bibitem [{\citenamefont {Rothman}\ and\ \citenamefont {Boughn}(2006)}]{Rothman:2006fp}%
  \BibitemOpen
  \bibfield  {author} {\bibinfo {author} {\bibfnamefont {T.}~\bibnamefont {Rothman}}\ and\ \bibinfo {author} {\bibfnamefont {S.}~\bibnamefont {Boughn}},\ }\href {\doibase 10.1007/s10701-006-9081-9} {\bibfield  {journal} {\bibinfo  {journal} {Found. Phys.}\ }\textbf {\bibinfo {volume} {36}},\ \bibinfo {pages} {1801} (\bibinfo {year} {2006})},\ \Eprint {http://arxiv.org/abs/gr-qc/0601043} {arXiv:gr-qc/0601043} \BibitemShut {NoStop}%
\bibitem [{\citenamefont {Capper}\ \emph {et~al.}(1974)\citenamefont {Capper}, \citenamefont {Duff},\ and\ \citenamefont {Halpern}}]{PhysRevD.10.461}%
  \BibitemOpen
  \bibfield  {author} {\bibinfo {author} {\bibfnamefont {D.~M.}\ \bibnamefont {Capper}}, \bibinfo {author} {\bibfnamefont {M.~J.}\ \bibnamefont {Duff}}, \ and\ \bibinfo {author} {\bibfnamefont {L.}~\bibnamefont {Halpern}},\ }\href {\doibase 10.1103/PhysRevD.10.461} {\bibfield  {journal} {\bibinfo  {journal} {Phys. Rev. D}\ }\textbf {\bibinfo {volume} {10}},\ \bibinfo {pages} {461} (\bibinfo {year} {1974})}\BibitemShut {NoStop}%
\bibitem [{\citenamefont {Boughn}\ and\ \citenamefont {Rothman}(2006)}]{Boughn:2006st}%
  \BibitemOpen
  \bibfield  {author} {\bibinfo {author} {\bibfnamefont {S.}~\bibnamefont {Boughn}}\ and\ \bibinfo {author} {\bibfnamefont {T.}~\bibnamefont {Rothman}},\ }\href {\doibase 10.1088/0264-9381/23/20/006} {\bibfield  {journal} {\bibinfo  {journal} {Class. Quant. Grav.}\ }\textbf {\bibinfo {volume} {23}},\ \bibinfo {pages} {5839} (\bibinfo {year} {2006})},\ \Eprint {http://arxiv.org/abs/gr-qc/0605052} {arXiv:gr-qc/0605052} \BibitemShut {NoStop}%
\bibitem [{\citenamefont {Hogan}(2008)}]{Hogan:2007pk}%
  \BibitemOpen
  \bibfield  {author} {\bibinfo {author} {\bibfnamefont {C.~J.}\ \bibnamefont {Hogan}},\ }\href {\doibase 10.1103/PhysRevD.77.104031} {\bibfield  {journal} {\bibinfo  {journal} {Phys. Rev. D}\ }\textbf {\bibinfo {volume} {77}},\ \bibinfo {pages} {104031} (\bibinfo {year} {2008})},\ \Eprint {http://arxiv.org/abs/0712.3419} {arXiv:0712.3419 [gr-qc]} \BibitemShut {NoStop}%
\bibitem [{\citenamefont {Dyson}(2013)}]{Dyson:2013hbl}%
  \BibitemOpen
  \bibfield  {author} {\bibinfo {author} {\bibfnamefont {F.}~\bibnamefont {Dyson}},\ }\href {\doibase 10.1142/S0217751X1330041X} {\bibfield  {journal} {\bibinfo  {journal} {Int. J. Mod. Phys. A}\ }\textbf {\bibinfo {volume} {28}},\ \bibinfo {pages} {1330041} (\bibinfo {year} {2013})}\BibitemShut {NoStop}%
\bibitem [{\citenamefont {Bringmann}\ \emph {et~al.}(2023)\citenamefont {Bringmann}, \citenamefont {Domcke}, \citenamefont {Fuchs},\ and\ \citenamefont {Kopp}}]{Bringmann:2023gba}%
  \BibitemOpen
  \bibfield  {author} {\bibinfo {author} {\bibfnamefont {T.}~\bibnamefont {Bringmann}}, \bibinfo {author} {\bibfnamefont {V.}~\bibnamefont {Domcke}}, \bibinfo {author} {\bibfnamefont {E.}~\bibnamefont {Fuchs}}, \ and\ \bibinfo {author} {\bibfnamefont {J.}~\bibnamefont {Kopp}},\ }\href {\doibase 10.1103/PhysRevD.108.L061303} {\bibfield  {journal} {\bibinfo  {journal} {Phys. Rev. D}\ }\textbf {\bibinfo {volume} {108}},\ \bibinfo {pages} {L061303} (\bibinfo {year} {2023})},\ \Eprint {http://arxiv.org/abs/2304.10579} {arXiv:2304.10579 [hep-ph]} \BibitemShut {NoStop}%
\bibitem [{\citenamefont {Gr{\"a}fe}\ \emph {et~al.}(2023)\citenamefont {Gr{\"a}fe}, \citenamefont {Adamietz},\ and\ \citenamefont {Sch{\"u}tzhold}}]{Grafe:2023ngy}%
  \BibitemOpen
  \bibfield  {author} {\bibinfo {author} {\bibfnamefont {J.}~\bibnamefont {Gr{\"a}fe}}, \bibinfo {author} {\bibfnamefont {F.}~\bibnamefont {Adamietz}}, \ and\ \bibinfo {author} {\bibfnamefont {R.}~\bibnamefont {Sch{\"u}tzhold}},\ }\href {\doibase 10.1103/PhysRevD.108.064056} {\bibfield  {journal} {\bibinfo  {journal} {Phys. Rev. D}\ }\textbf {\bibinfo {volume} {108}},\ \bibinfo {pages} {064056} (\bibinfo {year} {2023})},\ \Eprint {http://arxiv.org/abs/2302.14694} {arXiv:2302.14694 [quant-ph]} \BibitemShut {NoStop}%
\bibitem [{\citenamefont {Tobar}\ \emph {et~al.}(2024)\citenamefont {Tobar}, \citenamefont {Manikandan}, \citenamefont {Beitel},\ and\ \citenamefont {Pikovski}}]{Tobar:2023ksi}%
  \BibitemOpen
  \bibfield  {author} {\bibinfo {author} {\bibfnamefont {G.}~\bibnamefont {Tobar}}, \bibinfo {author} {\bibfnamefont {S.~K.}\ \bibnamefont {Manikandan}}, \bibinfo {author} {\bibfnamefont {T.}~\bibnamefont {Beitel}}, \ and\ \bibinfo {author} {\bibfnamefont {I.}~\bibnamefont {Pikovski}},\ }\href {\doibase 10.1038/s41467-024-51420-8} {\bibfield  {journal} {\bibinfo  {journal} {Nature Commun.}\ }\textbf {\bibinfo {volume} {15}},\ \bibinfo {pages} {7229} (\bibinfo {year} {2024})},\ \Eprint {http://arxiv.org/abs/2308.15440} {arXiv:2308.15440 [quant-ph]} \BibitemShut {NoStop}%
\bibitem [{\citenamefont {Carney}\ \emph {et~al.}(2024)\citenamefont {Carney}, \citenamefont {Domcke},\ and\ \citenamefont {Rodd}}]{Carney:2023nzz}%
  \BibitemOpen
  \bibfield  {author} {\bibinfo {author} {\bibfnamefont {D.}~\bibnamefont {Carney}}, \bibinfo {author} {\bibfnamefont {V.}~\bibnamefont {Domcke}}, \ and\ \bibinfo {author} {\bibfnamefont {N.~L.}\ \bibnamefont {Rodd}},\ }\href {\doibase 10.1103/PhysRevD.109.044009} {\bibfield  {journal} {\bibinfo  {journal} {Phys. Rev. D}\ }\textbf {\bibinfo {volume} {109}},\ \bibinfo {pages} {044009} (\bibinfo {year} {2024})},\ \Eprint {http://arxiv.org/abs/2308.12988} {arXiv:2308.12988 [hep-th]} \BibitemShut {NoStop}%
\bibitem [{\citenamefont {Arani}\ \emph {et~al.}(2026)\citenamefont {Arani}, \citenamefont {Lamine},\ and\ \citenamefont {Soda}}]{Arani:2026jyz}%
  \BibitemOpen
  \bibfield  {author} {\bibinfo {author} {\bibfnamefont {F.~S.}\ \bibnamefont {Arani}}, \bibinfo {author} {\bibfnamefont {B.}~\bibnamefont {Lamine}}, \ and\ \bibinfo {author} {\bibfnamefont {J.}~\bibnamefont {Soda}},\ }\href@noop {} {\  (\bibinfo {year} {2026})},\ \Eprint {http://arxiv.org/abs/2603.27687} {arXiv:2603.27687 [quant-ph]} \BibitemShut {NoStop}%
\bibitem [{\citenamefont {Sch{\"u}tzhold}(2025)}]{Schutzhold:2025vti}%
  \BibitemOpen
  \bibfield  {author} {\bibinfo {author} {\bibfnamefont {R.}~\bibnamefont {Sch{\"u}tzhold}},\ }\href {\doibase 10.1103/xd97-c6d7} {\bibfield  {journal} {\bibinfo  {journal} {Phys. Rev. Lett.}\ }\textbf {\bibinfo {volume} {135}},\ \bibinfo {pages} {171501} (\bibinfo {year} {2025})},\ \Eprint {http://arxiv.org/abs/2502.10221} {arXiv:2502.10221 [gr-qc]} \BibitemShut {NoStop}%
\bibitem [{\citenamefont {Pang}\ and\ \citenamefont {Chen}(2018)}]{Pang:2018eec}%
  \BibitemOpen
  \bibfield  {author} {\bibinfo {author} {\bibfnamefont {B.}~\bibnamefont {Pang}}\ and\ \bibinfo {author} {\bibfnamefont {Y.}~\bibnamefont {Chen}},\ }\href {\doibase 10.1103/PhysRevD.98.124006} {\bibfield  {journal} {\bibinfo  {journal} {Phys. Rev. D}\ }\textbf {\bibinfo {volume} {98}},\ \bibinfo {pages} {124006} (\bibinfo {year} {2018})},\ \Eprint {http://arxiv.org/abs/1808.09122} {arXiv:1808.09122 [quant-ph]} \BibitemShut {NoStop}%
\bibitem [{\citenamefont {Arani}\ \emph {et~al.}(2023)\citenamefont {Arani}, \citenamefont {Harouni}, \citenamefont {Lamine},\ and\ \citenamefont {Blanchard}}]{Arani:2021oim}%
  \BibitemOpen
  \bibfield  {author} {\bibinfo {author} {\bibfnamefont {F.~S.}\ \bibnamefont {Arani}}, \bibinfo {author} {\bibfnamefont {M.~B.}\ \bibnamefont {Harouni}}, \bibinfo {author} {\bibfnamefont {B.}~\bibnamefont {Lamine}}, \ and\ \bibinfo {author} {\bibfnamefont {A.}~\bibnamefont {Blanchard}},\ }\href {\doibase 10.1088/1402-4896/acb7ac} {\bibfield  {journal} {\bibinfo  {journal} {Phys. Scripta}\ }\textbf {\bibinfo {volume} {98}},\ \bibinfo {pages} {055004} (\bibinfo {year} {2023})},\ \Eprint {http://arxiv.org/abs/2110.10962} {arXiv:2110.10962 [gr-qc]} \BibitemShut {NoStop}%
\bibitem [{\citenamefont {Arani}\ \emph {et~al.}(2025{\natexlab{a}})\citenamefont {Arani}, \citenamefont {Harouni}, \citenamefont {Lamine},\ and\ \citenamefont {Blanchard}}]{Arani:2023ljw}%
  \BibitemOpen
  \bibfield  {author} {\bibinfo {author} {\bibfnamefont {F.~S.}\ \bibnamefont {Arani}}, \bibinfo {author} {\bibfnamefont {M.~B.}\ \bibnamefont {Harouni}}, \bibinfo {author} {\bibfnamefont {B.}~\bibnamefont {Lamine}}, \ and\ \bibinfo {author} {\bibfnamefont {A.}~\bibnamefont {Blanchard}},\ }\href {\doibase 10.1088/1475-7516/2025/05/032} {\bibfield  {journal} {\bibinfo  {journal} {JCAP}\ }\textbf {\bibinfo {volume} {05}},\ \bibinfo {pages} {032} (\bibinfo {year} {2025}{\natexlab{a}})},\ \Eprint {http://arxiv.org/abs/2312.00474} {arXiv:2312.00474 [gr-qc]} \BibitemShut {NoStop}%
\bibitem [{\citenamefont {Arani}\ \emph {et~al.}(2025{\natexlab{b}})\citenamefont {Arani}, \citenamefont {Lamine}, \citenamefont {Blanchard},\ and\ \citenamefont {Harouni}}]{Arani:2025nfk}%
  \BibitemOpen
  \bibfield  {author} {\bibinfo {author} {\bibfnamefont {F.~S.}\ \bibnamefont {Arani}}, \bibinfo {author} {\bibfnamefont {B.}~\bibnamefont {Lamine}}, \bibinfo {author} {\bibfnamefont {A.}~\bibnamefont {Blanchard}}, \ and\ \bibinfo {author} {\bibfnamefont {M.~B.}\ \bibnamefont {Harouni}},\ }\href {\doibase 10.1103/r2n8-pxds} {\bibfield  {journal} {\bibinfo  {journal} {Phys. Rev. D}\ }\textbf {\bibinfo {volume} {112}},\ \bibinfo {pages} {125032} (\bibinfo {year} {2025}{\natexlab{b}})},\ \Eprint {http://arxiv.org/abs/2506.06922} {arXiv:2506.06922 [gr-qc]} \BibitemShut {NoStop}%
\bibitem [{\citenamefont {Hong}\ \emph {et~al.}(1987)\citenamefont {Hong}, \citenamefont {Ou},\ and\ \citenamefont {Mandel}}]{Hong:1987}%
  \BibitemOpen
  \bibfield  {author} {\bibinfo {author} {\bibfnamefont {C.~K.}\ \bibnamefont {Hong}}, \bibinfo {author} {\bibfnamefont {Z.~Y.}\ \bibnamefont {Ou}}, \ and\ \bibinfo {author} {\bibfnamefont {L.}~\bibnamefont {Mandel}},\ }\href {\doibase 10.1103/PhysRevLett.59.2044} {\bibfield  {journal} {\bibinfo  {journal} {Phys. Rev. Lett.}\ }\textbf {\bibinfo {volume} {59}},\ \bibinfo {pages} {2044} (\bibinfo {year} {1987})}\BibitemShut {NoStop}%
\bibitem [{\citenamefont {Grishchuk}(1975)}]{Grishchuk:1974ny}%
  \BibitemOpen
  \bibfield  {author} {\bibinfo {author} {\bibfnamefont {L.~P.}\ \bibnamefont {Grishchuk}},\ }\href@noop {} {\bibfield  {journal} {\bibinfo  {journal} {Sov. Phys. JETP}\ }\textbf {\bibinfo {volume} {40}},\ \bibinfo {pages} {409} (\bibinfo {year} {1975})}\BibitemShut {NoStop}%
\bibitem [{\citenamefont {Albrecht}\ \emph {et~al.}(1994)\citenamefont {Albrecht}, \citenamefont {Ferreira}, \citenamefont {Joyce},\ and\ \citenamefont {Prokopec}}]{Albrecht:1992kf}%
  \BibitemOpen
  \bibfield  {author} {\bibinfo {author} {\bibfnamefont {A.}~\bibnamefont {Albrecht}}, \bibinfo {author} {\bibfnamefont {P.}~\bibnamefont {Ferreira}}, \bibinfo {author} {\bibfnamefont {M.}~\bibnamefont {Joyce}}, \ and\ \bibinfo {author} {\bibfnamefont {T.}~\bibnamefont {Prokopec}},\ }\href {\doibase 10.1103/PhysRevD.50.4807} {\bibfield  {journal} {\bibinfo  {journal} {Phys. Rev. D}\ }\textbf {\bibinfo {volume} {50}},\ \bibinfo {pages} {4807} (\bibinfo {year} {1994})},\ \Eprint {http://arxiv.org/abs/astro-ph/9303001} {arXiv:astro-ph/9303001} \BibitemShut {NoStop}%
\bibitem [{\citenamefont {Ricciardone}(2017)}]{Ricciardone:2016ddg}%
  \BibitemOpen
  \bibfield  {author} {\bibinfo {author} {\bibfnamefont {A.}~\bibnamefont {Ricciardone}},\ }\href {\doibase 10.1088/1742-6596/840/1/012030} {\bibfield  {journal} {\bibinfo  {journal} {J. Phys. Conf. Ser.}\ }\textbf {\bibinfo {volume} {840}},\ \bibinfo {pages} {012030} (\bibinfo {year} {2017})},\ \Eprint {http://arxiv.org/abs/1612.06799} {arXiv:1612.06799 [astro-ph.CO]} \BibitemShut {NoStop}%
\bibitem [{\citenamefont {Wang}(2024)}]{Wang:2024gko}%
  \BibitemOpen
  \bibfield  {author} {\bibinfo {author} {\bibfnamefont {D.}~\bibnamefont {Wang}},\ }\href@noop {} {\bibfield  {journal} {\bibinfo  {journal} {arxiv}\ } (\bibinfo {year} {2024})},\ \Eprint {http://arxiv.org/abs/2407.02714} {arXiv:2407.02714 [astro-ph.CO]} \BibitemShut {NoStop}%
\bibitem [{\citenamefont {Bass}\ \emph {et~al.}(2026)\citenamefont {Bass} \emph {et~al.}}]{Bass:2026wzo}%
  \BibitemOpen
  \bibfield  {author} {\bibinfo {author} {\bibfnamefont {S.~D.}\ \bibnamefont {Bass}} \emph {et~al.},\ }\href@noop {} {\bibfield  {journal} {\bibinfo  {journal} {arxiv}\ } (\bibinfo {year} {2026})},\ \Eprint {http://arxiv.org/abs/2603.25578} {arXiv:2603.25578 [astro-ph.CO]} \BibitemShut {NoStop}%
\bibitem [{\citenamefont {Abbott}\ \emph {et~al.}(2021)\citenamefont {Abbott} \emph {et~al.}}]{KAGRA:2021kbb}%
  \BibitemOpen
  \bibfield  {author} {\bibinfo {author} {\bibfnamefont {R.}~\bibnamefont {Abbott}} \emph {et~al.} (\bibinfo {collaboration} {KAGRA, Virgo, LIGO Scientific}),\ }\href {\doibase 10.1103/PhysRevD.104.022004} {\bibfield  {journal} {\bibinfo  {journal} {Phys. Rev. D}\ }\textbf {\bibinfo {volume} {104}},\ \bibinfo {pages} {022004} (\bibinfo {year} {2021})},\ \Eprint {http://arxiv.org/abs/2101.12130} {arXiv:2101.12130 [gr-qc]} \BibitemShut {NoStop}%
\bibitem [{\citenamefont {Chandran}\ and\ \citenamefont {Shankaranarayanan}(2024)}]{Chandran:2024utf}%
  \BibitemOpen
  \bibfield  {author} {\bibinfo {author} {\bibfnamefont {S.~M.}\ \bibnamefont {Chandran}}\ and\ \bibinfo {author} {\bibfnamefont {S.}~\bibnamefont {Shankaranarayanan}},\ }\href {\doibase 10.1142/S0218271824410098} {\bibfield  {journal} {\bibinfo  {journal} {Int. J. Mod. Phys. D}\ }\textbf {\bibinfo {volume} {33}},\ \bibinfo {pages} {2441009} (\bibinfo {year} {2024})},\ \Eprint {http://arxiv.org/abs/2405.08543} {arXiv:2405.08543 [astro-ph.CO]} \BibitemShut {NoStop}%
\bibitem [{\citenamefont {Chandran}\ \emph {et~al.}(2024)\citenamefont {Chandran}, \citenamefont {Rajeev},\ and\ \citenamefont {Shankaranarayanan}}]{Chandran:2023ogt}%
  \BibitemOpen
  \bibfield  {author} {\bibinfo {author} {\bibfnamefont {S.~M.}\ \bibnamefont {Chandran}}, \bibinfo {author} {\bibfnamefont {K.}~\bibnamefont {Rajeev}}, \ and\ \bibinfo {author} {\bibfnamefont {S.}~\bibnamefont {Shankaranarayanan}},\ }\href {\doibase 10.1103/PhysRevD.109.023503} {\bibfield  {journal} {\bibinfo  {journal} {Phys. Rev. D}\ }\textbf {\bibinfo {volume} {109}},\ \bibinfo {pages} {023503} (\bibinfo {year} {2024})},\ \Eprint {http://arxiv.org/abs/2307.13611} {arXiv:2307.13611 [gr-qc]} \BibitemShut {NoStop}%
\bibitem [{\citenamefont {Gupta}(1954)}]{Gupta:1954abc}%
  \BibitemOpen
  \bibfield  {author} {\bibinfo {author} {\bibfnamefont {S.~N.}\ \bibnamefont {Gupta}},\ }\href {\doibase 10.1103/PhysRev.96.1683} {\bibfield  {journal} {\bibinfo  {journal} {Phys. Rev.}\ }\textbf {\bibinfo {volume} {96}},\ \bibinfo {pages} {1683} (\bibinfo {year} {1954})}\BibitemShut {NoStop}%
\bibitem [{\citenamefont {Feynman}(1996)}]{Feynman:1996kb}%
  \BibitemOpen
  \bibfield  {author} {\bibinfo {author} {\bibfnamefont {R.~P.}\ \bibnamefont {Feynman}},\ }\href {\doibase 10.1201/9780429502859} {\emph {\bibinfo {title} {{Feynman lectures on gravitation}}}},\ edited by\ \bibinfo {editor} {\bibfnamefont {F.~B.}\ \bibnamefont {Morinigo}}, \bibinfo {editor} {\bibfnamefont {W.~G.}\ \bibnamefont {Wagner}}, \ and\ \bibinfo {editor} {\bibfnamefont {B.}~\bibnamefont {Hatfield}}\ (\bibinfo {year} {1996})\BibitemShut {NoStop}%
\bibitem [{\citenamefont {Saulson}(1997)}]{Saulson:1997vi}%
  \BibitemOpen
  \bibfield  {author} {\bibinfo {author} {\bibfnamefont {P.~R.}\ \bibnamefont {Saulson}},\ }\href {\doibase 10.1088/0264-9381/14/9/005} {\bibfield  {journal} {\bibinfo  {journal} {Class. Quant. Grav.}\ }\textbf {\bibinfo {volume} {14}},\ \bibinfo {pages} {2435} (\bibinfo {year} {1997})}\BibitemShut {NoStop}%
\bibitem [{\citenamefont {Ma}\ \emph {et~al.}(2015)\citenamefont {Ma}, \citenamefont {Blair}, \citenamefont {Zhao},\ and\ \citenamefont {Kells}}]{Ma:2014pqa}%
  \BibitemOpen
  \bibfield  {author} {\bibinfo {author} {\bibfnamefont {Y.}~\bibnamefont {Ma}}, \bibinfo {author} {\bibfnamefont {D.~G.}\ \bibnamefont {Blair}}, \bibinfo {author} {\bibfnamefont {C.}~\bibnamefont {Zhao}}, \ and\ \bibinfo {author} {\bibfnamefont {W.}~\bibnamefont {Kells}},\ }\href {\doibase 10.1088/0264-9381/32/1/015003} {\bibfield  {journal} {\bibinfo  {journal} {Class. Quant. Grav.}\ }\textbf {\bibinfo {volume} {32}},\ \bibinfo {pages} {015003} (\bibinfo {year} {2015})},\ \Eprint {http://arxiv.org/abs/1403.3186} {arXiv:1403.3186 [physics.optics]} \BibitemShut {NoStop}%
\bibitem [{\citenamefont {Satyanarayana}(1985)}]{mvs:prd32}%
  \BibitemOpen
  \bibfield  {author} {\bibinfo {author} {\bibfnamefont {M.~V.}\ \bibnamefont {Satyanarayana}},\ }\href {\doibase 10.1103/PhysRevD.32.400} {\bibfield  {journal} {\bibinfo  {journal} {Phys. Rev. D}\ }\textbf {\bibinfo {volume} {32}},\ \bibinfo {pages} {400} (\bibinfo {year} {1985})}\BibitemShut {NoStop}%
\bibitem [{\citenamefont {Unruh}(1980)}]{Unruh1980}%
  \BibitemOpen
  \bibfield  {author} {\bibinfo {author} {\bibfnamefont {W.~G.}\ \bibnamefont {Unruh}},\ }in\ \href@noop {} {\emph {\bibinfo {booktitle} {Gravitational Radiation, Collapsed Objects and Exact Solutions}}},\ \bibinfo {editor} {edited by\ \bibinfo {editor} {\bibfnamefont {C.}~\bibnamefont {Edwards}}}\ (\bibinfo  {publisher} {Springer Berlin Heidelberg},\ \bibinfo {address} {Berlin, Heidelberg},\ \bibinfo {year} {1980})\ pp.\ \bibinfo {pages} {385--426}\BibitemShut {NoStop}%
\bibitem [{\citenamefont {Miyauchi}\ \emph {et~al.}(2026)\citenamefont {Miyauchi}, \citenamefont {Omiya}, \citenamefont {Ota}, \citenamefont {Takeda},\ and\ \citenamefont {Tanaka}}]{Miyauchi:2026cnt}%
  \BibitemOpen
  \bibfield  {author} {\bibinfo {author} {\bibfnamefont {Y.}~\bibnamefont {Miyauchi}}, \bibinfo {author} {\bibfnamefont {H.}~\bibnamefont {Omiya}}, \bibinfo {author} {\bibfnamefont {A.}~\bibnamefont {Ota}}, \bibinfo {author} {\bibfnamefont {H.}~\bibnamefont {Takeda}}, \ and\ \bibinfo {author} {\bibfnamefont {T.}~\bibnamefont {Tanaka}},\ }\href@noop {} {\  (\bibinfo {year} {2026})},\ \Eprint {http://arxiv.org/abs/2608.08803} {arXiv:2608.08803 [hep-ph]} \BibitemShut {NoStop}%
\bibitem [{\citenamefont {Vermeulen}\ \emph {et~al.}(2021)\citenamefont {Vermeulen}, \citenamefont {Aiello}, \citenamefont {Ejlli}, \citenamefont {Griffiths}, \citenamefont {James}, \citenamefont {Dooley},\ and\ \citenamefont {Grote}}]{quest}%
  \BibitemOpen
  \bibfield  {author} {\bibinfo {author} {\bibfnamefont {S.~M.}\ \bibnamefont {Vermeulen}}, \bibinfo {author} {\bibfnamefont {L.}~\bibnamefont {Aiello}}, \bibinfo {author} {\bibfnamefont {A.}~\bibnamefont {Ejlli}}, \bibinfo {author} {\bibfnamefont {W.~L.}\ \bibnamefont {Griffiths}}, \bibinfo {author} {\bibfnamefont {A.~L.}\ \bibnamefont {James}}, \bibinfo {author} {\bibfnamefont {K.~L.}\ \bibnamefont {Dooley}}, \ and\ \bibinfo {author} {\bibfnamefont {H.}~\bibnamefont {Grote}},\ }\href {\doibase 10.1088/1361-6382/abe757} {\bibfield  {journal} {\bibinfo  {journal} {Class. Quant. Grav.}\ }\textbf {\bibinfo {volume} {38}},\ \bibinfo {pages} {085008} (\bibinfo {year} {2021})},\ \Eprint {http://arxiv.org/abs/2008.04957} {arXiv:2008.04957 [gr-qc]} \BibitemShut {NoStop}%
\bibitem [{\citenamefont {Vermeulen}\ \emph {et~al.}(2025)\citenamefont {Vermeulen}, \citenamefont {Cullen}, \citenamefont {Grass}, \citenamefont {MacMillan}, \citenamefont {Ramirez}, \citenamefont {Wack}, \citenamefont {Korzh}, \citenamefont {Lee}, \citenamefont {Zurek}, \citenamefont {Stoughton},\ and\ \citenamefont {McCuller}}]{gquest}%
  \BibitemOpen
  \bibfield  {author} {\bibinfo {author} {\bibfnamefont {S.~M.}\ \bibnamefont {Vermeulen}}, \bibinfo {author} {\bibfnamefont {T.}~\bibnamefont {Cullen}}, \bibinfo {author} {\bibfnamefont {D.}~\bibnamefont {Grass}}, \bibinfo {author} {\bibfnamefont {I.~A.~O.}\ \bibnamefont {MacMillan}}, \bibinfo {author} {\bibfnamefont {A.~J.}\ \bibnamefont {Ramirez}}, \bibinfo {author} {\bibfnamefont {J.}~\bibnamefont {Wack}}, \bibinfo {author} {\bibfnamefont {B.}~\bibnamefont {Korzh}}, \bibinfo {author} {\bibfnamefont {V.~S.~H.}\ \bibnamefont {Lee}}, \bibinfo {author} {\bibfnamefont {K.~M.}\ \bibnamefont {Zurek}}, \bibinfo {author} {\bibfnamefont {C.}~\bibnamefont {Stoughton}}, \ and\ \bibinfo {author} {\bibfnamefont {L.}~\bibnamefont {McCuller}},\ }\href {\doibase 10.1103/PhysRevX.15.011034} {\bibfield  {journal} {\bibinfo  {journal} {Phys. Rev. X}\ }\textbf {\bibinfo {volume} {15}},\ \bibinfo {pages} {011034} (\bibinfo {year} {2025})}\BibitemShut {NoStop}%
\bibitem [{\citenamefont {Kanno}\ \emph {et~al.}(2026)\citenamefont {Kanno}, \citenamefont {Soda},\ and\ \citenamefont {Taniguchi}}]{Kanno:2025how}%
  \BibitemOpen
  \bibfield  {author} {\bibinfo {author} {\bibfnamefont {S.}~\bibnamefont {Kanno}}, \bibinfo {author} {\bibfnamefont {J.}~\bibnamefont {Soda}}, \ and\ \bibinfo {author} {\bibfnamefont {A.}~\bibnamefont {Taniguchi}},\ }\href {\doibase 10.1103/kv1t-j27m} {\bibfield  {journal} {\bibinfo  {journal} {Phys. Rev. Lett.}\ }\textbf {\bibinfo {volume} {136}},\ \bibinfo {pages} {061404} (\bibinfo {year} {2026})},\ \Eprint {http://arxiv.org/abs/2508.17947} {arXiv:2508.17947 [gr-qc]} \BibitemShut {NoStop}%
\bibitem [{\citenamefont {Aasi}\ \emph {et~al.}(2015)\citenamefont {Aasi} \emph {et~al.}}]{LIGOScientific:2014pky}%
  \BibitemOpen
  \bibfield  {author} {\bibinfo {author} {\bibfnamefont {J.}~\bibnamefont {Aasi}} \emph {et~al.} (\bibinfo {collaboration} {LIGO Scientific}),\ }\href {\doibase 10.1088/0264-9381/32/7/074001} {\bibfield  {journal} {\bibinfo  {journal} {Class. Quant. Grav.}\ }\textbf {\bibinfo {volume} {32}},\ \bibinfo {pages} {074001} (\bibinfo {year} {2015})},\ \Eprint {http://arxiv.org/abs/1411.4547} {arXiv:1411.4547 [gr-qc]} \BibitemShut {NoStop}%
\bibitem [{\citenamefont {Acernese}\ \emph {et~al.}(2015)\citenamefont {Acernese} \emph {et~al.}}]{VIRGO:2014yos}%
  \BibitemOpen
  \bibfield  {author} {\bibinfo {author} {\bibfnamefont {F.}~\bibnamefont {Acernese}} \emph {et~al.} (\bibinfo {collaboration} {VIRGO}),\ }\href {\doibase 10.1088/0264-9381/32/2/024001} {\bibfield  {journal} {\bibinfo  {journal} {Class. Quant. Grav.}\ }\textbf {\bibinfo {volume} {32}},\ \bibinfo {pages} {024001} (\bibinfo {year} {2015})},\ \Eprint {http://arxiv.org/abs/1408.3978} {arXiv:1408.3978 [gr-qc]} \BibitemShut {NoStop}%
\bibitem [{\citenamefont {Aso}\ \emph {et~al.}(2013)\citenamefont {Aso}, \citenamefont {Michimura}, \citenamefont {Somiya}, \citenamefont {Ando}, \citenamefont {Miyakawa}, \citenamefont {Sekiguchi}, \citenamefont {Tatsumi},\ and\ \citenamefont {Yamamoto}}]{Aso:2013eba}%
  \BibitemOpen
  \bibfield  {author} {\bibinfo {author} {\bibfnamefont {Y.}~\bibnamefont {Aso}}, \bibinfo {author} {\bibfnamefont {Y.}~\bibnamefont {Michimura}}, \bibinfo {author} {\bibfnamefont {K.}~\bibnamefont {Somiya}}, \bibinfo {author} {\bibfnamefont {M.}~\bibnamefont {Ando}}, \bibinfo {author} {\bibfnamefont {O.}~\bibnamefont {Miyakawa}}, \bibinfo {author} {\bibfnamefont {T.}~\bibnamefont {Sekiguchi}}, \bibinfo {author} {\bibfnamefont {D.}~\bibnamefont {Tatsumi}}, \ and\ \bibinfo {author} {\bibfnamefont {H.}~\bibnamefont {Yamamoto}} (\bibinfo {collaboration} {KAGRA}),\ }\href {\doibase 10.1103/PhysRevD.88.043007} {\bibfield  {journal} {\bibinfo  {journal} {Phys. Rev. D}\ }\textbf {\bibinfo {volume} {88}},\ \bibinfo {pages} {043007} (\bibinfo {year} {2013})},\ \Eprint {http://arxiv.org/abs/1306.6747} {arXiv:1306.6747 [gr-qc]} \BibitemShut {NoStop}%
\bibitem [{\citenamefont {Schnabel}\ \emph {et~al.}(2010)\citenamefont {Schnabel}, \citenamefont {Mavalvala}, \citenamefont {Mcclelland},\ and\ \citenamefont {Lam}}]{Schnabel:2010rha}%
  \BibitemOpen
  \bibfield  {author} {\bibinfo {author} {\bibfnamefont {R.}~\bibnamefont {Schnabel}}, \bibinfo {author} {\bibfnamefont {N.}~\bibnamefont {Mavalvala}}, \bibinfo {author} {\bibfnamefont {D.~E.}\ \bibnamefont {Mcclelland}}, \ and\ \bibinfo {author} {\bibfnamefont {P.~K.}\ \bibnamefont {Lam}},\ }\href {\doibase 10.1038/ncomms1122} {\bibfield  {journal} {\bibinfo  {journal} {Nature Commun.}\ }\textbf {\bibinfo {volume} {1}},\ \bibinfo {pages} {121} (\bibinfo {year} {2010})},\ \Eprint {http://arxiv.org/abs/2411.07313} {arXiv:2411.07313 [quant-ph]} \BibitemShut {NoStop}%
\bibitem [{\citenamefont {Dimopoulos}\ \emph {et~al.}(2008)\citenamefont {Dimopoulos}, \citenamefont {Graham}, \citenamefont {Hogan}, \citenamefont {Kasevich},\ and\ \citenamefont {Rajendran}}]{Dimopoulos:2008sv}%
  \BibitemOpen
  \bibfield  {author} {\bibinfo {author} {\bibfnamefont {S.}~\bibnamefont {Dimopoulos}}, \bibinfo {author} {\bibfnamefont {P.~W.}\ \bibnamefont {Graham}}, \bibinfo {author} {\bibfnamefont {J.~M.}\ \bibnamefont {Hogan}}, \bibinfo {author} {\bibfnamefont {M.~A.}\ \bibnamefont {Kasevich}}, \ and\ \bibinfo {author} {\bibfnamefont {S.}~\bibnamefont {Rajendran}},\ }\href {\doibase 10.1103/PhysRevD.78.122002} {\bibfield  {journal} {\bibinfo  {journal} {Phys. Rev. D}\ }\textbf {\bibinfo {volume} {78}},\ \bibinfo {pages} {122002} (\bibinfo {year} {2008})},\ \Eprint {http://arxiv.org/abs/0806.2125} {arXiv:0806.2125 [gr-qc]} \BibitemShut {NoStop}%
\bibitem [{\citenamefont {Badurina}\ \emph {et~al.}(2020)\citenamefont {Badurina} \emph {et~al.}}]{Badurina:2019hst}%
  \BibitemOpen
  \bibfield  {author} {\bibinfo {author} {\bibfnamefont {L.}~\bibnamefont {Badurina}} \emph {et~al.},\ }\href {\doibase 10.1088/1475-7516/2020/05/011} {\bibfield  {journal} {\bibinfo  {journal} {JCAP}\ }\textbf {\bibinfo {volume} {05}},\ \bibinfo {pages} {011} (\bibinfo {year} {2020})},\ \Eprint {http://arxiv.org/abs/1911.11755} {arXiv:1911.11755 [astro-ph.CO]} \BibitemShut {NoStop}%
\bibitem [{\citenamefont {Glauber}(1963)}]{Glauber:1963tqt}%
  \BibitemOpen
  \bibfield  {author} {\bibinfo {author} {\bibfnamefont {R.~J.}\ \bibnamefont {Glauber}},\ }\href {\doibase 10.1103/PhysRev.130.2529} {\bibfield  {journal} {\bibinfo  {journal} {Phys. Rev.}\ }\textbf {\bibinfo {volume} {130}},\ \bibinfo {pages} {2529} (\bibinfo {year} {1963})}\BibitemShut {NoStop}%
\bibitem [{\citenamefont {Loudon}(2000)}]{Loudon:tqt}%
  \BibitemOpen
  \bibfield  {author} {\bibinfo {author} {\bibfnamefont {R.}~\bibnamefont {Loudon}},\ }\href {\doibase 10.1093/oso/9780198501770.001.0001} {\emph {\bibinfo {title} {The Quantum Theory of Light}}}\ (\bibinfo  {publisher} {Oxford University Press},\ \bibinfo {year} {2000})\BibitemShut {NoStop}%
\bibitem [{\citenamefont {Fox}(2006)}]{Fox:2006qo}%
  \BibitemOpen
  \bibfield  {author} {\bibinfo {author} {\bibfnamefont {M.}~\bibnamefont {Fox}},\ }\href {\doibase 10.1093/oso/9780198566724.001.0001} {\emph {\bibinfo {title} {Quantum Optics: An Introduction}}}\ (\bibinfo  {publisher} {Oxford University Press},\ \bibinfo {year} {2006})\BibitemShut {NoStop}%
\bibitem [{\citenamefont {Crowder}\ and\ \citenamefont {Cornish}(2005)}]{Crowder:2005nr}%
  \BibitemOpen
  \bibfield  {author} {\bibinfo {author} {\bibfnamefont {J.}~\bibnamefont {Crowder}}\ and\ \bibinfo {author} {\bibfnamefont {N.~J.}\ \bibnamefont {Cornish}},\ }\href {\doibase 10.1103/PhysRevD.72.083005} {\bibfield  {journal} {\bibinfo  {journal} {Phys. Rev. D}\ }\textbf {\bibinfo {volume} {72}},\ \bibinfo {pages} {083005} (\bibinfo {year} {2005})},\ \Eprint {http://arxiv.org/abs/gr-qc/0506015} {arXiv:gr-qc/0506015} \BibitemShut {NoStop}%
\bibitem [{\citenamefont {Chen}\ \emph {et~al.}(2006)\citenamefont {Chen}, \citenamefont {Pai}, \citenamefont {Somiya}, \citenamefont {Kawamura}, \citenamefont {Sato}, \citenamefont {Kokeyama},\ and\ \citenamefont {Ward}}]{Chen:2006zra}%
  \BibitemOpen
  \bibfield  {author} {\bibinfo {author} {\bibfnamefont {Y.}~\bibnamefont {Chen}}, \bibinfo {author} {\bibfnamefont {A.}~\bibnamefont {Pai}}, \bibinfo {author} {\bibfnamefont {K.}~\bibnamefont {Somiya}}, \bibinfo {author} {\bibfnamefont {S.}~\bibnamefont {Kawamura}}, \bibinfo {author} {\bibfnamefont {S.}~\bibnamefont {Sato}}, \bibinfo {author} {\bibfnamefont {K.}~\bibnamefont {Kokeyama}}, \ and\ \bibinfo {author} {\bibfnamefont {R.~L.}\ \bibnamefont {Ward}},\ }\href {\doibase 10.1103/PhysRevLett.97.151103} {\bibfield  {journal} {\bibinfo  {journal} {Phys. Rev. Lett.}\ }\textbf {\bibinfo {volume} {97}},\ \bibinfo {pages} {151103} (\bibinfo {year} {2006})},\ \Eprint {http://arxiv.org/abs/gr-qc/0603054} {arXiv:gr-qc/0603054} \BibitemShut {NoStop}%
\bibitem [{\citenamefont {Sato}\ \emph {et~al.}(2007)\citenamefont {Sato}, \citenamefont {Kawamura}, \citenamefont {Kokeyama}, \citenamefont {Ward}, \citenamefont {Chen}, \citenamefont {Pai},\ and\ \citenamefont {Somiya}}]{Sato:2006gk}%
  \BibitemOpen
  \bibfield  {author} {\bibinfo {author} {\bibfnamefont {S.}~\bibnamefont {Sato}}, \bibinfo {author} {\bibfnamefont {S.}~\bibnamefont {Kawamura}}, \bibinfo {author} {\bibfnamefont {K.}~\bibnamefont {Kokeyama}}, \bibinfo {author} {\bibfnamefont {R.~L.}\ \bibnamefont {Ward}}, \bibinfo {author} {\bibfnamefont {Y.}~\bibnamefont {Chen}}, \bibinfo {author} {\bibfnamefont {A.}~\bibnamefont {Pai}}, \ and\ \bibinfo {author} {\bibfnamefont {K.}~\bibnamefont {Somiya}},\ }\href {\doibase 10.1103/PhysRevLett.98.141101} {\bibfield  {journal} {\bibinfo  {journal} {Phys. Rev. Lett.}\ }\textbf {\bibinfo {volume} {98}},\ \bibinfo {pages} {141101} (\bibinfo {year} {2007})},\ \Eprint {http://arxiv.org/abs/gr-qc/0608095} {arXiv:gr-qc/0608095} \BibitemShut {NoStop}%
\bibitem [{\citenamefont {Liu}\ and\ \citenamefont {Gong}(2020)}]{Liu:2020xcv}%
  \BibitemOpen
  \bibfield  {author} {\bibinfo {author} {\bibfnamefont {M.}~\bibnamefont {Liu}}\ and\ \bibinfo {author} {\bibfnamefont {B.}~\bibnamefont {Gong}},\ }\href {\doibase 10.1038/s41598-020-72850-6} {\bibfield  {journal} {\bibinfo  {journal} {Sci. Rep.}\ }\textbf {\bibinfo {volume} {10}},\ \bibinfo {pages} {16285} (\bibinfo {year} {2020})},\ \Eprint {http://arxiv.org/abs/2001.06129} {arXiv:2001.06129 [gr-qc]} \BibitemShut {NoStop}%
\bibitem [{\citenamefont {Pan}\ \emph {et~al.}(2012)\citenamefont {Pan}, \citenamefont {Chen}, \citenamefont {Lu}, \citenamefont {Weinfurter}, \citenamefont {Zeilinger},\ and\ \citenamefont {{\.Z}ukowski}}]{Pan:2011xkw}%
  \BibitemOpen
  \bibfield  {author} {\bibinfo {author} {\bibfnamefont {J.-W.}\ \bibnamefont {Pan}}, \bibinfo {author} {\bibfnamefont {Z.-B.}\ \bibnamefont {Chen}}, \bibinfo {author} {\bibfnamefont {C.-Y.}\ \bibnamefont {Lu}}, \bibinfo {author} {\bibfnamefont {H.}~\bibnamefont {Weinfurter}}, \bibinfo {author} {\bibfnamefont {A.}~\bibnamefont {Zeilinger}}, \ and\ \bibinfo {author} {\bibfnamefont {M.}~\bibnamefont {{\.Z}ukowski}},\ }\href {\doibase 10.1103/RevModPhys.84.777} {\bibfield  {journal} {\bibinfo  {journal} {Rev. Mod. Phys.}\ }\textbf {\bibinfo {volume} {84}},\ \bibinfo {pages} {777} (\bibinfo {year} {2012})},\ \Eprint {http://arxiv.org/abs/0805.2853} {arXiv:0805.2853 [quant-ph]} \BibitemShut {NoStop}%
\bibitem [{\citenamefont {Fisher}(1922)}]{Fisher1922}%
  \BibitemOpen
  \bibfield  {author} {\bibinfo {author} {\bibfnamefont {R.~A.}\ \bibnamefont {Fisher}},\ }\href {\doibase 10.1098/rsta.1922.0009} {\bibfield  {journal} {\bibinfo  {journal} {Philosophical Transactions of the Royal Society of London, Series A: Containing Papers of a Mathematical or Physical Character}\ }\textbf {\bibinfo {volume} {222}},\ \bibinfo {pages} {309} (\bibinfo {year} {1922})},\ \Eprint {http://arxiv.org/abs/https://royalsocietypublishing.org/rsta/article-pdf/222/594-604/309/265367/rsta.1922.0009.pdf} {https://royalsocietypublishing.org/rsta/article-pdf/222/594-604/309/265367/rsta.1922.0009.pdf} \BibitemShut {NoStop}%
\bibitem [{\citenamefont {Lyons}\ \emph {et~al.}(2018)\citenamefont {Lyons}, \citenamefont {Knee}, \citenamefont {Bolduc}, \citenamefont {Roger}, \citenamefont {Leach}, \citenamefont {Gauger},\ and\ \citenamefont {Faccio}}]{Lyon:sciadv.aap9416}%
  \BibitemOpen
  \bibfield  {author} {\bibinfo {author} {\bibfnamefont {A.}~\bibnamefont {Lyons}}, \bibinfo {author} {\bibfnamefont {G.~C.}\ \bibnamefont {Knee}}, \bibinfo {author} {\bibfnamefont {E.}~\bibnamefont {Bolduc}}, \bibinfo {author} {\bibfnamefont {T.}~\bibnamefont {Roger}}, \bibinfo {author} {\bibfnamefont {J.}~\bibnamefont {Leach}}, \bibinfo {author} {\bibfnamefont {E.~M.}\ \bibnamefont {Gauger}}, \ and\ \bibinfo {author} {\bibfnamefont {D.}~\bibnamefont {Faccio}},\ }\href {\doibase 10.1126/sciadv.aap9416} {\bibfield  {journal} {\bibinfo  {journal} {Science Advances}\ }\textbf {\bibinfo {volume} {4}},\ \bibinfo {pages} {eaap9416} (\bibinfo {year} {2018})},\ \Eprint {http://arxiv.org/abs/https://www.science.org/doi/pdf/10.1126/sciadv.aap9416} {https://www.science.org/doi/pdf/10.1126/sciadv.aap9416} \BibitemShut {NoStop}%
\bibitem [{\citenamefont {Rao}(1992)}]{Rao1992}%
  \BibitemOpen
  \bibfield  {author} {\bibinfo {author} {\bibfnamefont {C.~R.}\ \bibnamefont {Rao}},\ }\enquote {\bibinfo {title} {Information and the accuracy attainable in the estimation of statistical parameters},}\ in\ \href {\doibase 10.1007/978-1-4612-0919-5_16} {\emph {\bibinfo {booktitle} {Breakthroughs in Statistics: Foundations and Basic Theory}}},\ \bibinfo {editor} {edited by\ \bibinfo {editor} {\bibfnamefont {S.}~\bibnamefont {Kotz}}\ and\ \bibinfo {editor} {\bibfnamefont {N.~L.}\ \bibnamefont {Johnson}}}\ (\bibinfo  {publisher} {Springer New York},\ \bibinfo {address} {New York, NY},\ \bibinfo {year} {1992})\ pp.\ \bibinfo {pages} {235--247}\BibitemShut {NoStop}%
\bibitem [{\citenamefont {Cram{\'e}r}(1999)}]{cramer1999mathematical}%
  \BibitemOpen
  \bibfield  {author} {\bibinfo {author} {\bibfnamefont {H.}~\bibnamefont {Cram{\'e}r}},\ }\href {https://books.google.co.in/books?id=CRTKKaJO0DYC} {\emph {\bibinfo {title} {Mathematical Methods of Statistics}}},\ Princeton Landmarks in Mathematics and Physics\ (\bibinfo  {publisher} {Princeton University Press},\ \bibinfo {year} {1999})\BibitemShut {NoStop}%
\bibitem [{\citenamefont {Zhong}\ \emph {et~al.}(2018)\citenamefont {Zhong} \emph {et~al.}}]{Zhong:2018jgi}%
  \BibitemOpen
  \bibfield  {author} {\bibinfo {author} {\bibfnamefont {H.-S.}\ \bibnamefont {Zhong}} \emph {et~al.},\ }\href {\doibase 10.1103/PhysRevLett.121.250505} {\bibfield  {journal} {\bibinfo  {journal} {Phys. Rev. Lett.}\ }\textbf {\bibinfo {volume} {121}},\ \bibinfo {pages} {250505} (\bibinfo {year} {2018})},\ \Eprint {http://arxiv.org/abs/1810.04823} {arXiv:1810.04823 [quant-ph]} \BibitemShut {NoStop}%
\bibitem [{\citenamefont {Erhard}\ \emph {et~al.}(2020)\citenamefont {Erhard}, \citenamefont {Krenn},\ and\ \citenamefont {Zeilinger}}]{Erhard:2019bke}%
  \BibitemOpen
  \bibfield  {author} {\bibinfo {author} {\bibfnamefont {M.}~\bibnamefont {Erhard}}, \bibinfo {author} {\bibfnamefont {M.}~\bibnamefont {Krenn}}, \ and\ \bibinfo {author} {\bibfnamefont {A.}~\bibnamefont {Zeilinger}},\ }\href {\doibase 10.1038/s42254-020-0193-5} {\bibfield  {journal} {\bibinfo  {journal} {Nature Rev. Phys.}\ }\textbf {\bibinfo {volume} {2}},\ \bibinfo {pages} {365} (\bibinfo {year} {2020})},\ \Eprint {http://arxiv.org/abs/1911.10006} {arXiv:1911.10006 [quant-ph]} \BibitemShut {NoStop}%
\bibitem [{\citenamefont {Maggiore}(2007)}]{Maggiore:2007ulw}%
  \BibitemOpen
  \bibfield  {author} {\bibinfo {author} {\bibfnamefont {M.}~\bibnamefont {Maggiore}},\ }\href {\doibase 10.1093/acprof:oso/9780198570745.001.0001} {\emph {\bibinfo {title} {{Gravitational Waves. Vol. 1: Theory and Experiments}}}}\ (\bibinfo  {publisher} {Oxford University Press},\ \bibinfo {year} {2007})\BibitemShut {NoStop}%
\bibitem [{\citenamefont {Sathyaprakash}\ and\ \citenamefont {Schutz}(2009)}]{Sathyaprakash:2009xs}%
  \BibitemOpen
  \bibfield  {author} {\bibinfo {author} {\bibfnamefont {B.~S.}\ \bibnamefont {Sathyaprakash}}\ and\ \bibinfo {author} {\bibfnamefont {B.~F.}\ \bibnamefont {Schutz}},\ }\href {\doibase 10.12942/lrr-2009-2} {\bibfield  {journal} {\bibinfo  {journal} {Living Rev. Rel.}\ }\textbf {\bibinfo {volume} {12}},\ \bibinfo {pages} {2} (\bibinfo {year} {2009})},\ \Eprint {http://arxiv.org/abs/0903.0338} {arXiv:0903.0338 [gr-qc]} \BibitemShut {NoStop}%
\bibitem [{\citenamefont {Jin}\ and\ \citenamefont {Qiao}(2025)}]{Jin:2024mma}%
  \BibitemOpen
  \bibfield  {author} {\bibinfo {author} {\bibfnamefont {H.-B.}\ \bibnamefont {Jin}}\ and\ \bibinfo {author} {\bibfnamefont {C.-F.}\ \bibnamefont {Qiao}},\ }\href {\doibase 10.1007/s11433-024-2519-6} {\bibfield  {journal} {\bibinfo  {journal} {Sci. China Phys. Mech. Astron.}\ }\textbf {\bibinfo {volume} {68}},\ \bibinfo {pages} {220414} (\bibinfo {year} {2025})},\ \Eprint {http://arxiv.org/abs/2405.03492} {arXiv:2405.03492 [gr-qc]} \BibitemShut {NoStop}%
\bibitem [{\citenamefont {Caves}(1981)}]{Caves:1981hw}%
  \BibitemOpen
  \bibfield  {author} {\bibinfo {author} {\bibfnamefont {C.~M.}\ \bibnamefont {Caves}},\ }\href {\doibase 10.1103/PhysRevD.23.1693} {\bibfield  {journal} {\bibinfo  {journal} {Phys. Rev. D}\ }\textbf {\bibinfo {volume} {23}},\ \bibinfo {pages} {1693} (\bibinfo {year} {1981})}\BibitemShut {NoStop}%
\bibitem [{\citenamefont {Martynov}\ \emph {et~al.}(2017)\citenamefont {Martynov} \emph {et~al.}}]{Martynov:2017ufx}%
  \BibitemOpen
  \bibfield  {author} {\bibinfo {author} {\bibfnamefont {D.~V.}\ \bibnamefont {Martynov}} \emph {et~al.},\ }\href {\doibase 10.1103/PhysRevA.95.043831} {\bibfield  {journal} {\bibinfo  {journal} {Phys. Rev. A}\ }\textbf {\bibinfo {volume} {95}},\ \bibinfo {pages} {043831} (\bibinfo {year} {2017})},\ \Eprint {http://arxiv.org/abs/1702.03329} {arXiv:1702.03329 [physics.optics]} \BibitemShut {NoStop}%
\bibitem [{\citenamefont {Yu}\ \emph {et~al.}(2020)\citenamefont {Yu} \emph {et~al.}}]{LIGOScientific:2020luc}%
  \BibitemOpen
  \bibfield  {author} {\bibinfo {author} {\bibfnamefont {H.}~\bibnamefont {Yu}} \emph {et~al.} (\bibinfo {collaboration} {LIGO Scientific}),\ }\href {\doibase 10.1038/s41586-020-2420-8} {\bibfield  {journal} {\bibinfo  {journal} {Nature}\ }\textbf {\bibinfo {volume} {583}},\ \bibinfo {pages} {43} (\bibinfo {year} {2020})},\ \Eprint {http://arxiv.org/abs/2002.01519} {arXiv:2002.01519 [quant-ph]} \BibitemShut {NoStop}%
\end{thebibliography}
%\input{refs-arxiv-ver.bbl} 
 
%%%%%%%%%%%%%%%%%%%%%%%%%%%%%%%%%%%%%%%%%%%%%%%%%%%%%%%%%%%%%%%%%%%%%%%%%%%%%%

\end{document}